\documentclass[10pt,english,draftclsnofoot,onecolumn]{IEEEtran}
\usepackage[T1]{fontenc}
\usepackage[latin9]{inputenc}
\usepackage{amstext}
\usepackage{amssymb}
\usepackage{graphicx}
\usepackage{caption}
\usepackage{subfigure}
\usepackage[lined,boxed,commentsnumbered]{algorithm2e}
\usepackage{array}
\newcolumntype{L}[1]{>{\raggedright\let\newline\\\arraybackslash\hspace{0pt}}m{#1}}
\newcolumntype{C}[1]{>{\centering\let\newline\\\arraybackslash\hspace{0pt}}m{#1}}
\newcolumntype{R}[1]{>{\raggedleft\let\newline\\\arraybackslash\hspace{0pt}}m{#1}}

\makeatletter

\providecommand{\tabularnewline}{\\}

\usepackage{myPack}\usepackage{myMathPack}\usepackage{Steven}

\usepackage{myPack}\usepackage{myMathPack}

\cset{\Integers}{Z}
\cset{\Reals}{R}
\newcommand{\PosReals}{\Reals_{\geq 0}}
\cset{\Matroids}{M}

\newcommand{\IntsUpToN}{[[N]]}

\vect{\matroidRankVec}{r}
\scalar{\matroidRankFunc}{r}

\makeatother
\newtheorem{observation}{Observation}
\usepackage{babel}
\begin{document}

\title{Multilevel Diversity Coding Systems: Rate Regions, Codes, Computation, \& Forbidden Minors}

\author{%
Congduan Li,~\IEEEmembership{Student~Member, IEEE,} %
Steven Weber,~\IEEEmembership{Senior~Member, IEEE,} and %

John MacLaren Walsh, ~\IEEEmembership{Member, IEEE.}
\thanks{%
This work was funded by the National Science Foundation Award CCF--1016580. 
}%
\thanks{%
C.~Li, S.~Weber and J.~W.~Walsh are with the Department of Electrical and Computer Engineering, Drexel University, Philadelphia, PA USA (email: \textsf{congduan.li@drexel.edu}, \textsf{sweber@coe.drexel.edu}, and \textsf{jwalsh@coe.drexel.edu}).
Preliminary results were presented at Allerton 2012 \cite{CongduanAllerton2012} and NetCod 2013 \cite{CongduanNetCod2013}.
}%
}

\maketitle
\begin{abstract}
The rate regions of multilevel diversity coding systems (MDCS), a sub-class of the broader family of multi-source multi-sink networks with special structure, are investigated. After showing how to enumerate all non-isomorphic MDCS instances of a given size, the Shannon outer bound and several achievable inner bounds based on linear codes are given for the rate region of each non-isomorphic instance. For thousands of MDCS instances, the bounds match, and hence exact rate regions are proven. Results gained from these computations are summarized in key statistics involving aspects such as the sufficiency of scalar binary codes, the necessary size of vector binary codes, etc. Also, it is shown how to generate computer aided human readable converse proofs, as well as how to construct the codes for an achievability proof. Based on this large repository of rate regions, a series of results about general MDCS cases that they inspired are introduced and proved. In particular, a series of embedding operations that preserve the property of sufficiency of scalar or vector codes are presented. The utility of these operations is demonstrated by boiling the thousands of MDCS instances for which binary scalar codes are insufficient down to 12 forbidden smallest embedded MDCS instances.
\end{abstract}


\section{Introduction}

For more than a decade, network coding problems have drawn a substantial amount of attention because of the ability of a network to support more traffic when network coding is utilized instead of routing \cite{NetworkInfoFlow2000}. For a single-source multicast network, the rate region can be characterized by the max-flow min-cut of the network graph. However, the capacity regions of more general multi-source multicast networks and multiple unicast networks under network coding are still open. In \cite{YanYeungTranIT2012}, an  implicit characterization of the achievable rate region for acylic multi-source multicast networks was given in terms of $\Gamma_N^*$, the fundamental region of entropic vectors \cite{Yeungbook}.


While it is known that $\bar{\Gamma}_N^*$ is a convex cone, the exact characterization of it
is still an open problem for $N\geq4$. In fact, it has been shown that the problem of determining
the capacity regions of all networks under network coding is equivalent to that of determining $\bar{\Gamma}_N^*$ \cite{Chan2007ISIT,YanYeungTranIT2012}.
However, one can use outer
or inner bounds on $\bar{\Gamma}_{N}^{*}$ to replace it in the rate
region expression in \cite{YanYeungTranIT2012} in order to obtain outer bounds and inner bounds
for the network coding capacity region. When these bounds match for the
network in question, the capacity region has been determined exactly. For
example, in our previous work \cite{CongduanNetCod2013,CongduanAllerton2012},
we presented algorithms combining inner bounds obtained from representable
matroids, together with the Shannon outer bound, to determine coding rate
regions for multi-terminal coded networks. A merit of the method of
multiple source multicast rate region computation presented there
is that the representable matroid inner bound utilized naturally corresponds
to the use of linear codes over a specified field size, and these
codes can be reconstructed from the rate region \cite{CongduanNetCod2013}.

Recently, many researchers in the network coding community have shown interest in distributed storage systems, where network codes are used to store data on disks on different network computers in a manner which enables them to recover the system after some disk or computer failures \cite{Dimakis2013VLDB}. After a failure, the recovered disk can either contain exactly the same contents as before, in which case the problem is referred to as an exact repair problem \cite{TianISIT2013,Tian433Journalversion}, or a possibly different encoding of the data that still enables the computer array to have the same storage and recovery capabilities as before, which is known as the functional repair problem \cite{DimakisTranIT2010}. Work on these problems has shown that the fundamental design tradeoffs between storage and repair bandwidth are consequences of network coding rate regions \cite{DimakisTranIT2010,Tian433Journalversion}.

Some of the earliest models that inspired network coding and distributed storage were multi-level diversity coding systems (MDCS) \cite{HauThesis1995,RocheThesis}.  Despite being some of the oldest models for distributed storage of interest, and some of the simplest, for unsymmetric cases \cite{Roche3levelSMDCS,YeungZhangSMDCS1999}, they remain largely unsolved \cite{HauThesis1995}.  Similarly, network coding and distributed storage solutions remain highly dependent on symmetries \cite{DimakisTranIT2010,ThakorISIT2013,TianISIT2013} which will in many instances not be present in real life.  Since MDCS cases are some of the simplest asymetric networks of interest, it makes sense to study their rate regions first as one moves towards the direction of more general, and more difficult network coding and distributed storage rate region problems.


The concept of MDCS was introduced by 
Yeung \cite{Yeung1995MDC}, inspired by diversity coding systems \cite{RocheThesis}. The multiple sources in a MDCS have ordered importance with the most important sources having priority for decoding. The  sources are accessible and encoded by multiple encoders. Every decoder has access to a different subset of the encoders and wishes to reproduce the  $k$ highest priority sources, where $k$ depends on the decoder. As in most network coding and distributed storage problems, the sources are assumed to be independent. Note that this definition of MDCS is slightly different from that in the later paper \cite{FuYeungTranIT2002}, where the source variables represent some possibly correlated random variables, there is one decoder for every subset of encoders, and every decoder wishes to reproduce one source. However, the model in this paper can be taken as a special case of \cite{FuYeungTranIT2002}.

Some small MDCS instances were studied in \cite{HauThesis1995,HauYeungISIT1997} where the number of encoders and number of sources were up to 3. Additionally, in \cite{HauThesis1995,HauYeungISIT1997}, the sufficiency of superposition codes was studied, where superposition coding means encoders can encode sources separately and then simply concatenate the coded messages. It is shown in \cite{HauThesis1995,HauYeungISIT1997} that for most 2-level and 3-level 3-encoder MDCS instances, superposition coding is sufficient to obtain the same rate region as all possible codes. For the instances where superposition coding does not suffice, linear codes are constructed that achieve the parts of the fundamental rate region  that superposition coding cannot achieve.

One special type of MDCS, known as symmetrical MDCS (SMDCS), was studied in \cite{Roche3levelSMDCS,YeungZhangSMDCS1999}. In a SMDCS problem, there are $K$ sources, $K$ encoders and $2^K-1$ decoders where every decoder has access to one unique non-empty subset of the encoders. All decoders who have access to the same number of encoders will reproduce exactly the same prioritized sources. For instance, all $\left(\begin{array}{cc}|\Emc|\\ l \end{array}\right)$ decoders which have access to the size-$l$ subsets of encoders respectively, must be able to reproduce first $l$ sources, $\Xbf_{1:l}$. Due to the special structure of SMDCS, their rate regions are characterized exactly in \cite{YeungZhangSMDCS1999}. In addition, it is shown that superposition coding suffices for all SMDCS instances.

Another special type of MDCS, known as asymmetrical MDCS (AMDCS), was studied in \cite{MohTianAMDCS2010}, and is a different special case of the MDCS definition in \cite{FuYeungTranIT2002} than the one considered here. In an AMDCS problem in \cite{MohTianAMDCS2010}, $2^K-1$ sources with an ordered importance are encoded into $K$ messages. The $2^K-1$ decoders, where each has access to a non-empty subset of the $K$ messages, are mapped to levels from $1$ to $2^K-1$, i.e., are reproducing $X_1,\Xbf_{1:2},\ldots, \Xbf_{1:(2^K-1)}$, respectively. It is shown in \cite{MohTianAMDCS2010} that superposition (source separation coding) does not suffice for AMDCS when $K=3$ and coding between sources is necessary to achieve the entire rate region.

Though for special MDCS problems described above, their exact rate regions have been derived with analytical expressions, the exact rate regions of MDCS instances in general are still open.  Furthermore, in the general case, it is not clear if the rate regions obtained from the Shannon outer bound are achievable, and, if so, by simple linear codes. This work calculates bounds on rate regions of thousands of MDCS instances, and investigates the sufficiency of simple codes as well. 

\textbf{Contributions:} The primary contributions of this paper include: $i.)$ an algorithm to enumerate MDCS instances is given; $ii.)$ an analytical expression for the rate region of MDCS is obtained by adapting \cite{YanYeungTranIT2012}; $iii.)$ an indirect way to calculate MDCS rate regions by bounding them using Shannon outer bound and representable matroid inner bound on region of entropic vectors is proposed; $iv.)$ the construction of simple linear codes for achievable rate regions is given; $v.)$ rate regions and their achievability by various classes of codes for thousands of MDCS instances are proven and discussed; $vi.)$ several embedding operations are defined for which it is proved that non-sufficiency of $\Fbb_q$ linear codes is inherited by a larger MDCS instance from its smaller embedded MDCS instance; $vii.)$ using these operations, the thousands of MDCS instances for which scalar binary codes do not suffice are boiled down to 12 forbidden embedded instances; $viii.)$ an algorithm to generate converse proofs for rate regions automatically with the help of computer is given.

\textbf{Notation:} A capital letter is used to represent a random variable. For example, $X_i$ is a random variable with index $i$. A vector is represented by a solid bolded capital letter. For instance, $\mathbf{X}_{1:k}=(X_1,\ldots,X_k).$ Another commonly used notation for index set is $[[K]]=1:K$, where $1:K=\{1,2,\ldots,K\}$. A set is usually denoted by calligraphic letters like $\mathcal{D},\mathcal{E}$, etc. We use hollow bolded capital letters, e.g $\Abb$, to denote matrices with the exception that we still use $\mathbb{F},\mathbb{R}$ to represent a field and real numbers respectively.

\textbf{Organization:} The rest of this paper is organized as follows. \S\ref{sec:mdcs} states the problem model, then introduces an algorithm for enumerating MDCS instances. After that, inspired by the notion of forbidden minors from matroid theory \cite{OxleyMatroidBook}, some operations which define a notion of embedding between different MDCS instances are defined that preserve insufficiency of a class of codes. The analytical form of the rate regions of MDCS and the way to calculate it by utilizing bounds on region of entropic vectors are presented in \S\ref{sec:rateReg}. A construction method for simple codes to achieve the inner bounds is provided in \S\ref{sec:codeCons}.  In \S\ref{sec:embedding}, we investigate the sufficiency of certain class of simple linear codes, e.g, binary and ternary codes, superposition coding.   Rate regions of thousands of bigger MDCS instances that were not studied before are presented in \S\ref{sec:results}.  Based on the operations defined in \S\ref{sec:embedding}, we find the smallest MDCS instances which, when embedded in a larger MDCS instance, imply the insufficiency of $\Fbb_q$ linear codes and superposition coding.  In \S\ref{sec:cap} we describe how to generate a converse proof for a MDCS instance with a computer.   \S\ref{sec:conclusion} concludes the paper and states the directions for the future work.

\section{MDCS and Background}

\label{sec:mdcs}
\begin{figure}[t]

\centering
\includegraphics[scale=0.6]{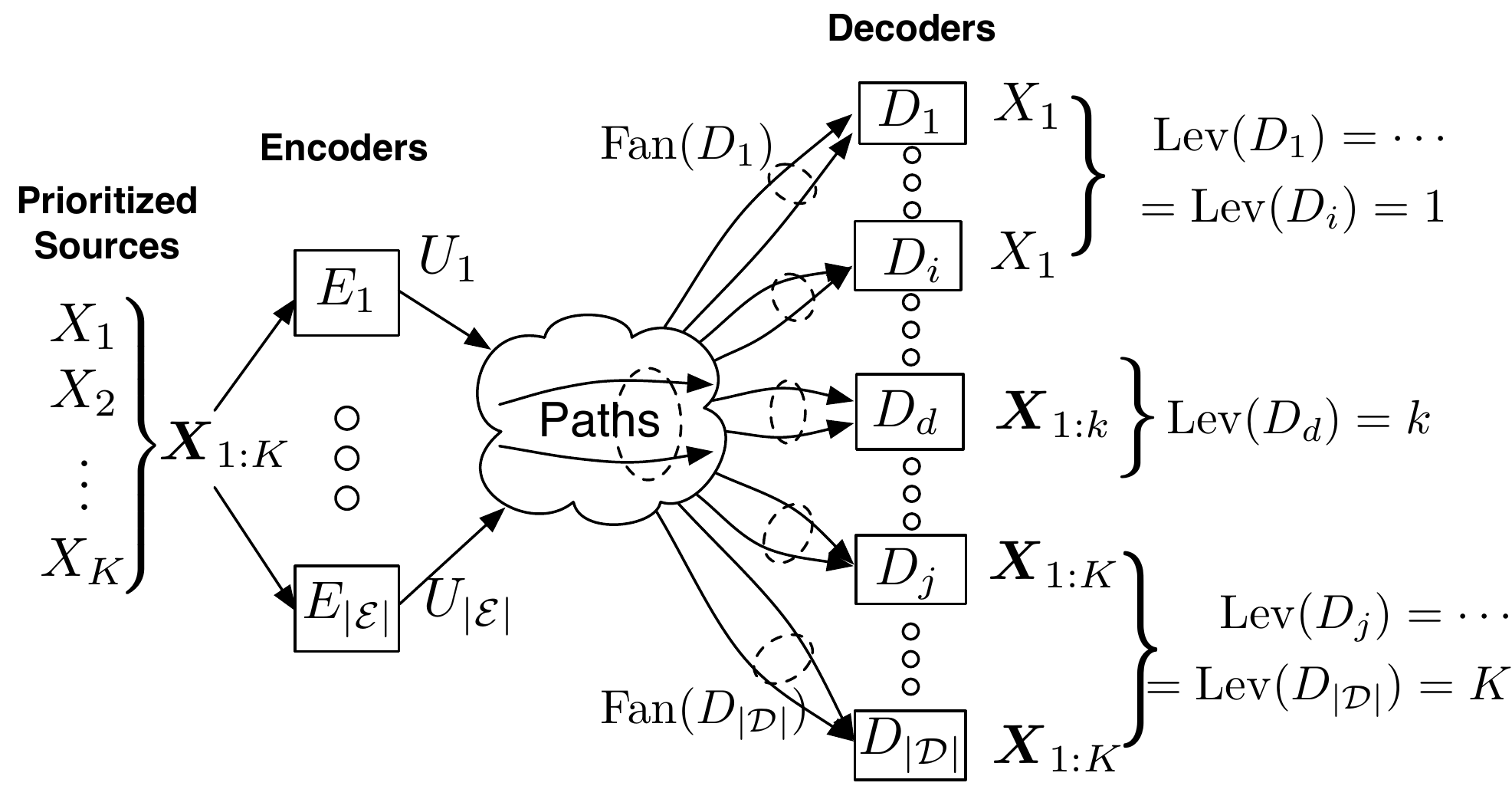}
\caption{\label{generalMDCS}Diagram of a general MDCS instance $\Asf$. Independent sources $\mathbf{X}_{1:K}$ are available to every encoder in $\mathcal{E}$. A decoder $D_d$ has access to encoders in ${\rm Fan}(D_d)$ and is able to recover the first $\mathrm{Lev}(D_d)$ sources, $\mathbf{X}_{1:\mathrm{Lev}(D_d)}$.}
\end{figure}

In a MDCS instance, as shown in Fig.\ \ref{generalMDCS} and denoted as $\Asf$, there are $K$ \emph{independent}
sources $\boldsymbol{X}_{1:K}\equiv(X_{1},\ldots,X_{K})$ where source $k$ has
support $\Xmc_{k}$, and the sources are prioritized into $K$ levels
with $X_{1}$ ($X_{K}$) the highest (lowest) priority source, respectively.
As is standard in source coding, each source $X_{k}$ is in fact an i.i.d.
sequence of random variables $\{X_{k}^{t},t=1,2,\ldots\}$ 
in $t$, and $X_{k}$ is a representative random variable with
this distribution.


All sources are made available to each of a collection of encoders 
indexed by the finite set $\Emc$. The output ${\rm Out}(E_e)$ of an encoder $E_e$ is description/message
variable $U_{e},e\in \Emc$. The message variables are mapped to a collection
of decoders  indexed by the set $\mathcal{D}$ that are classified into $K$ levels,
where a level $k$ decoder must losslessly (in the typical Shannon sense) recover source variables $\Xbf_{1:k}=(X_{1},\ldots,X_{k})$, for
each $k\in\{1,\ldots,K\}$. The mapping of encoders to decoders dictates
the description variables that are available for each decoder in this
recovery. The collection of mappings is denoted as a set $\Gmc$, $\Gmc\subseteq \Emc\times\Dmc$ of edges where $(E_e,D_d)\in \Gmc$ if $E_e$ is accessible by $D_d$. The set of encoders mapped to a particular decoder $D_d$ is called the {\it fan} of $D_d$, and is denoted as ${\rm Fan}(D_d)=\{E_e|(E_e,D_d)\in \Gmc\}$. Similarly, the set of decoders connected to a particular encoder $E_e$ is called the {\it fan} of $E_e$, and is denoted by ${\rm Fan}(E_e)=\{D_d|(E_e,D_d)\in \Gmc\}$. A decoder $D_d$ is said to be a level $k$ decoder, denoted by $\mathrm{Lev}(D_d)=k$, if it wishes to recover exclusively the first $k$ sources $\mathbf{X}_{1:k}$. Different level $k$ decoders must recover the same
subset of source variables using distinct subsets of description
variables (encoders), among which one must not be subset of another. If we denote the input and output of a level $k$ decoder $D_d$ as ${\rm In}(D_d)$ and ${\rm Out}(D_d)$, respectively, we have ${\rm In}(D_d)=\{U_e|E_e\in {\rm Fan}(D_d), \forall e\in \Emc\}$ and ${\rm Out}(D_d)=\Xbf_{1:k}$. 

We say that a MDCS instance is valid if it obeys the following constraints:
\begin{enumerate}

\item [(\textbf{C1}.)] If $\mathrm{Lev}(D_i)=\mathrm{Lev}(D_j)$, then $\mathrm{Fan}(D_i)\not\subseteq \mathrm{Fan}(D_j)$ and $\mathrm{Fan}(D_j)\not\subseteq \mathrm{Fan}(D_i)$;

\item [(\textbf{C2}.)] If $\mathrm{Lev}(D_i)>\mathrm{Lev}(D_j)$, then $\mathrm{Fan}(D_i)\not\subseteq \mathrm{Fan}(D_j)$;
\item [(\textbf{C3}.)] $\bigcup_{i\in\Dmc} {\rm Fan}(D_i)=\Emc$;
\item [(\textbf{C4}.)] There $\nexists\, k,l\in \Emc$ such that ${\rm Fan}(E_k)={\rm Fan}(E_l)$.
\item [(\textbf{C5}.)] $\forall k\in [[K]],\  \exists\, d\in \Dmc$ such that ${\rm Lev}(D_d)=k$.

\end{enumerate}

The first condition (\textbf{C1}) indicates that the fan of a decoder cannot be a subset of the fan of another decoder in the same level, for otherwise the decoder with access to more encoders would be redundant. The condition (\textbf{C2}) says that the fan of a higher level decoder cannot be a subset of the fan of a lower level decoder, for otherwise there exists a contradiction in their decoding capabilities. The condition (\textbf{C3}) requires that every encoder must be contained in the fan of at least one decoder. The condition (\textbf{C4}) requires that no two encoders have exactly the same fan, for otherwise the two encoders can be combined together. The condition (\textbf{C5}) ensures that there exists at least one decoder for every level.

\subsection{Representation of MDCS instances}
A MDCS instance with $K$ sources and $|\Emc|$ encoders, shorted as a $(K,|\Emc|)$ MDCS instance, mainly specifies the relationships between encoders and decoders, since we assume that each encoder has access to all sources. One representation of a MDCS instance is to list the fan of each decoder. Since  ${\rm Fan}(D_d)\subseteq \mathcal{E}, \forall d\in \Dmc$, one can represent ${\rm Fan}(D_d)$ using a $|\Emc|$-bit vector or a corresponding integer value, where the entries of the vector from left to right are mapped to $E_{|\Emc|},\ldots,E_1$.  With this encoding, a MDCS instance can be easily represented by a matrix where the row indices represent the level of decoders and entries are integers representing the fan of each decoder.  When some of the levels have fewer decoders than other levels, the row vector includes zeros to make them the same length.  For example, the configuration matrix for the MDCS instance shown in Fig.\ \ref{fig:main} is
\begin{equation*}
\left[
\begin{array}{cc}
0 & 1\\
0 & 3\\
5 & 6
\end{array}
\right].
\end{equation*}
This encoding refers to a $(3,3)$ MDCS instance where there is one level-1 decoder which has access to $E_1$ ($(001)_2=1$), one level-2 decoder which has access to $\{E_1,E_2\}$ ($(011)_2=3$), and two level-3 decoders which have access to $\{E_1,E_3\}$ ($(101)_2=5$) and $\{E_2,E_3\}$ ($(110)_2=6$) respectively.

Another notation for a MDCS instance that we will use extensively in this paper is the tuple $(\Xbf_{[[K]]},\Emc,\Dmc,\Lbf,\Gmc)$ where $\Xbf_{[[K]]}=(X_1,\ldots,X_K)$ represents the $K$ sources, $\Emc$ is the encoder set, $\Dmc$ is the decoder set with corresponding levels in the vector $\Lbf$, and $\Gmc$ is the set of edges between encoders and decoders which indicates the accesses of each decoder: if $D_d$ has access to $E_e$, the edge $(E_e,D_d)\in \Gmc$. For instance, the MDCS in Fig.\ \ref{fig:main} can be represented using the tuple $(\Xbf_{[[K]]},\Emc,\Dmc,\Lbf,\Gmc)$ with $K=3$, $\Emc=\{E_1,E_2,E_3\}$, $\Dmc=\{D_1,D_2,D_3,D_4\}$, $\Lbf=[1\ 2\ 3\ 3]$ and $\Gmc=\{(E_1,D_1),(E_1,D_2),(E_1,D_3),(E_2,D_2),(E_2,D_3),(E_2,D_4),(E_3,D_4)\}$.

Note that not all pairs of integers $(K,|\Emc|)$ correspond to a possible MDCS instance.  To see why, observe that if a MDCS instance has $|\Emc|$ encoders, there are at most $2^{|\Emc|}-1$ possible decoders since the fan of each decoder is a distinct subset of encoders, and there are at most $2^{|\Emc|}-1$ non-empty such subsets. Furthermore, since it is required that there exists at least one decoder for every level as shown in (\textbf{C5}), the number of sources or levels $K$ is a lower bound on the number of decoders. Thus, we have
\begin{equation} \label{ineq:up}
K\leq 2^{|\Emc|}-1.
\end{equation}
Theoretically speaking, there exist MDCS instances for any $(K,{|\Emc|})$ pair satisfying inequality \eqref{ineq:up}. 
Hence, the next natural question, which will be discussed in next subsection, is how many instances there are for a valid $(K,{|\Emc|})$ pair.

\subsection{Enumeration of non-isomorphic MDCS instances}


When counting the number of possible MDCS instances, we may not wish to distinguish two instances that are symmetric to one another.  In particular, though all sources are prioritized, one can permute the encoder variables and associated fan of decoders in a valid MDCS instance to get another symmetric valid MDCS instance. Two such instances are said to be isomorphic to one another.

\begin{definition}[Isomorphic MDCS instances]
Suppose there are two $(K,{|\Emc|})$ MDCS instances denoted as $\Asf,\Asf'$ with $\Asf=(\{X_1,\ldots,X_K\},\Emc,\Dmc,\Lbf,\Gmc)$ and $\Asf'=(\{X_1,\ldots,X_K\},\Emc',\Dmc',\Lbf',\Gmc')$ respectively. $\Asf$ and $\Asf'$ are {\it isomorphic}, denoted as $\Asf\cong \Asf'$, if and only if there exist a permutation of encoders $\Emc$, $\pi:\Emc\rightarrow \Emc'$ such that $\Emc'=\pi(\Emc)$, $\Dmc'=\Dmc$, $\Lbf'=\Lbf$, $\Gmc'=\{(\pi(E_e),D_d)|(E_e,D_d)\in \Gmc\}$. Equivalently, $\Asf'$ can be obtained by permuting encoders of $\Asf$ in the configuration.
\end{definition}

Since the isomorphism merely permutes the encoders, to study all possible MDCS instances, it suffices to consider one representative in each isomorphism class, i.e., only consider non-isomorphic MDCS instances.

The easiest way to obtain the list of non-isomorphic MDCS instances is to remove isomorphism from the list of all MDCS instances. In order to obtain all MDCS instances, we observe that the fans of decoders at the same level is a {\it Sperner family} \cite{Sperner1928} of the encoders set $\Emc$ without consideration of empty set, as required by condition (\textbf{C1}). A Sperner family of $\Emc$, sometimes also called an {\it independent system} or {\it clutter}, is a collection of subsets of $\Emc$ such that no element is contained in another. 

After including consideration for the conditions (\textbf{C2})--(\textbf{C5}), an algorithm to enumerate isomorphic and non-isomorphic MDCS instances is given in Algorithm \ref{alg:enum}, where the algorithm to augment a MDCS instance for level $l$ with a collection of Sperner families is shown in Algorithm \ref{alg:aug}. The enumeration process works as follows.

\begin{algorithm}
\SetAlgoLined
 \KwIn{Encoder index set $\Emc$, Number of sources $K$} 
 \KwOut{All non-isomorphic MDCS instances $\Mmc$, all MDCS instances with isomorphism $\Mmc'$}
 \BlankLine
 \textbf{Initialization:} List all Sperner families ${\rm Sper}(\Emc)$ of $\Emc$,  $pool={\rm Sper}(\Emc)$, $\Asf=(\Xbf_{[[K]]},\Emc,\emptyset,\emptyset,\emptyset$), $\Mmc'=\Asf$\;
 \For{$i=1:K$}{
 	$\Mmc''=\Mmc',\Mmc'=\emptyset$\;
	\For{every $\Asf\in \Mmc''$}{
		$pool={\rm Sper}(\Emc)\setminus \{\Imc\in{\rm Sper}(\Emc)|Aug(\Asf,\Imc,i)\text{ is invalid MDCS instance from conditions (\textbf{C1})--(\textbf{C5})}\}$, $\Amc=Aug(\Asf,pool,i),\Mmc'=\Mmc'\cup \Amc$\;
			}
			}			
$index=1$\;
\While{$\Mmc'\neq \emptyset$}{
$\Mmc(index)=\Mmc'(1)$;
$\Mmc'=\Mmc'\setminus\{\Imc\in\Mmc'|\Imc\cong \Mmc'(1)\}$\;
$index=index+1$;
}

 \caption{Enumerate isomorphic and non-isomorphic $(K,{|\Emc|})$ MDCS instances\label{alg:enum}}
\end{algorithm}

\begin{algorithm}
\SetAlgoLined
 \KwIn{MDCS instance $\Asf=(\Xbf_{[[K]]},\Emc,\Dmc,\Gmc)$, Sperner families $pool$, Level $l$} 
 \KwOut{All MDCS instances with elements in $pool$ as level $l$ configuration of $\Asf$: $\Amc=Aug(\Asf,pool,l)$}
 $\Amc=\emptyset$\;
 \For{every $\Imc\in pool$}{
 $\Dmc'=\Dmc\cup \{|\Dmc|+1,\ldots,|\Dmc|+|\Imc|\},\Lbf'=[\Lbf,l,\ldots,l]$ and Length($\Lbf'$)=$|\Lbf|+|\Imc|$\;
 \For{$i=1:|\Imc|$}{
$ \Gmc'=\Gmc\cup \{(E_e,D_i)|E_e\in\Imc(i)\}$\;
 }
$ \Amc=\Amc \cup (\Xbf_{[[K]]},\Emc,\Dmc',\Lbf',\Gmc');$
}
\caption{Augment a MDCS instance with Sperner families}
\label{alg:aug}
\end{algorithm}

\begin{enumerate}
\item List all Sperner families, ${\rm Sper}(\Emc)$ of the encoders set $\Emc$ (required by (\textbf{C1}));
\item All Sperner families are possible configurations for level 1, except the empty set and the whole set (when $K>1$, from (\textbf{C5}));
\item For every level $l,\ l>1$ (required by (\textbf{C5})), consider current configurations up to level $l-1$ one by one. For each configuration, remove the Sperner families from the list of all Sperner families containing at least one element that is subset of an element that is already selected in the current configuration (required by (\textbf{C2})). The remaining Sperner families are possible valid configurations for this current configuration at level $l$;
\item When $l=K$, if all encoders have been assigned access to at least one decoder (required by (\textbf{C3})), and no two encoders have the same fan (required by (\textbf{C4})), a valid MDCS instance has been obtained;
\item After step $4)$, all MDCS instances are obtained with isomorphism. Then we remove isomorphism by keeping one instance in every isomorphism class, where the MDCS instances can be obtained by permuting encoders from one another, and remove the others in the isomorphism class from the list of all MDCS instances.

\end{enumerate}

This algorithm lists both all isomorphic and all non-isomorphic $K$-level $|\Emc|$-encoder MDCS instances. Note that because the list of Sperner families is fixed, one can use the indices of this list to speed up the enumeration process. Also, two lookup tables for every Sperner family may help in the enumeration regarding the running time. One table records Sperner families who are permutations of a particular Sperner family and the other records those Sperner families which contain a subset of one of its elements. This preprocessing aids rapid isomorphism removal.

The numbers of MDCS instances for some $(K,|\Emc|)$ pairs are listed in Table \ref{MDCSlist}. 
Hau \cite{HauThesis1995} enumerated $31$ distinct MDCS instances
for the case of $K=2$ levels and
$|\Emc|=3$ encoders, and $69$ distinct instances for the case of $K=3$ levels and $|\Emc|=3$ encoders, after
symmetries are removed. However, we found that he had three redundant $(K,|\Emc|)=(2,3)$ MDCS instances because these three actually should belong to $(K,|\Emc|)=(2,2)$ MDCS instances. In addition, he missed two $(K,|\Emc|)=(3,3)$ instances and counted one $(3,2)$ instance as a valid $(3,3)$ MDCS instance. Furthermore, since we are requiring MDCS instances to satisfy (\textbf{C4}), 5 (2) instances in $(2,3)$ ($(3,3)$) that violate this condition are found. Therefore, there are $23$ and $68$ non-isomorphic instances for $(K,|\Emc|)=(2,3)$ and $(K,|\Emc|)=(3,3)$  respectively in our lists.

\begin{table}
\caption{\label{MDCSlist}List of numbers of MDCS configurations. $|{\rm Sper}(\Emc)|$ represents the number of all Sperner families of $\Emc$, $|\Mmc'|$ is the number of all isomorphic configurations, and $|\Mmc|$ is the number of all non-isomorphic configurations.}
\begin{center}
\begin{tabular}{|c|c|c|c|c|c|c|c|c|c|}
\hline
$(K,|\Emc|)$& \multicolumn{3}{c|}{$2$} & \multicolumn{3}{c|}{$3$}& \multicolumn{3}{c|}{$4$} \\ \cline{2-10}
 & $|{\rm Sper}(\Emc)|$\textsuperscript{*}& $|\Mmc'|$ & $|\Mmc|$ & $|{\rm Sper}(\Emc)|$\textsuperscript{*}& $|\Mmc'|$ & $|\Mmc|$& $|{\rm Sper}(\Emc)|$\textsuperscript{*}& $|\Mmc'|$ & $|\Mmc|$\\ \hline
$1$ & 4 & 1 & 1 & 18 & 5& 3& 166 & 76 & 13 \\ \hline
$2$ & 4 & 5 & 3 & 18 & 96& 23\textsuperscript{**} & 166 & 6145 & 445 \\ \hline
$3$ & 4 & 2& 1 & 18 & 325& 68\textsuperscript{***} & 166 & 128388 & 6803\\ \hline
\multicolumn{10}{l}{\textsuperscript{*}\footnotesize{Empty set is not considered in the Sperner family}}\\
\multicolumn{10}{l}{\textsuperscript{**}\footnotesize{\cite{HauThesis1995} counted 3 cases in (2,2) as valid (2,3) MDCS instances and listed 5 instances not satisfying (\textbf{C4})}}\\
\multicolumn{10}{l}{\textsuperscript{***}\footnotesize{\cite{HauThesis1995} counted the case in (3,2) as a valid (3,3) MDCS instance and missed two. It also listed 2 instances not satisfying (\textbf{C4}) }}
\end{tabular}
\end{center}
\end{table}

\section{Analytical form of the rate region}
\label{sec:rateReg}
In this section, we present an analytical form of the coding rate (capacity) region of MDCS instances in terms of $\Gamma_N^*$, which has similar formulation of the rate region of general multi-source multi-sink acyclic networks \cite{YanYeungTranIT2012}. In addition, we review some useful bounds on $\Gamma_N^*$ that can be used to compute the rate region.
\subsection{Expression of rate region}
As shown in Fig.\ \ref{generalMDCS}, we suppose that the source variable generated at source $k$ is $X_k$ with support alphabet $\Xmc_k$ and normalized entropy $H(X_k)=\sum_{x\in\Xmc_k}-p_x\log_q(p_x)$, for every $k\in [[K]]$. Let $\Rbf_{\Emc}=(R_{1},\ldots,R_{|\Emc|})\in\Rbb_{+}^{|\Emc|}$ be the rate vector
for the encoders, measured in number of $\Fbb_q$ digits. An $(n,\Rbf)$ block code in $\Fbb_q$ is defined as follows. For each blocklength $n\in\Nbb$
we consider a collection of $|\Emc|$ block encoders, where encoder $E_e$
maps each block of $n$ variables from each of the $K$ sources to
one of $q^{nR_{e}}$ different descriptions,
\begin{equation}
f_{e}^{(n)}:\prod_{k=1}^{K}\Xmc_{k}^{n}\to\{1,\ldots,q^{nR_{e}}\},~e\in \Emc.
\end{equation}
 The encoder outputs are indicated by $U_{e}=f_{e}^{(n)}(\Xbf_{1:K}^{1:n})$
for $e\in \Emc$. The $|\Dmc|$ decoders are characterized by their priority
level and their available descriptions. Specifically, decoder $D_d$
has an associated priority level $k\in[[K]]$ and an available subset
of the descriptions, i.e., input of fan of decoder $D_d$. In other words, decoder $D_d$ has input from its fan
$\Ubf_{{\rm In}(D_d)}=(U_{e},E_e\in {\rm Fan}(D_{d}))$ and must asymptotically losslessly
recover source variables $\Xbf_{1:k}$, 
\begin{equation} \label{eq:decfun}
g_{d}^{(n)}:\prod_{e : E_e\in {\rm Fan}(D_{d})}\{1,\ldots,q^{nR_{e}}\}\to\prod_{i=1}^{k}\Xmc_{i}^{n},~d\in \Dmc.
\end{equation}
 The rate region $\Rmc_{\Emc}$ for the configuration specified by $K$,
$\mathcal{E}$, and the encoder to decoder mappings $({\rm Fan}(D_{d}),~d\in\Dmc)$ consists
of all rate vectors $\Rbf_{\Emc}=(R_{1},\ldots,R_{|\Emc|})$ such that there
exist sequences of encoders $f^{n}=(f_{e}^{n},e\in \Emc)$ and decoders
$g^{n}=(g_{d}^{n},d\in \Dmc)$ for which the asymptotic probability of
error goes to zero in $n$ at all decoders. Specifically, define the
probability of error for each level $l_d={\rm Lev}(D_d)$ decoder $D_d$ as 
\begin{equation}
p_{d,l_d}^{n,{\rm err}}(\Rbf)=\Pbb(g_{d}(f_{e}(\Xbf_{1:K}^{1:n}),E_e\in{\rm Fan}(D_{d}))\neq \Xbf_{1:l_d}^{1:n}),
\end{equation}
and the maximum over these as 
\begin{equation}
p^{n,{\rm err}}(\Rbf)=\max_{k\in [[K]]}\max_{d : l_d=k}p_{d,k}^{n,err}.
\end{equation}
A rate vector is in the rate region, $\Rbf_{\Emc}\in\Rmc_{\Emc}$, provided there
exists $\{f_{e}^{n}\}$ and $\{g_{d}^{n}\}$ such that $p^{n,{\rm err}}(\Rbf)\to0$
as $n\to\infty$.

The rate region $\mathcal{R}_{\Emc}$ can be expressed in terms of the region
of entropic vectors, $\Gamma_{N}^{*}$ \cite{YanYeungTranIT2012}. For the
MDCS problem, collect all of the involved random variables into the set 
$\mathcal{N}=\{Y_{k},k\in [[K]]\}\cup\{U_{e},e\in \Emc\}$ and define $N=|\mathcal{N}|$,
where $Y_{k},k\in [[K]]$ is the auxiliary random variable associated with
$X_{k},k\in[[K]]$, and $U_e$ is the random variable associated with
encoder $e\in\Emc$. The rate region $\mathcal{R}_{\Emc}$ is the set of rate vectors $\mathbf{R}_{\Emc}$ such that
there exists $\mathbf{h}\in\Gamma_{N}^{*}$ satisfying the following
(see \cite{YanYeungTranIT2012}) 
\begin{eqnarray}
h_{\Ybf_{1:K}} & = & \sum_{k=1}^{K}h_{Y_{k}}\label{eq:indepSources}\\
h_{U_{e}|\Ybf_{1:K}} & = & 0,e\in \Emc\label{eq:EncFunc}\\
h_{Y_{k}} & \geq & H(X_{k}),k\in [[K]]\label{eq:sourceEnt}\\
h_{\Ybf_{1:{\rm Lev}(D_d)}|\Ubf_{{\rm In}(D_d)}} & = & 0,\, \forall d\in \Dmc \label{eq:DecFunc}\\
R_{e} & \geq & h_{U_{e}},e\in\Emc\label{eq:RateCons}
\end{eqnarray}
where the conditional entropies $h_{A|B}$ are naturally equivalent
to $h_{AB}-h_{B}$. These constraints can be interpreted as follows:
(\ref{eq:indepSources}) represents that sources are independent;
(\ref{eq:EncFunc}) represents that each description is a function
of all sources available; \eqref{eq:sourceEnt} represents that  (\ref{eq:DecFunc}) represents that the source rate constraints;
recovered source messages at a decoder are a function of the input
descriptions available to it; and (\ref{eq:RateCons}) represents
the coding rate constraints.

Define the sets  
\begin{eqnarray}
\mathcal{L}_{1} & = & \{\mathbf{h}\in\mathbb{R}_{+}^{2^{N}-1}:h_{\Ybf_{1:K}}=\sum_{i\in[[K]]}h_{Y_{i}}\}\\
\mathcal{L}_{2} & = & \{\mathbf{h}\in\mathbb{R}_{+}^{2^{N}-1}:h_{U_{e}|\Ybf_{1:K}}=0,e\in\Emc\}\\
\mathcal{L}_{4} & = & \{\mathbf{h}\in\mathbb{R}_{+}^{2^{N}-1}:h_{Y_{k}}\geq H(X_{k}),k\in[[K]]\} \label{eq:Lsourcerate}\\
\mathcal{L}_{5} & = & \{\mathbf{h}\in\mathbb{R}_{+}^{2^{N}-1}:h_{\Ybf_{1:k}|\Ubf_{{\rm In}(D_d)}}=0,d\in \Dmc\} \label{eq:dec}
\end{eqnarray}
Note that the sets $\mathcal{L}_{1},\mathcal{L}_{2},\mathcal{L}_{4},\mathcal{L}_{5}$ are corresponding to
the constraints (\ref{eq:indepSources}), (\ref{eq:EncFunc}), \eqref{eq:sourceEnt}, (\ref{eq:DecFunc}), respectively. Define 
\begin{equation}
\mathcal{R}'_{\Emc}=\mathrm{Ex}(\mathrm{Proj}_{h_{U_{e},e\in\Emc}}(\overline{\rm{con}(\Gamma_{N}^{*}\cap\mathcal{L}_{12})}\cap \mathcal{L}_4\cap\mathcal{L}_5))\label{eq:generalrateregion}
\end{equation}
 where $\mathcal{L}_{12}=\mathcal{L}_{1} \cap \mathcal{L}_{2}$, $\mathrm{Proj}_{h_{U_{e},e\in\Emc}}(\Bmc)$
is the projection of the set $\Bmc$ on the coordinates $(h_{U_{e}},e\in\Emc)$,
and $\mathrm{Ex}(\Bmc)=\{\mathbf{h}\in\mathbb{R}_{+}^{|\Emc|}:\mathbf{h}\geq\mathbf{h}'\textrm{ for some }\mathbf{h}'\in\Bmc\}$,
for $\Bmc\subset\mathbb{R}_{+}^{|\Emc|}$.  Then, $\Rmc'_{\Emc}$ is equivalent to the rate region $\Rmc_{\Emc}$.

\begin{theorem}
\label{thm:rateregion}
\begin{equation}
\mathcal{R}_{\Emc}=\mathcal{R}'_{\Emc}.
\end{equation}

\end{theorem}
In pursuing the proof of Theorem \ref{thm:rateregion}, a brief review of \cite{YanYeungTranIT2012} is presented here. In \cite{YanYeungTranIT2012}, an implicit characterization of the achievable rate region for a general acyclic multi-source multi-sink network coding problem is obtained in terms of $\Gamma_N^*$, the fundamental region of entropic vectors. The achievable rate region $\Rmc_{\Smc}$ consists of all feasible source rate vectors $\Rbf_{\Smc}$, whose part is played by the source entropies $H(X_k), k \in [[K]]$ in our formulation, for given network link capacities $R_e,e\in \Emc$, whose part is played by the encoder rates in our formulation.  In particular, the network coding capacity region expression in \cite{YanYeungTranIT2012} assumes a series of fixed network link capacities $R_e,e \in\Emc$, and calculates from this a series of possible source rates.  Here, we will no longer view the link capacities as fixed, and instead aim to obtain a series of inequalities, forming a convex cone, which link the rates $R_e, e\in\Emc$ and the source entropies $H(X_k), k \in [[K]]$ to describe the rate region for all possible rates and all possible source entropies.  To do this, we will first show how to modify the approach from \cite{YanYeungTranIT2012} to get the region of possible link capacities/ encoder rates $R_e,e\in\Emc$ for a fixed series of source entropies, then show how to further modify the expressions to get a series of inequalities which treat both $R_e, e \in \Emc$ and $H(X_k), k\in [[K]]$ as variables.

It is not difficult to find that MDCS is a special case of the general model in \cite{YanYeungTranIT2012}. The degenerated points include:
\begin{enumerate}
\item No intermediate nodes are considered in the MDCS model, while intermediate nodes are included in the model in \cite{YanYeungTranIT2012}.
\item In \cite{YanYeungTranIT2012}, the output of a decoder is an arbitrary collection of sources. However, in MDCS, the output of a decoder is specified in a particular manner such that a level $l$ decoder requires $X_1,\ldots,X_l$, the first $l$ sources.
\end{enumerate}

Due to the peculiarities of MDCS, if we delete $\Lmc_3$ in \cite{YanYeungTranIT2012}, which corresponds to the encoding function for intermediate nodes, and change their decoding constraints and functions to \eqref{eq:dec} and \eqref{eq:decfun} respectively, we can define a source rate region $\Rmc'_{\Smc}$ of MDCS for channel capacities (encoding rate capacities) $R_e,e\in \Emc$ as
\begin{equation}
\mathcal{R}'_{\Smc}=\Lambda(\mathrm{Proj}_{h_{Y_{k},k\in[[K]]}}(\overline{\rm{con}(\Gamma_{N}^{*}\cap\mathcal{L}_{12})}\cap \mathcal{L}'_4\cap\mathcal{L}_5)),\label{eq:rateregionYeung}
\end{equation}
where $\Lambda(\Amc')=\{\hbf\in\Rbb_+^{K}:\hbf\leq \hbf' \text{ for some }\hbf'\in\Amc'\}$ and $\Lmc'_4=\{\mathbf{h}\in\mathbb{R}_{+}^{2^{N}-1}:h_{U_{e}}\leq R_e,e\in\Emc\}$ corresponding to the rate constraints in \eqref{eq:RateCons}.

Then we can derive the following Corollary from Theorem 1 in \cite{YanYeungTranIT2012}.
\begin{corollary}[\cite{YanYeungTranIT2012} Theorem 1 applied to MDCS] \label{cor:Y}
\begin{equation}
\mathcal{R}_{\Smc}=\mathcal{R}'_{\Smc}.
\end{equation}

\end{corollary}

\begin{IEEEproof}
The only difference between this corollary and Theorem 1
in \cite{YanYeungTranIT2012} is that we do not have the constraint $\mathcal{L}_3$ which corresponds to the functions of intermediate nodes.
\end{IEEEproof}

Based on Corollary \ref{cor:Y}, we provide the proof of Theorem \ref{thm:rateregion}.
\begin{IEEEproof}[Proof of Theorem \ref{thm:rateregion}]

\textbf{Converse:} We need to show that for any point $\Rbf_{\Emc}=(R_1,\ldots,R_{|\Emc|})$ in the rate region $\Rmc_{\Emc}$ respect to source rates $H(X_1),\ldots,H(X_K)$, we have $\Rbf_{\Emc}\in \Rmc'_{\Emc}$, i.e., $\Rmc_{\Emc}\subseteq \Rmc'_{\Emc}$.

Since $\Rbf_{\Emc}$ is achievable, there exist encoding and decoding functions at encoders and decoders such that all decoding requirements are satisfied with source rates at $H(X_1),\ldots,H(X_K)$ and coding rates $H(U_e)\leq R_e,e\in \Emc$. In other words, if $\Rbf_{\Emc}$ is given as capacities of the encoders in question, the same encoding and decoding functions make the source rate tuple $\Rbf_{\Smc}=(H(X_1),\ldots,H(X_K))$ achievable, i.e., $\Rbf_{\Smc}\in \Rmc_{\Smc}$. From Corollary \ref{cor:Y} we see that $\Rbf_{\Smc}\in \Rmc'_{\Smc}$. Then there exists a vector $\hbf\in \overline{\rm{con}(\Gamma_{N}^{*}\cap\mathcal{L}_{12})}\cap \mathcal{L}'_4\cap\mathcal{L}_5$ such that  $\Rbf_{\Smc}={\rm Proj}_{Y_{k},k\in[[K]]}(\hbf)$. We see that $\hbf\in \Lmc_4$ since $h_{Y_k}=H(X_k),k\in [[K]]$. Therefore, we have $\hbf\in \overline{\rm{con}(\Gamma_{N}^{*}\cap\mathcal{L}_{12})}\cap \mathcal{L}_4\cap\mathcal{L}_5$. Also we know that $\hbf\in \Lmc'_4$, $h_{U_e}\leq R_e,e\in\Emc$. Then we can get $\Rbf_{\Emc}=(R_e,e\in\Emc)\in {\rm Ex}({\rm Proj}_{h_{U_e},e\in \Emc}(\overline{\rm{con}(\Gamma_{N}^{*}\cap\mathcal{L}_{12})}\cap \mathcal{L}_4\cap\mathcal{L}_5))$. Therefore, $\Rbf_{\Emc}\in \Rmc'_{\Emc}$ and further $\Rmc_{\Emc}\subseteq \Rmc'_{\Emc}$.

\textbf{Achievability:} We need to show that for any point $\Rbf'_{\Emc}\in \Rmc'_{\Emc}$, there exists a code to achieve it. Formally, $\forall \Rbf'_{\Emc}\in \Rmc'_{\Emc}$, we have $\Rbf'_{\Emc}\in \Rmc_{\Emc}$.

Suppose $\Rbf'_{\Emc}\in \Rmc'_{\Emc}$. Then there exists a vector $\hbf'\in \overline{\rm{con}(\Gamma_{N}^{*}\cap\mathcal{L}_{12})}\cap \mathcal{L}_4\cap\mathcal{L}_5$ such that  $\Rbf'_{\Emc}={\rm Proj}_{U_{e},e\in\Emc}(\hbf')$. We see that $\hbf'\in \Lmc'_4$ since $h_{U_e}= H(U_e),e\in \Emc$. Therefore, we have $\hbf'\in \overline{\rm{con}(\Gamma_{N}^{*}\cap\mathcal{L}_{12})}\cap \mathcal{L}'_4\cap\mathcal{L}_5$. According to Corollary \ref{cor:Y}, we get ${\rm Proj}_{H(X_k),k\in[[K]]}\hbf'\in \Rmc_{\Smc}$. By the proof of Corollary \ref{cor:Y} in \cite{YanYeungTranIT2012}, $\hbf'$ is achievable by some codes with capacities of $\Rbf'_{\Emc}$. Also we know that $\hbf'\in \Lmc_4$, $h_{Y_k}\geq H(X_k),k\in[[K]]$. Then we can get $\Rbf_{\Smc}=(H(X_k),k\in[[K]])\leq {\rm Proj}_{h_{Y_k},k\in [[K]]}(\hbf')$ is also achievable with capacities of $R_e=h'_{U_e},e\in \Emc$. Equivalently, if we set source rates as $H(X_k),k\in [[K]]$, there exists a code to achieve the coding rates $R_e,e\in \Emc$ using the same code. Therefore, $\Rbf'_{\Emc}$ is achievable and further $\Rmc'_{\Emc}\subseteq \Rmc_{\Emc}$.
\end{IEEEproof}


\subsection{Computation of rate region}
\begin{figure}
\center
\subfigure [When constraints cut off gap (after projection) between outer and inner bound, we have $\mathcal{R}=\mathcal{R}_{{\rm out}}=\mathcal{R}_{{\rm in}}$] {\includegraphics[scale =0.5] {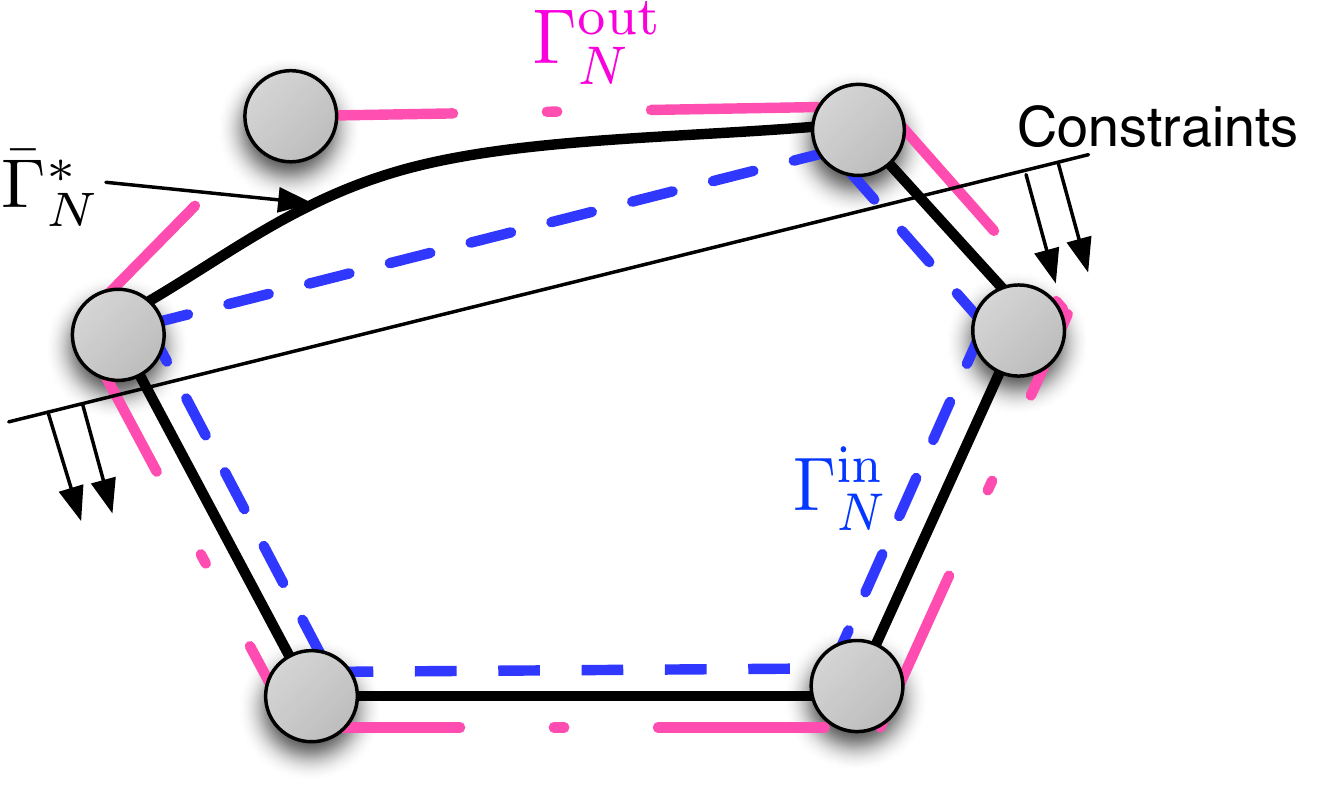}}
\subfigure [When constraints do NOT cut off gap (after projection) between outer and inner bound, we have $ \mathcal{R}_{{\rm in}} \subseteq \mathcal{R} \subseteq \mathcal{R}_{{\rm out}}$] {\includegraphics[scale =0.5] {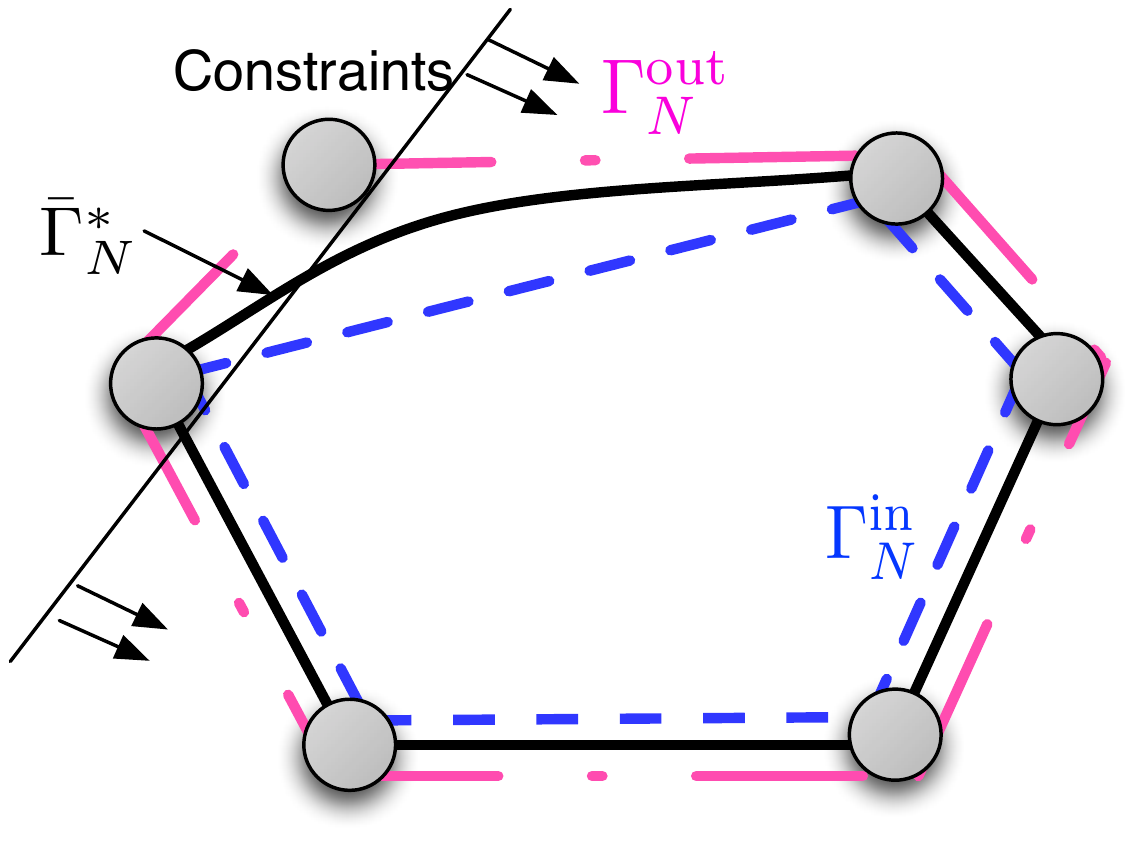}}

\caption{Illustration of computing rate region via bounding the region of entropic vectors} \label{fig:compRR}
\end{figure}

While the analytical formation gives a possible way in principle to calculate the rate region of any MDCS instance, we still have some problems. We know that $\Gamma_{N}^{*}$ is unknown and even not polyhedral for $N\geq4$.
Thus, the direct calculation of rate regions from (\ref{eq:generalrateregion})
for a MDCS instance with more than $4$ variables is infeasible. However,
replacing $\Gamma_{N}^{*}$ with polyhedral inner and outer bounds
allows (\ref{eq:generalrateregion}) to become a polyhedral computation,
which involves applying some constraints onto a polyhedra and then
projecting down onto some coordinates. This inspires us to substitute
$\Gamma_{N}^{*}$ with its closed polyhedral outer $\Gamma_N^{\rm out}$ and inner bounds $\Gamma_N^{\rm in}$
respectively, and get an outer 
\begin{equation}\label{eq:regionout}
\mathcal{R}_{{\rm out}}=\mathrm{Ex}(\mathrm{Proj}_{h_{U_{e},e\in\Emc}}(\Gamma_{N}^{{\rm out}}\cap\mathcal{L}_{1245}))
\end{equation}
and inner bound
\begin{equation}\label{eq:regionin}
\mathcal{R}_{{\rm in}}=\mathrm{Ex}(\mathrm{Proj}_{h_{U_{e},e\in\Emc}}(\Gamma_{N}^{{\rm in}}\cap\mathcal{L}_{1245}))
\end{equation}
on the rate region.
If $\mathcal{R}_{{\rm out}}=\mathcal{R}_{{\rm in}}$, we know $\mathcal{R}=\mathcal{R}_{{\rm out}}=\mathcal{R}_{{\rm in}}$. Otherwise, tighter bounds are necessary.
Fig.\ \ref{fig:compRR} illustrates these two situations.

As described previously, it is desirable to specify a rate region as a series of inequalities linking sources entropies and encoder rates. However, the equations \eqref{eq:generalrateregion}, \eqref{eq:regionout} and \eqref{eq:regionin} are functions of the given source entropies $H(X_k),k\in [[K]]$, as can be seen from the constraints $\Lmc_4$ in \eqref{eq:Lsourcerate}. 
With a slight abuse of notation, if we define
\begin{equation}
\Lmc''_4=\{(\mathbf{h}^T,\Rbf^T)^T\in\mathbb{R}_{+}^{2^N-1+|\Emc|}:R_e\geq h_{U_{e}},e\in\Emc\}, \label{eq:Lratefree}
\end{equation} 
the rate regions in \eqref{eq:generalrateregion}, \eqref{eq:regionout} and \eqref{eq:regionin}, which are expressed exclusively in terms of the variables $R_e,H(X_k),e\in\Emc,k\in [[K]]$, will be
\begin{eqnarray}
\Rmc_{\Emc}&=&\mathrm{Proj}_{R_{e},e\in\mathcal{E},H(X_k),k\in [[K]]}(\overline{\rm{con}(\Gamma_{N}^{*}\cap\mathcal{L}_{12})}\cap\mathcal{L}_5\cap \mathcal{L}''_4)\label{eq:generalrateregionfree}\\
\mathcal{R}_{{\rm out}}&=&\mathrm{Proj}_{R_{e},e\in\mathcal{E},H(X_k),k\in [[K]]}(\Gamma_{N}^{{\rm out}}\cap\mathcal{L}_{125}\cap\Lmc''_4)\label{eq:regionoutfree}\\
\mathcal{R}_{{\rm in}}&=&\mathrm{Proj}_{R_{e},e\in\mathcal{E},H(X_k),k\in [[K]]}(\Gamma_{N}^{{\rm in}}\cap\mathcal{L}_{125}\cap\Lmc''_4).\label{eq:regioninfree}
\end{eqnarray}
Here, the projection on $H(X_k)$ is in fact projection on $\hbf_{Y_k}$ since we let $H(X_k)=\hbf_{Y_k},k\in [[K]]$. Note that, $\Lmc''_4$ introduces new variables and is a cone in $2^N-1+|\Emc|$ dimensional space. In addition, since $\bar{\Gamma}_N^*,\Lmc_{1,2,5}$ do not have dimensions on $R_e,e\in \Emc$, we add these free dimensions in the intersection.

In this work, we will follow \eqref{eq:regionoutfree} and \eqref{eq:regioninfree} to calculate the rate region. Typically, the Shannon outer bound $\Gamma_N$ and some inner bounds obtained from matroids, especially representable matroids, are used. We will introduce these bounds in the next subsection. For details on the polyhedral computation methods used to obtain  these bounds, interested readers are referred to \cite{CongduanNetCod2013,CongduanAllerton2012,JayantISIT2014}. 

\subsection{Construction of bounds on rate region}

We pass now to discussing the bounds on the region of entropic vectors
utilized in our work. We would like to first review the region of entropic vectors.

\subsubsection{Region of entropic vectors $\Gamma_{N}^{*}$}

Consider an arbitrary collection $\mathbf{X}=(X_{1},\ldots,X_{N})$
of $N$ discrete random variables with joint probability mass function
$p_{X}$. To each of the $2^{N}-1$ non-empty subsets of the collection
of random variables, $X_{\Amc}:=(X_{i}\left|i\in\Amc\right.)$
with $\Amc\subseteq\{1,\ldots,N\}\equiv[[N]]$, there is associated
a joint Shannon entropy $H(X_{\Amc})$. Stacking these subset
entropies for different subsets into a $2^{N}-1$ dimensional vector
we form an entropy vector 
\begin{equation}
\mathbf{h}=\left[H(X_{\Amc})\left|\Amc\subseteq[[N]],\Amc\neq\emptyset\right.\right]
\end{equation}
 By virtue of having been created in this manner, the vector $\mathbf{h}$
must live in some subset of $\mathbb{R}_{+}^{2^{N}-1}$, and is said
to be \emph{entropic} due to the existence of $p_{X}$. However, not
every point in $\mathbb{R}_{+}^{2^{N}-1}$ is entropic since for some
many points, there does not exist an associated valid distribution $p_{X}$.
The set of all entropic vectors form a region, denoted as $\Gamma_{N}^{*}$.  It is known that the closure of the region of entropic vectors $\bar{\Gamma}_{N}^{*}$ is a convex cone \cite{YanYeungTranIT2012}.

\subsubsection{Shannon outer bound $\Gamma_{N}$}

Next observe that elementary properties of Shannon entropies indicates
that $H(X_{\Amc})$ is a non-decreasing submodular function,
so that $\forall\Amc\subseteq\Bmc\subseteq[[N]],\forall\mathcal{C},\mathcal{D}\subseteq[[N]]$
\begin{eqnarray}
H(X_{\Amc}) & \leq & H(X_{\Bmc})\label{eq:nonDec}\\
H(X_{\mathcal{C}\cup\mathcal{D}})+H(X_{\mathcal{C}\cap\mathcal{D}}) & \leq & H(X_{\mathcal{C}})+H(X_{\mathcal{D}}).\label{eq:submod}
\end{eqnarray}
 Since they are true for any collection of subset entropies, these
linear inequalities (\ref{eq:nonDec}), (\ref{eq:submod}) can be
viewed as supporting halfspaces for $\Gamma_{N}^{*}$.

Thus, the intersection of all such inequalities form a polyhedral
outer bound $\Gamma_{N}$ for $\Gamma_{N}^{*}$ and $\bar{\Gamma}_{N}^{*}$,
where 
\[
\Gamma_{N}:=\left\{ \mathbf{h}\in\mathbb{R}_{\geq 0}^{2^{N}-1}\left|\begin{array}{c}
h_{\Amc}\leq h_{\Bmc}\quad\forall\Amc\subseteq\Bmc\\
h_{\mathcal{C}\cup\mathcal{D}}+h_{\mathcal{C}\cap\mathcal{D}}\leq h_{\mathcal{C}}+h_{\mathcal{D}}\quad\forall\mathcal{C},\mathcal{D}
\end{array}\right.\right\} .
\]
 This outer bound $\Gamma_{N}$ is known as the \emph{Shannon outer
bound}, as it can be thought of as the set of all inequalities resulting
from the positivity of Shannon's information measures among the random
variables. Fujishige observed in 1978 \cite{Fujishige_IC_78} that
the entropy function for a collection of random variables $(X_{i},i\in[N])$
viewed as a set function is a polymatroid rank function, where a set
function $\rho:2^{\mathcal{S}}\to\Rbb_{+}$ is a rank function of a polymatroid
if it obeys the following axioms: 
\begin{enumerate}
\item Normalization: $\rho(\emptyset)=0$; 
\item Monotonicity: if $\Amc\subseteq \Bmc\subseteq \mathcal{S}$ then $\rho(\Amc)\leq\rho(\Bmc)$; 
\item Submodularity: if $\Amc,\Bmc\subseteq \mathcal{S}$ then $\rho(\Amc\cup \Bmc)+\rho(\Amc\cap \Bmc)\leq\rho(\Amc)+\rho(\Bmc)$. 
\end{enumerate}
While $\Gamma_{2}=\Gamma_{2}^{*}$ and $\Gamma_{3}=\bar{\Gamma}_{3}^{*}$,
$\bar{\Gamma}_{N}^{*}\subsetneq\Gamma_{N}$ for all $N\geq4$ \cite{YanYeungTranIT2012},
and indeed it is known \cite{Matus_ISIT_2007} that $\bar{\Gamma}_{N}^{*}$
is not even polyhedral for $N\geq4$.

\subsubsection{\label{sec:Matroid-theory}Matroid basics}

Matroid theory \cite{OxleyMatroidBook} is an abstract generalization
of the independence in the context of linear algebra to the more general
setting of set systems, i.e., collections of subsets of a ground set
obeying certain axioms. The ground set of size $N$ is without loss
of generality $\mathcal{S}=[[N]]$, and in our context each element of the ground
set will correspond to a random variable. There are numerous equivalent
definitions of matroids; we first present one commonly used in terms
of independent sets. 
\begin{definition} \cite{OxleyMatroidBook}
A matroid $M$ is an ordered pair $(\mathcal{S},\mathcal{I})$ consisting of
a finite set $\mathcal{S}$ (the ground set) and a collection $\mathcal{I}$
(called independent sets) of subsets of $S$ obeying: 
\begin{enumerate}
\item Normalization: $\emptyset\in\mathcal{I}$; 
\item Heredity: If $I\in\mathcal{I}$ and $I'\subseteq I$, then $I'\in\mathcal{I}$; 
\item Independence augmentation: If $I_{1}\in\mathcal{I}$ and $I_{2}\in\mathcal{I}$
and $|I_{1}|<|I_{2}|$, then there is an element $e\in I_{2}-I_{1}$
such that $I_{1}\cup\{e\}\in\mathcal{I}$. 
\end{enumerate}
\end{definition} 
Another common (and equivalent) definition of matroids utilizes {\em{rank
functions}}. For a matroid $M=(\mathcal{S},\Imc)$ with $|\mathcal{S}|=N$ the rank function $r:2^{\mathcal{S}}\to\{0,\ldots,N\}$ is defined as the size of the
largest independent set contained in each subset of $\mathcal{S}$, i.e., $r_M(\Amc)=\max_{\Bmc\subseteq \Amc}\{|\Bmc|:\Bmc\in\Imc\}$.
The rank of a matroid, $r_{M}$, is the rank of the ground set, $r_{M}=r_M(\mathcal{S})$.
The rank function of a matroid can be shown to obey the following
properties. In fact these properties may instead be viewed as an alternate
definition of a matroid in that any set function obeying these axioms
is the rank function of a matroid. 
\begin{definition} \label{def:mtrk}
A set function $r:2^{\mathcal{S}}\to\{0,\ldots,N\}$ is a rank function of a matroid if it obeys the following axioms: 
\begin{enumerate}
\item Cardinality: $r_M(\Amc)\leq|\Amc|$; 
\item Monotonicity: if $\Amc\subseteq \Bmc\subseteq \mathcal{S}$ then $r_M(\Amc)\leq r_M(\Bmc)$; 
\item Submodularity: if $\Amc,\Bmc\subseteq \mathcal{S}$ then $r_M(\Amc\cup \Bmc)+r_M(\Amc\cap \Bmc)\leq r_M(\Amc)+r_M(\Bmc)$. 
\end{enumerate}
\end{definition} 
There are many operations on matroids, such as {\em{contraction}}
and {\em{deletion}}. Details about these operations can be found
in \cite{OxleyMatroidBook}. Next we give the definition of the 
important concept of a matroid {\em{minor}} based on these two operations.

\begin{definition} 
If $M$ is a matroid on $\mathcal{S}$ and $\mathcal{T}\subseteq \mathcal{S}$,
a matroid $M'$ on $\mathcal{T}$ is called a \textit{minor }of $M$ if $M'$
is obtained by any combination of deletion ($\backslash$) and contraction
($/$) of $M$. 
\end{definition} 

The operations of deletion and contraction
mentioned in the definition yield new matroids with new rank functions for the minors.
Specifically, let $M/\mathcal{T}$ denote the matroid obtained by contraction
of $M$ on $\mathcal{T}\subset \mathcal{S}$, and let $M\setminus \mathcal{T}$ denote the matroid
obtained by deletion from $M$ of $\mathcal{T}\subset \mathcal{S}$. Then, by \cite{OxleyMatroidBook}
(3.1.5,7), $\forall \mathcal{X}\subseteq \mathcal{S}\setminus \mathcal{T}$ 
\begin{equation}
\begin{array}{rcl}
r_{M/\mathcal{T}}(\mathcal{X}) & = & r_{M}(\mathcal{X}\cup \mathcal{T})-r_{M}(\mathcal{T})\\
r_{M\setminus \mathcal{T}}(\mathcal{X}) & = & r_{M}(\mathcal{X})
\end{array}\label{eq:condel}
\end{equation}

To each set function $\matroidRankFunc_{N}:2^{\IntsUpToN}\rightarrow\PosReals$
we will associate a vector $\matroidRankVec_{N}\in\PosReals^{2^{N}-1}$
formed by stacking the various values of the function $\matroidRankFunc_{N}$
into a vector, e.g.., in a manner associated with a binary counter, such as
\begin{equation}
\matroidRankVec=\left[\begin{array}{c}
\matroidRankFunc(\{1\})\ 
\matroidRankFunc(\{2\})\
\matroidRankFunc(\{1,2\})\
\matroidRankFunc(\{3\})\
\matroidRankFunc(\{1,3\})\
\matroidRankFunc(\{2,3\})\
\matroidRankFunc(\{1,2,3\})\
\matroidRankFunc(\{4\})\
\cdots
\end{array}\right]^T.
\end{equation}
 For any such function, $\matroidRankFunc_{N}$, and hence for any
such vector $\matroidRankVec_{N}$, for any pair of sets $\Amc,\Bmc\subseteq\IntsUpToN$
we will define the \emph{minor} associated with deleting $\IntsUpToN\setminus(\Amc\cup\Bmc)$
and contracting on $\Amc$ as 
\begin{equation}
\matroidRankFunc_{\Bmc|\Amc}(\mathcal{C}):=\matroidRankFunc(\mathcal{C}\cup\Amc)-\matroidRankFunc(\Amc)\quad\forall\mathcal{C}\subseteq\Bmc
\end{equation}
 and $\matroidRankVec_{\Bmc|\Amc}$ will denote the
associated vector, ordered again by a binary counter whose bit positions
are created by enumerating again the elements of $\Bmc$ (i.e.
keeping the same order). If $\matroidRankFunc$ is the rank function
of a matroid, this definition is consistent with the definition of
the matroid operations of taking a minor, deleting and contracting,
although we will apply them to any real valued set function here.

Though there are many classes of matroids, we are especially interested
in one of them, {\em{representable matroids}}, because they can
be related to linear codes to solve network coding problems, as discussed
in \cite{CongduanNetCod2013,CongduanAllerton2012}.

\subsubsection{Representable matroids}

Representable matroids are an important class of matroids which connect
the independent sets to the conventional notion of independence in
a vector space. 

\begin{definition} A matroid $M$ with ground set
$S$ of size $|S|=N$ and rank $r_{M}=r$ is representable over a
field $\mathbb{F}$ if there exists a matrix $\Abb\in\mathbb{F}^{r\times N}$
such that for each independent set $I\in\mathcal{I}$ the corresponding
columns in $\Abb$, viewed as vectors in $\mathbb{F}^{r}$, are linearly
independent. 
\end{definition} 

There has been significant effort towards
characterizing the set of matroids that are representable over various
field sizes, with a complete answer only available for fields of sizes
two, three, and four. For example, the characterization of binary
representable matroids due to Tutte is: 
A matroid $M$ is binary representable (representable over a binary
field) iff it does not have the matroid $U_{2,4}$ as a minor. 
Here, $U_{k,N}$ is the {\em uniform} matroid on the ground set
$S=[N]$ with independent sets $\mathcal{I}$ equal to all subsets
of $[N]$ of size at most $k$. For example, $U_{2,4}$ has as its
independent sets 
\begin{equation}
\mathcal{I}=\{\emptyset,1,2,3,4,\{1,2\},\{1,3\},\{1,4\},\{2,3\},\{2,4\},\{3,4\}\}.
\end{equation}

Another important observation is that the first non-representable
matroid is {\em{Vámos}} matroid, a well known matroid on ground
set of size $8$. That is to say, all matroids are representable,
at least in some field, for $N\leq7$.

\subsubsection{Inner bounds from representable matroids}

Suppose a matroid $M$ with ground set $\Smc$ of size $|\Smc|=N$
and rank $r_M(\Smc)=k$ is representable over the finite field $\Fbb_{q}$
of size $q$ and the representing matrix is $\Abb\in\Fbb_{q}^{k\times N}$
such that $\forall \Bmc\subseteq\Smc$ $r_M(\Bmc)=\textrm{rank}(\Abb_{:,\Bmc})$,
the matrix rank of the columns of $\Abb$ indexed by $B$. Let $\Gamma_{N}^{q}$
be the conic hull of all rank functions of matroid with $N$ elements
and representable in $\Fbb_{q}$. This provides an inner bound $\Gamma_{N}^{q}\subseteq\bar{\Gamma}_{N}^{*}$,
because any extremal rank function $r$ of $\Gamma_{N}^{q}$ is by
definition representable and hence is associated with a matrix representation
$\Abb\in\Fbb_{q}^{k\times N}$, from which we can create the random
variables 
\begin{equation}
(X_{1},\ldots,X_{N})=\ubf\Abb,~~\ubf\sim\mathcal{U}(\Fbb_{q}^{k}).\label{eq:entVecRepMat}
\end{equation}
 whose elements have entropy for each subset $\Amc$ of $h_{\Amc}=r_M(\Amc)\log_{2}\ q$, for $\Amc\subseteq\Smc$.
Hence, all extreme rays of $\Gamma_{N}^{q}$ are entropic, and $\Gamma_{N}^{q}\subseteq\bar{\Gamma}_{N}^{*}$. Further, as will be discussed in \S\ref{sec:codeCons}, if a vector in the rate region of a network is (projection of) a $\Fbb_q$-representable matroid rank, the representation $\Abb$ can be used as a linear code to achieve that rate vector and this code is denoted as a {\it scalar $\Fbb_q$ code}.

One can further generalize the relationship between representable matroids and entropic vectors established by (\ref{eq:entVecRepMat}) by partitioning $\Smc=\{1,\ldots,N'\}$ up into $N$ disjoint sets, $\Smc_{1},\ldots,\Smc_{N}$ and defining for $n\in\{1,\ldots,N\}$ the new vector-valued, random variables $\Xbf'_{n}=[X_{n}|n\in \Smc_{n'}]$. The associated entropic vector will have entropies $h_{A}=r_M(\cup_{n\in A}\Smc_{n})\log_{2}q$, and is thus proportional to a \emph{projection} of the original rank vector $\rbf$ keeping only those elements corresponding to all elements in a set in the partition appearing together. Thus, such a projection of $\Gamma_{N'}^{q}$ forms an inner bound to $\bar{\Gamma}_{N}^{*}$, which we will refer to as a \emph{vector representable matroid inner bound} $\Gamma_{N,N'}^{q}$. As $N' \to \infty$, $\Gamma_{N,\infty}^q$ is the conic hull of all ranks of subspaces on $\Fbb_q$. The union over all field sizes for $\Gamma_{N,\infty}^q$ is the conic hull of the set of ranks of subspaces. Similarly, if a vector in the rate region of a network is (projection of) a vector $\Fbb_q$-representable matroid rank, the representation $\Abb$ can be used as a linear code to achieve that rate vector and this code is denoted as a {\it vector $\Fbb_q$ code}, as we will explain in next section.


Each of the bounds discussed in this section could be used in equation \eqref{eq:regionoutfree} and \eqref{eq:regioninfree} to calculate bounds on the rate region for an MDCS instance $\Asf$. If we substitute the Shannon outer bound $\Gamma_N$ into \eqref{eq:regionoutfree}, we get
\begin{equation}
\mathcal{R}_{{\rm out}}(\Asf)=\mathrm{Proj}_{R_{e},e\in\mathcal{E},H(X_k),k\in [[K]]}(\Gamma_{N}\cap\mathcal{L}_{125}\cap\Lmc''_4).\label{eq:Shannonfree}
\end{equation}

Similarly, when representable matroid inner bound $\Gamma_N^q$ and the vector representable matroid inner bound $\Gamma_{N,\infty}^q$ are substituted into \eqref{eq:regioninfree}, we get
\begin{eqnarray}
\mathcal{R}_{s,q}(\Asf)&=&\mathrm{Proj}_{R_{e},e\in\mathcal{E},H(X_k),k\in [[K]]}(\Gamma_{N}^{q}\cap\mathcal{L}_{125}\cap\Lmc''_4),\label{eq:rsq}\\
\mathcal{R}_{q}(\Asf)&=&\mathrm{Proj}_{R_{e},e\in\mathcal{E},H(X_k),k\in [[K]]}(\Gamma_{N,\infty}^{q}\cap\mathcal{L}_{125}\cap\Lmc''_4).
\end{eqnarray}

If $\Gamma_{N,N'}^q$ is substituted in \eqref{eq:regioninfree}, the inner bound obtained on the rate region is denoted as $\Rmc_q^{N,N'}(\Asf)$.

\subsubsection{Superposition Coding Rate Region}
Another important inner bound for the rate region is the superposition coding rate region. In superposition coding, in each encoder, sources are coded independently from one another, and the output of each encoder is the concatenation of the separately coded messages across the different sources.  Since sources are coded separately, the coding rate of each encoder is simply the sum of its separated coding rates for each source. The minimum separated coding rate for each source is determined by the max-flow min-cut bound \cite{Yeungbook}, since this is equivalent to single-source multicast network coding. In particular, suppose a decoder $D_d$ has input $\{U_e|e\in{\rm Fan}(D_d)\}$ and recovers $X_1,\ldots,X_k$, then we have
\begin{eqnarray}
R_e&=&\sum_{i=1}^K R_e^{X_i},e\in{\rm Fan}(D_d),\label{eq:sp1}\\
\sum_{e\in {\rm Fan}(D_d)} R_e^{X_i}&\geq& H(X_i),i=1,\ldots,k.\label{eq:sp2}
\end{eqnarray}

Let $\rbf=(R_e,H(X_k),R_e^{X_k}|e\in\Emc,k\in[[K]])$ be the stacking of all of the involved variables in these inequalities. After considering all decoders in $\Dmc$, we project the polyhedron defined by \eqref{eq:sp1}, \eqref{eq:sp2} onto the dimensions of $R_e,H(X_k),e\in \Emc,k\in[[K]]$ to get the superposition coding rate region. That is, the superposition coding rate region $\Rmc_{sp}(\Asf)$ for a MDCS instance $\Asf$ is
\begin{equation}
\Rmc_{sp}(\Asf):={\rm Proj}_{R_e,H(X_k),e\in\Emc,k\in[[K]]}\left\{\rbf\in\Rbb_{\geq 0}^{|\Emc|+K+K|\Emc|}\left| \begin{array}{c}
\displaystyle R_e=\sum_{i=1}^K R_e^{X_i},e\in\Emc,\\
\displaystyle\sum_{e\in {\rm Fan}(D_d)} R_e^{X_i}\geq H(X_i),i=1,\ldots,{\rm Lev}(D_d), \forall d\in\Dmc.
\end{array}\right.
\right\}.\label{eq:speff}
\end{equation}

While equation \eqref{eq:speff} is the form of the superposition rate region most amenable to computation, we will also give a different but equivalent form for the ease of proving Theorem \ref{thm:EncUniCont} and Theorem \ref{thm:EncUnif} in \S\ref{sec:embedding}. First, we modify the decoding constraints $\Lmc_5$ in equation \eqref{eq:dec} to be
\begin{equation}
\Lmc'_5=\left\{(\hbf,R_e^{X_k}| e\in\Emc,k\in[[K]])\in\Rbb_{\geq 0}^{|2^N-1+K|\Emc|}\left| \begin{array}{c}
\displaystyle \hbf_{U_e}=\sum_{i=1}^K R_e^{X_i},e\in\Emc,\\
\displaystyle\sum_{e\in {\rm Fan}(D_d)} R_e^{X_i}\geq H(X_i),i=1,\ldots,{\rm Lev}(D_d), \forall d\in\Dmc.
\end{array}\right.
\right\}.\label{eq:decsp}
\end{equation}

Then, an alternate form of the superposition coding rate region is
\begin{equation}
\Rmc_{sp}(\Asf)=\mathrm{Proj}_{R_{e},e\in\mathcal{E},H(X_k),k\in [[K]]}(\overline{\rm{con}(\Gamma_{N}^{*}\cap\mathcal{L}_{12})}\cap\mathcal{L}'_5\cap \mathcal{L}''_4).\label{eq:generalrateregionfreesp}
\end{equation}

If the outer bound matches with any of the inner bounds presented above, the exact rate region $\Rmc(\Asf)$ has been obtained. We will compute the rate regions and some bounds on more than 7000 non-isomorphic MDCS instances in \S \ref{sec:results}. But first, in the next section, we will show how to construct the codes achieving the points in the inner bounds presented in this section.

\section{Constructing Linear Codes to Achieve the Inner Bounds}

\label{sec:codeCons} 

This section gives explicit constructions for codes achieving all the points in the inner bounds presented in the previous section. 
We will explain how a point in the rate region obtained
from the representable matroid inner bound $\Gamma_N^q$ can be achieved by linear codes which are constructed from the matrix representations of representable
matroids. We will then explain how to extend this to vector codes associated with the vector inner bounds $\Gamma_{N,N'}^q$. Some examples will be shown for illustration.

We begin by observing that for any MDCS instance $\Asf$, the scalar inner bound $\Rmc_{s,q}(\Asf)$ defined in \eqref{eq:rsq} is a convex cone whose dimensions are associated with the variables $\left[h_{X_{k}},R_{e},k\in[[K]],e\in \Emc\right]$.
For any polyhedral cone $\Pmc$, let $\text{Extr}(\Pmc)$ denote the set of representative vectors of its extreme rays.
\begin{theorem}
\label{thm:coniccombination}Let $\Rbf\in\Rmc_{s,q}$, there exists a $\Rbf'\in\Rmc_{s,q}$ with $R'_e\leq R_e,e\in\Emc$ and $H'(X_k)=H(X_k),k\in [[K]]$, and $\Tmc\subseteq\textrm{Extr}(\Gamma_{N}^{q}),\alpha_{i}\geq 0$ such that $\Rbf'=\sum_{\rbf_{i}\in\Tmc}\alpha_{i}\text{Proj}_{H(X_{k}),H(U_{e}),k\in[[K]],e\in \Emc}\rbf_{i}$.
\end{theorem}

\begin{IEEEproof} Let $\Rbf\in\Rmc_{s,q}$, from \eqref{eq:regioninfree} we see that there exists a point $\Rbf'\in\Rmc_{s,q}$, and random variables $U_e,e\in\Emc$ with $R_e\geq R'_e=H(U_e),e\in\Emc$ and $H'(X_k)=H(X_k),k\in [[K]]$. \eqref{eq:regioninfree} also shows that there exists a $\hbf\in \Gamma_N^q$ such that $\Rbf'={\rm Proj}_{H(X_k),H(U_e),k\in[[K]],e\in \Emc}\hbf$, and $\hbf$ satisfies the network constraints. Next observe that, all network constraints are Shannon-type inequalities set equal to zero. For instance, $\Lmc_1$, representing the source independence constraint, is the Shannon inequality $\sum_{k=1}^K H(X_k)-H(\Xbf_{[[K]]})\geq 0$ set to zero. $\Lmc_2$, representing the encoding constraints, is the Shannon inequalities $H(U_e|\Xbf_{[[K]]]})\geq 0,e\in\Emc$ set to zeros. $\Lmc_5$, representing the decoding functions, is the Shannon inequalities $H(\Xbf_{1:{\rm Lev}(D_d)}|{\rm In}(D_d))\geq 0,d\in \Dmc$ set to zeros. Since $\Gamma_N^q\subseteq \Gamma_N$ and all points in $\Gamma_N^q$ must obey Shannon-type inequalities, no point in $\Gamma_N^q$ can lie in the negative side of the network constraints, hence, $\Tmc={\rm Extr}(\Gamma_N^q\cap \Lmc_{1,2,5})\subseteq {\rm Extr}(\Gamma_N^q)$.  Therefore, $\hbf$ can be expressed as a conic combination of extreme rays in $\Tmc\subseteq\text{Extr}(\Gamma_N^q)$, i.e., $\hbf=\sum_{\rbf_{i}\in\Tmc}\alpha_{i}\rbf_{i}$ with $\alpha_{i}\geq 0$. Then 
\begin{eqnarray}
\Rbf'&=&{\rm Proj}_{H(X_k),H(U_e),k\in[[K]],e\in \Emc}\hbf\\
&=& {\rm Proj}_{H(X_k),H(U_e),k\in[[K]],e\in \Emc}\sum_{\rbf_{i}\in\Tmc}\alpha_{i}\rbf_{i}\\
&=&\sum_{\rbf_{i}\in\Tmc}\alpha_{i}{\rm Proj}_{H(X_k),H(U_e),k\in[[K]],e\in \Emc}\rbf_{i}.
\end{eqnarray}\end{IEEEproof}
Before we show the construction of a code to achieve an arbitrary point
in the rate region, it is necessary to show that a rank function of
$\Fbb_{q}$-representable matroid $M$ is associated with a linear
network code in $\Fbb_{q}$.

Given a particular MDCS instance, we first define a \emph{network-$\Fbb_{q}$-matroid
mapping}, by loosening some conditions in \cite{DFZMatroidNetworks},
to be $f:\Xbf_{1:K}\cup\{U_{e},e\in \Emc\}\to\Smc'$, which associates each
source variable and encoded message variable with a collection of elements
forming one set in a partition $\Smc'$ of a ground set $\Smc$ of
a $\Fbb_{q}$-representable matroid $M$ with rank function $r_M$, such that: 
\begin{enumerate}

\item $f$ is an  one to one map; 

\item $r_M(\bigcup_{k=1}^{K}f(X_{k}))=\sum_{i=1}^{K}r_M(f(X_{k}))$,

\item $r_M(f(\text{In}(v)))=r_M(f(\text{In}(v))\cup f(\text{Out}(v))),\forall v\in \Emc\cup \Dmc$,
due to the encoder and decoder functions. Here, $\text{In}(v)$ is
a collection of input variables to $v$ and $\text{Out}(v)$ is a
collection of output variables from $v$. 
\end{enumerate}

If all of the elements in $\Smc'$ are singletons, then $f$ is an
one-to-one mapping to $\Smc$, and the matrix representation of $M$
can be used as linear code in $\mathbb{F}_{q}$ for this MDCS instance,
since the mapping guarantees the MDCS network constraints are obeyed. Such a
coding solution is called a \emph{basic scalar solution}. If $\Smc'$
contains some elements that have cardinalities greater than 1, the
representation of $M$ is interpreted as a collection of bases of
$|\Smc'|$ subspaces, which can also be used as a linear code and
such a solution is called a \emph{basic vector solution}.

We first construct the code for basic solutions, where entropies of network variables are ranks of associated elements in the matroid ($H_q(X_k)=r_M(f(X_k)),k\in [[K]]$ and $H(U_e)=r_M(f(U_e)),e\in \Emc$).  In particular, there are $\sum_{k\in[[K]]}r_M(f(X_{k}))$ $q$-ary digits $\Xbf_{k\in \Kmc^1}r_M(f(X_{k}))$, where $\Kmc^1=\{k\in [[K]]|H(X_k)\neq 0\}$. There exists
a representation $\bar\Cbb$ with dimension $(\sum_{k\in[[K]]}r_M(f(X_{k})))\times(\sum_{k\in[[K]]}r_M(f(X_{k}))+\sum_{e\in \Emc}r(f(U_e)))$
associated with the rank function $r_M$ of $M$, where $\bar\Cbb=[\mathbb{I}_{\sum_{k\in[[K]]}r_M(f(X_{k}))}\ \Cbb]$
and the identity matrix $\mathbb{I}_{\sum_{k\in[[K]]}r_M(f(X_{k}))}$
is mapped to the source digits with non-zero entropies and the rest $\Cbb$ is mapped to
coded messages such that $U_{e}=\Xbf_{k\in \Kmc^1}\Cbb_{:,I(U_{e})},e\in \Emc$, where
$I(U_{e})$ indicates the columns mapped to message $U_{e}$
($\Cbb_{:,I(U_{e})}=\mathbb{O}_{\sum_{k\in[[K]]}r_M(f(X_{k}))\times1}$
if $H(U_{e})=0$). $\Cbb$ is a \emph{semi-simplified solution}
by deleting rows which are associated with sources with zero entropy
but keeping the column size as $\sum_{e\in\Emc}r_M(f(U_e))$. In the following context, basic
solutions are semi-simplified.

Now let us consider a point $\Rbf\in\Rmc_{q}$ associated with source
entropies $H_{q}(X_{k}),k\in[[K]]$. Suppose $\Rbf=\sum_{\rbf_{i}\in\Tmc}\alpha_{i}\text{Proj}_{h_{X_{k}},h_{U_e},k\in[[K]],e\in \Emc}\rbf_{i},\alpha_{i}\geq0$,
where $\Tmc\subseteq\textrm{Extr}(\Gamma_{N}^{q})$, and for every $\rbf_{i}\in\Tmc$,
there exists an associated semi-simplified basic scalar solution $\Cbb^{(i)}$
for the network.

Let $H_{q}(X_{k,i}),k\in[[K]]$ be the source entropies, $R_{e,i},e\in \Emc,i=1,\ldots,|\Tmc|$
be the rates associated with $\rbf_{i}$. According to Theorem \ref{thm:coniccombination},
$H_q(X_{k})=\sum_{i=1}^{|\Tmc|}\alpha_{i}H_{q}(X_{k,i})$ and $R_{e}=\sum_{i=1}^{|\Tmc|}\alpha_{i}R_{e,i}$,
where $H_{q}(X_{k,i})\in\{0,1\}$ and $R_{e,i}\in\{0,1\}$ because
they are from matroids. 

The construction of a code to achieve $\Rbf$ is as follows. 
\begin{enumerate}
\item Find rational numbers $\tilde{\alpha}_{i}=\frac{t_{i}}{n_{i}}\approx\alpha_{i},\ i=1,\ldots,|\Tmc|$,
then $\tilde{H}_q(X_{k})=\sum_{i=1}^{|\Tmc|}\tilde{\alpha}_{i}H_q(X_{k,i})$
and $\tilde{R}_{e}=\sum_{i=1}^{|\Tmc|}\tilde{\alpha}_{i}R_{e,i}$
are the approximation, which can be arbitrarily close, of source entropies
and rates, respectively; 

\item Let $L=\text{LCM}(\{n_{i}\})$ be the block length; 

\item Suppose $L$ blocks of all $K$ source variables $\Xbf_{1:K}^{1:L}$
are losslessly converted to uniformly distributed $q$-ary digits
by some fix-length source code using a sufficiently large number of
outer blocks. We gather these $q$-ary digits formed by individually
compressing the original source variables into a row vector $\tilde{\mathbf{X}}$,
$\textrm{length}(\tilde{\mathbf{X}})=L\sum_{k=1}^{K}\tilde{H}_q(X_{k})$. 

\item Let $\tilde{t}_{i}=L\tilde{\alpha}_{i}$ be the number of times
we will use code $\Cbb^{(i)}$. For every time we use $\Cbb^{(i)}$,
the number of $q$-ary digits encoded is equal to the number of rows
in $\Cbb^{(i)}$ (note that $\Cbb^{(i)}$ is semi-simplified). So there
exists a partition of $\tilde{\mathbf{X}}$ consisting of $\sum_{i=1}^{|\Tmc|}\tilde{t}_{i}$
elements in total and all $\tilde{t}_{i}$ elements mapped with $\Cbb^{(i)}$
have the same cardinality which is the number of rows in $\Cbb^{(i)},\forall i=1,\ldots,|\Tmc|$.
More specifically, we are drawing $\tilde{t}_{i}H_q(X_{k,i})$ samples
from $X_{k}$'s buffer for the $\tilde{t}_{i}$ repetitions of the
basic solution $\Cbb^{(i)}$. 

\item Let $\tilde{\Xbf}'=\tilde{\Xbf}\Gbb$ ($\Gbb$ is a shuffled
identity matrix to relocate the $q$-ary digits in $\tilde{\Xbf}$)
be a rearrangement of $\tilde{\Xbf}$ such that the source digits
are mapped in the same order as the basic solutions in the constructed
code $\tilde{\Cbb}$ which repeats $\Cbb^{(i)}$ for $\tilde{t}_{i}$
times, $\forall i\in\{1,2,\ldots,|\Tmc|\}$ in the way as follows.
\begin{equation}
\tilde{\mathbf{U}}=\tilde{\Xbf}'\times\textrm{BlkDiag}(\underbrace{\Cbb^{(1)},\ldots}_{\tilde{t}_{1}\ \textrm{times}},\underbrace{\Cbb^{(i)},\ldots}_{\tilde{t}_{i}\ \textrm{times}},\underbrace{\Cbb^{(m)},\ldots}_{\tilde{t}_{m}\ \textrm{times}})
\end{equation}
 where $\text{BlkDiag}(\cdot)$ is a block diagonalizing function. 

\item Note that all $\Cbb^{(i)}$ have the same column size and the
column indices are mapped to $e\in \Emc$. Therefore, we can rearrange
the columns in $\tilde{\Cbb}$ to group all columns containing $\Cbb^{i}_{:,I_i(U_{e})},i=1,\ldots,|\Tmc|$
to be an encoding function for $e$. That is, $\Cbb=\textrm{concatenation}(\Cbb_{:,I(U_{e})}),\ \Cbb_{:,I(U_{e})}=\tilde{\Cbb}_{:,I\left(e+|\Emc|\cdot\left[0:\sum_{i=1}^{|\Tmc|}\tilde{t}_{i}-1\right]\right)}$.
$\Cbb$ can be further simplified by deleting all-zero columns. 
\end{enumerate}


Indeed, we can see that the code constructed this way can achieve the point $\Rbf\in \Rmc_{q}$ by examining
\begin{eqnarray}
H_{q}(\tilde{U}_{e}) & = & \text{rank}(\Cbb_{:,I(U_{e})})\\
 & = & \sum_{i=1}^{|\Tmc|}\tilde{t}_{i}\text{rank}(\Cbb^{(i)}_{:,I(U_{e})})\\
 & = & \sum_{i=1}^{|\Tmc|}\tilde{t}_{i}R_{i,e}\\
 & = & L\sum_{i=1}^{|\Tmc|}\tilde{\alpha}_{i}R_{i,e}\\
 & = & L\tilde{R}_{e}.
\end{eqnarray}
Therefore, the actual rate per source variable is
\begin{equation}
\hat{R}_e = \frac{H_q(\tilde{U}_e)}{L}=\tilde{R}_e\approx R_e,
\end{equation}
with arbitrarily small offset if the fraction approximations are arbitrarily close. If $\Gamma_N^q$ is used in obtaining the rate region, $\Cbb^{(i)},\forall i=1,\ldots,|\Tmc|$ are basic scalar solution, we call the constructed code a \emph{scalar representation solution}. Similarly, if $\Gamma_{N,N'}^{q}$ is used in obtaining the rate region, some basic vector solutions $\Cbb^{(i)},\textrm{ some }i=1,\ldots,|\Tmc|$ may be needed in constructing the code $\Cbb$. We call such a code involving basic vector solution(s) a \emph{vector representation solution}.


\begin{figure}
\centering
 \includegraphics[scale=0.65]{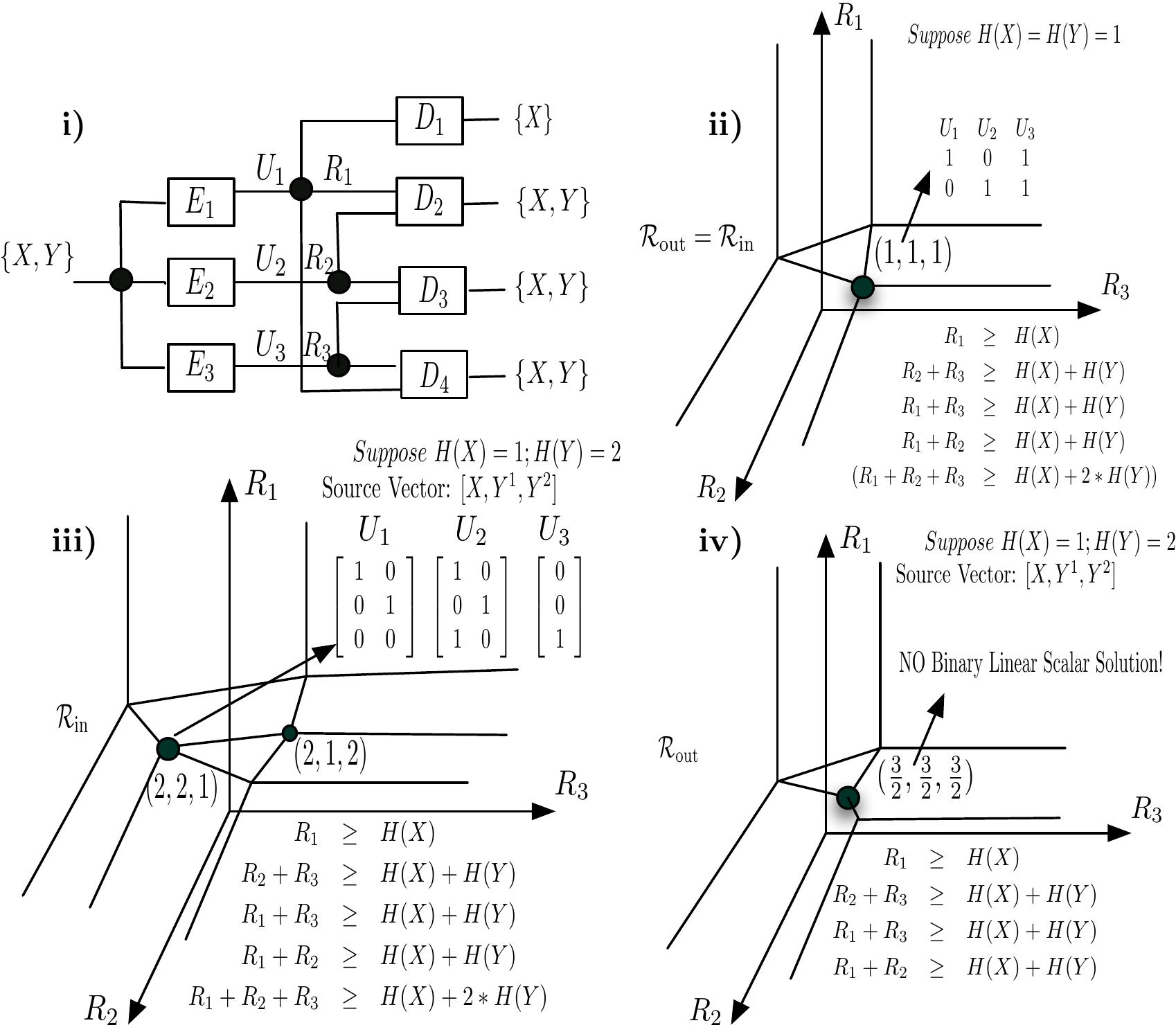} 
 \caption{ \textbf{i)} Example MDCS instance. \textbf{ii)} Rate region and corresponding
codes. \textbf{iii)} Scalar codes for inner bound. \textbf{iv)} No
scalar code for outer bound. 
}
\label{fig:main} 
\end{figure}
\begin{example}
Consider a 2-level-3-encoder MDCS instance, shown in Fig.\ \ref{fig:main}.

There are two sources $X,Y$, three encoders $E_{1},E_{2},E_{3}$ with corresponding coded message variables $U_{i},i=1,2,3$ and rate constraints $R_{i},i=1,2,3$, four decoders classified into two levels with access to encoders as shown in the following table. Level 1 means this decoder recovers $X$, while level 2 means it recovers $X,Y$. 

\centerline{
\begin{tabular}{c|c}
level 1  & level 2\tabularnewline \hline 
\{(1)\}  & \{(1, 2), (1, 3), (2, 3)\}\tabularnewline
\end{tabular}}
The outer $\mathcal{R}_{\text{out}}$ and scalar binary inner bound $\mathcal{R}_{s,2}$ on rate region obtained from Shannon outer bound and binary inner bound are: 
\begin{eqnarray}
\Rmc_{\rm out} &=& \left\{ \Rbf : 
\begin{array}{rcl}
R_1 & \geq & H(X) \\ 
R_2+R_3 & \geq & H(X)+H(Y) \\ 
R_1+R_3 & \geq & H(X)+H(Y) \\ 
R_1+R_2 & \geq & H(X)+H(Y) 
\end{array} 
\right\} \\
\Rmc_{s,2} &=& \Rmc_{\rm out} \cap \{ \Rbf : R_1+R_2+R_3 \geq H(X)+2 H(Y) \} \nonumber 
\end{eqnarray}
where $\Rbf = (R_1,R_2,R_3)$.  Note that when $H(X)=H(Y)$, $\mathcal{R}_{\text{out}}=\mathcal{R}_{s,2}$.  This is a case where the inner and outer bounds match, and thus we are able to find a coding solution for this network.  The scalar binary code 
\vspace{-1mm}
\begin{equation}
U_{1}U_{2}U_{3}=XY\times
\left[\begin{array}{ccc}
1 & 0 & 1\\
0 & 1 & 1
\end{array}\right]
\end{equation} 
\vspace{-3mm}

\noindent achieves the extreme point $(1,1,1)$ in $\Rmc$ when $H(X)=H(Y)=1$, as shown in Fig.\ \ref{fig:main}.

If $H(X)=H(Y)=h$, the corresponding extreme point $(h,h,h)$ will
be achieved by repetition of the basic solution:
\begin{equation}
U_{1}U_{2}U_{3}=X^{1}\ldots X^{h}Y^{1}\ldots Y^{h}\times\left[\begin{array}{ccccc}
\mathbb{I}_{h} & \vdots & \mathbb{O}_{h} & \vdots & \mathbb{I}_{h}\\
\mathbb{O}_{h} & \vdots & \mathbb{I}_{h} & \vdots & \mathbb{I}_{h}
\end{array}\right].
\end{equation}

If $h$ is not integer, we can easily approximate it with arbitrarily precision using fractions $\frac{t}{n}\approx h$ and then decide the block size as $n$ to construct the block code by repeating the above basic solution for $t$ times in block diagonal manner. Then, on average we will achieve $(h,h,h)$.

However, $\mathcal{R}_{\text{out}}\neq\mathcal{R}_{s,2}$ in general for this example, if $H(X)\neq H(Y)$.  For $H(X)=1,H(Y)=2$ there is a gap between the inner and outer bounds. For the inner bound, we can find a scalar code solution. Fig.\ \ref{fig:main} shows a binary code to achieve the inner bound extreme point $\Rbf = (2,2,1)$, which is a conic combination of two basic solutions. 
Let $H(U_{1})=R_{1},H(U_{2})=R_{2},H(U_{3})=R_{3}$, the point $(H(X),H(Y),H(U_{1}),H(U_{2}),H(U_{3}))=(1,2,2,2,1)=(1,1,1,1,1)+(0,1,1,1,0)$.
The solution corresponding to $(0,1,1,1,0)$ is 
\begin{equation}
U_{1}^{1}U_{2}^{1}U_{3}^{1}=Y^1\times[1\ 1\ 0]
\end{equation}
and the solution corresponding to $(1,1,1,1,1)$ is 
\begin{equation}
U_{1}^{2}U_{2}^{2}U_{3}^{2}=XY^{2}\times\left[\begin{array}{ccc}
1 & 1 & 0\\
0 & 1 & 1
\end{array}\right].
\end{equation}
The final scalar representation solution with shuffling of columns is shown in Fig.\ \ref{fig:main}.

For $\mathcal{R}_{{\rm out}}$, we know there does not exist scalar binary coding solution for some extreme point. For example, as shown in Fig.\ \ref{fig:main}, there is no scalar solution for the outer bound extreme point $(\frac{3}{2},\frac{3}{2},\frac{3}{2})$. However, $\Gamma_{5,6}^{2}$ makes up the gap and we know there must exist a solution to achieve this point. Actually, we can find a binary vector representation solution for this point. Note that we only need to group two outcomes of source variables and encode them together. Suppose we have source vector $\vbf=[X_{1},X_{2},Y_{1}^{1},Y_{1}^{2},Y_{2}^{1},Y_{2}^{2}]$ where the lower index indicates two outcomes in time while upper index indicates the position in one outcome. One vector representation coding solution (with columns shuffled) is
\begin{equation}
U_{1}U_{2}U_{3}=\vbf\times
\left[\begin{array}{ccc|ccc|ccc}
1 & 0 & 0 & 0 & 0 & 1 & 0 & 0 & 0\\
0 & 1 & 0 & 0 & 0 & 0 & 1 & 0 & 0\\
0 & 0 & 0 & 1 & 0 & 0 & 1 & 0 & 0\\
0 & 0 & 1 & 0 & 0 & 0 & 0 & 1 & 0\\
0 & 0 & 0 & 0 & 0 & 1 & 0 & 0 & 1\\
0 & 0 & 1 & 0 & 1 & 0 & 0 & 0 & 0
\end{array}\right],
\end{equation}
which can also be expressed as a conic combination of two basic solutions. Let $H(U_{1})=R_{1},H(U_{2})=R_{2},H(U_{3})=R_{3}$, the point $2\times (H(X),H(Y),H(U_{1}),H(U_{2}),H(U_{3}))=(2,4,3,3,3)=(1,1,1,1,1)+(1,3,2,2,2)$.
The solution corresponding to $(1,1,1,1,1)$ is 
\begin{equation}
U_{1}^{1}U_{2}^{1}U_{3}^{1}=X_1Y_2^1\times \left[\begin{array}{ccc}
1& 1 & 0\\
0 & 1 &1
\end{array}
\right],
\end{equation}
and the solution corresponding to $(1,3,2,2,2)$ is 
\begin{equation}
U_{1}^{2}U_{2}^{2}U_{3}^{2}=X_2Y_1^{1}Y_1^2Y_2^2\times\left[\begin{array}{cc|cc|cc}
1 & 0 & 0 & 0 & 1 & 0\\
0 & 0 & 1 & 0 & 1 & 0\\
0 & 1 & 0 & 0 & 0 & 1\\
0 & 0 & 0 & 1& 0 & 1
\end{array}\right].
\end{equation}

\end{example}

Having provided code constructions in these examples, we now pass to investigating embedding operations for smaller MDCS instances into larger MDCS instances such that the larger MDCS instances inherits the insufficiency of a class of codes from the smaller MDCS instances.

\section{Embedded MDCS Instances and the Preservation of Coding Class Sufficiency}
\label{sec:embedding}
In \cite{HauThesis1995}, a definition of {\it embedded MDCS} instances was given for $(2,3)$ and $(3,3)$ MDCS instances where a $(2,3)$ MDCS instance $\Asf$ is embedded in a $(3,3)$ MDCS instance $\Asf'$ if it can be obtained by deleting one source variable in $\Asf'$, as we will define in Definition \ref{def:sourdel}. We would like to extend the definition of embedded MDCS instances, because we are interested in the relationships between different $(K,|\Emc|)$ MDCS problems with respect to sufficiency of certain linear codes, as will be shown in \S\ref{sec:results}. We would like to show that the insufficiency of certain classes of codes will be preserved when one extends a smaller MDCS instance to a bigger one. For that, we first define some operations on MDCS instances that can obtain a smaller MDCS instance from a bigger one.

\subsection{Embedding Operation Definitions} \label{sec:embmdcs}

We generalize the definition of source deletion first. When a source is deleted, the decoders that demand it will no longer demand it after deletion.
\begin{definition}[Source Deletion $\Asf\backslash X_k$]\label{def:sourdel}
Suppose a MDCS instance $\Asf=(\{X_1,\ldots,X_K\},\Emc,\Dmc,\Lbf',\Gmc')$. When a source $X_k$ is deleted, denoted as $\Asf\backslash X_k$, in the new MDCS instance $\Asf'=(\{X_1,\ldots,X_K\}\setminus X_k,\Emc',\Dmc',\Lbf',\Gmc')$, we will have:
\begin{enumerate}
\item $\Emc'=\Emc$;
\item $\Dmc'=\Dmc\setminus \{D_j|\exists D_i\in \Dmc, i\neq j,{\rm Out}_{\Asf}(D_i)={\rm Out}_{\Asf}(D_j)\setminus X_k$, ${\rm Fan}_{\Asf}(D_i)\subseteq {\rm Fan}_{\Asf}(D_j)\}$;
\item For $\Lbf'$, ${\rm Lev}_{\Asf'}(D_d)={\rm Lev}_{\Asf}(D_d)-1,\forall D_d$ such that $X_k\in {\rm Out}_{\Asf}(D_d)$;
\item $\Gmc'=\{(E_i,D_d)|E_i\in \Emc', D_d\in \Dmc', (E_i,D_d)\in \Gmc\}$.
\end{enumerate}

\end{definition}

\begin{figure}
\centering 
\subfigure [\label{fig:examplee}Demonstration of source deletion: when source $Z$ is deleted, decoders that previously required it will no longer require it.]{\includegraphics[scale=0.60]{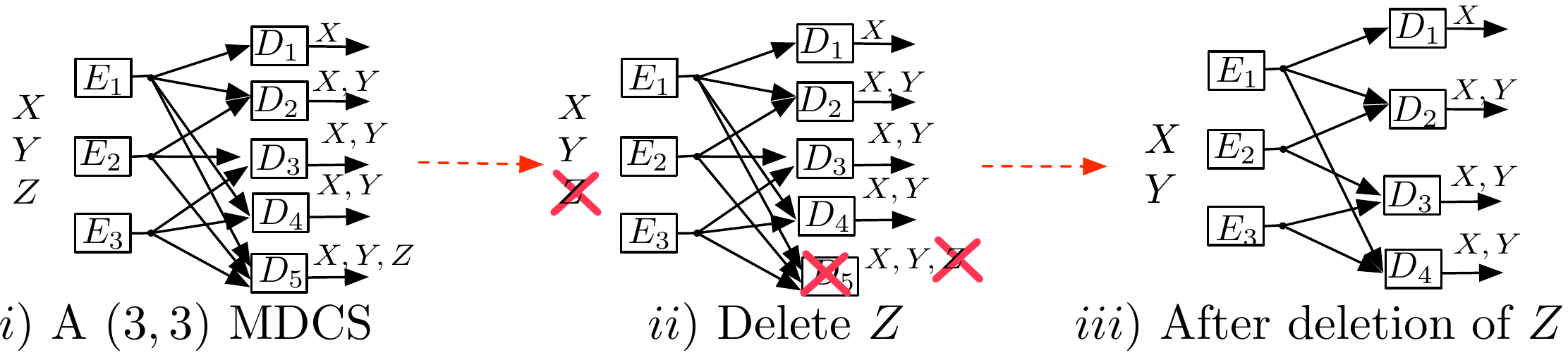} }
\subfigure [\label{fig:examplea}Demonstration of encoder contraction: when $E_4$ is contracted, the fan of it will be deleted.]
{\includegraphics[scale=0.60]{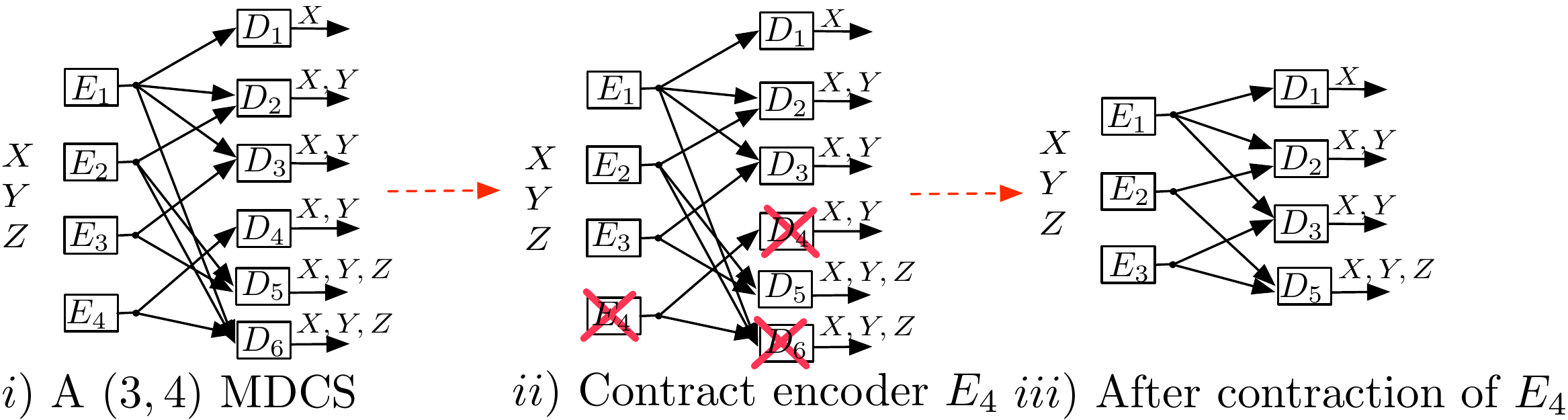} }
\subfigure [\label{fig:exampleb}Demonstration of encoder deletion: when $E_4$ is deleted, the fan of it will keep the same decoding abilities and lower-level decoders will be superseded if they have same fan as the fan of $E_4$ after deletion.]{\includegraphics[scale=0.60]{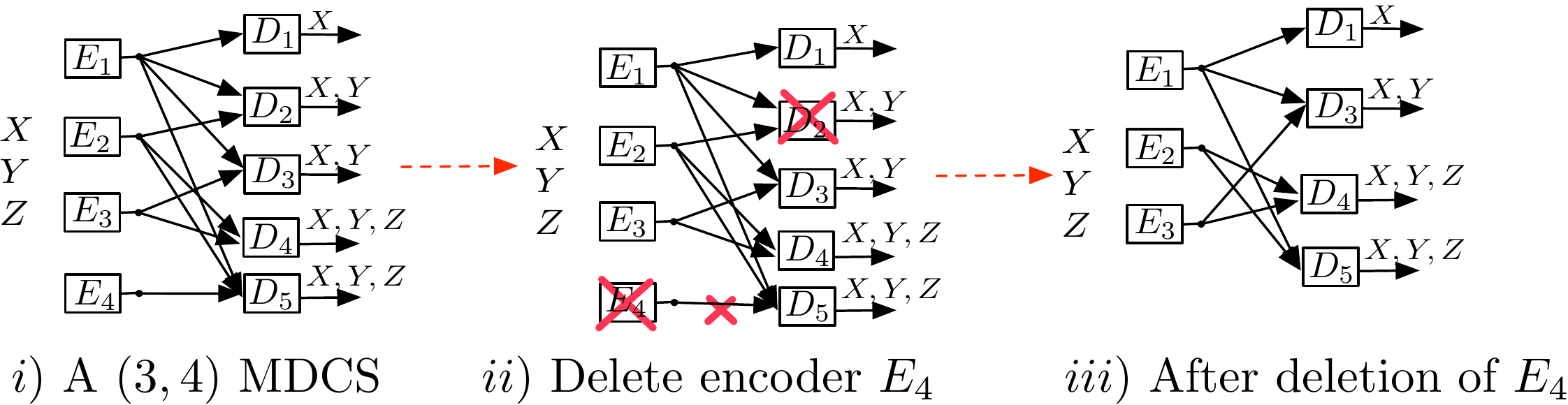} }
\subfigure [\label{fig:exampled}Demonstration of encoder unification: when $E_4$ is unified with $E_3=E_4$, the fan of $E_4$ will also have access to $E_3$ after $E_4$ is removed. If conflicts occur, existing decoders become redundant. For instance, $D_3$ is superseded because $D_4$ only has access to $E_3$ but is able to decode $X,Y$.]{\includegraphics[scale=0.60]{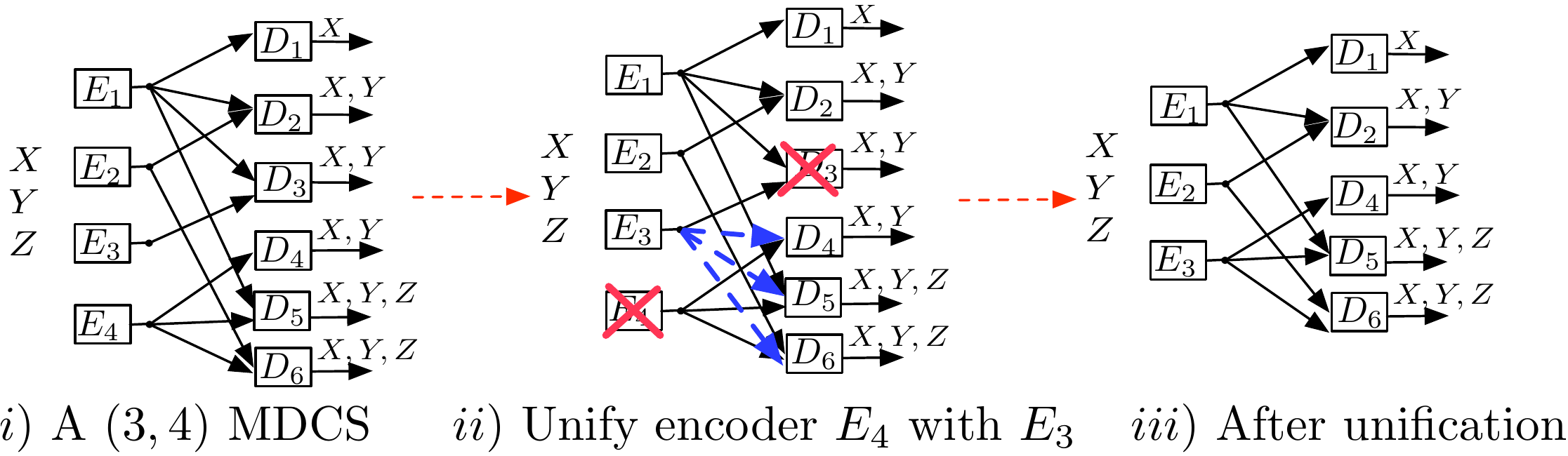} }

\caption{Demonstration of operations on MDCS instances}
\label{fig:embedexample} 
\end{figure}

This is straightforward because the deletion of a source just changes the decoding requirements of decoders. 
Fig.\ \ref{fig:examplee} demonstrates the deletion of a source. When source $Z$ is deleted, $D_5$ will no longer require $Z$ and thus becomes a level-2 decoder. However, since $D_2$ only has access to $E_1,E_2$ but is also a level-2 decoder, $D'_5$ becomes redundant and is deleted.

Next, we consider the operation of contracting an encoder. When an encoder is contracted, all of the decoders in its fan will be deleted, as well as all the edges associated with the contracted decoders.
\begin{definition}[Encoder Contraction $(\Asf\slash E_e)$]
Suppose a MDCS instance $\mathsf{A}=(\{X_1,\ldots,X_K\},\Emc,\Dmc,\Lbf,\Gmc)$. A smaller MDCS instance $\Asf'=(\{X_1,\ldots,X_K\},\Emc',\Dmc',\Lbf',\Gmc')$ is obtained by contracting $E_e,e\in \Emc$, denoted by $\Asf\slash E_e$, if:
\begin{enumerate}
\item $\Emc'=\Emc\setminus E_e$;
\item $\Dmc'=\Dmc\setminus \{D_d | D_d\in {\rm Fan}_{\Asf}(E_e)\}$;
\item For $\Lbf'$, ${\rm Lev}_{\Asf'}(D_i)={\rm Lev}_{\Asf}(D_i),\forall D_i\in \Dmc'$;
\item $\Gmc'=\Gmc\setminus \{(E_i,D_d) | E_i\in \Emc, D_d\in {\rm Fan}_{\Asf}(E_e), (E_i,D_d)\in \Gmc\}$.
\end{enumerate}
\end{definition}

This operation assumes that when an encoder is contracted, its fan will directly have access to its input, all the sources, which makes the decoding requirements obviously satisfied. Fig.\ \ref{fig:examplea} demonstrates the contraction of an encoder. As it shows, when encoder $E_4$ is contracted, all decoders which have access to $E_4$, i.e., fan of $E_4$, become redundant and are deleted.

Next, we define deletion of an encoder as follows.

\begin{definition}[Encoder Deletion $(\Asf \backslash E_e)$]\label{def:EncCont}
Suppose a MDCS instance $\Asf=(\{X_1,\ldots,X_K\},\Emc,\Dmc,\Lbf,\Gmc)$. When an encoder $E_e\in \mathcal{E}$ is deleted, denoted as $\Asf\backslash E_e$, in the new MDCS instance $\Asf'=(\{X_1,\ldots,X_K\},\Emc',\Dmc',\Lbf',\Gmc')$, we will have:
\begin{enumerate}
\item $\Emc'=\Emc\setminus E_e$;
\item $\Dmc'=\Dmc\setminus \{D_j \in \Dmc| \exists D_i\in {\rm Fan}_{\Asf}(E_e)\setminus D_j, {\rm Fan}_{\Asf}(D_i)\setminus E_e\subseteq {\rm Fan}_{\Asf}(D_j)$ and $\mathrm{Lev}_{\Asf}(D_j)\leq \mathrm{Lev}_{\Asf}(D_i)\}$;
\item For $\Lbf'$, ${\rm Lev}_{\Asf'}(D_d)={\rm Lev}_{\Asf}(D_d),\forall D_d\in \Dmc'$;
\item $\Gmc'=\{(E_i,D_d)|E_i\in \Emc', D_d\in \Dmc', (E_i,D_d)\in \Gmc\}$.
\end{enumerate}

\end{definition}

The essence of this definition is to keep the dependence relationship between input and output of the decoders when an encoder is deleted. In other words, when an encoder is deleted, the decoders that have access to it should function as before. Note that, if there exists a decoder $D_i$ that only has access to $E_e$, after the deletion of $E_e$, it has access to no encoders but needs to keep the same decoding capability, which means the sources $X_1,\ldots, X_{{\rm Lev}(D_i)}$ will also be deleted.

Fig.\ \ref{fig:exampleb} demonstrates the deletion of an encoder. When encoder $E_4$ is deleted, $D_6$ no longer has access to $E_4$ but still has access to $E_1,E_2$. Note that, since $D_2$ also has access to $E_1,E_2$ but is a level-2 decoder, $D_2$ becomes redundant and is deleted.

When the reverse operation of encoder contraction is considered, a MDCS instance can be extended by adding a new encoder with some new decoders that must talk with the new encoder obeying (\textbf{C1})--(\textbf{C5}). If some class of codes does not suffice in the smaller network, it will not suffice in the bigger one either. Similarly, the insufficiency can be preserved by considering to extend a smaller MDCS instance by adding some arbitrary redundant encoder and decoders, which is the reverse of encoder deletion. There exists some other ways to preserve the non-sufficiency of certain class of codes. For instance, if a class of codes does not suffice for a smaller network, there does not exist a construction of codes for at least one encoder to satisfy all the network constraints. If a new encoder dependent on that encoder and some other redundant decoding requirements are constructed obeying the conditions (\textbf{C1})--(\textbf{C5}), that class codes still cannot be sufficient for the new network. We define another operation based on this intuition as follows.

\begin{definition}[Encoder Unification $(\Asf \cap\{E_i=E_j\}$]\label{def:encuni}
Suppose a MDCS instance $\Asf=(\{X_1,\ldots,X_K\},\Emc,\Dmc,\Lbf,\Gmc)$. When an encoder $E_j\in \mathcal{E}$ is unified with $E_i=E_j,i\neq j$, denoted as $\Asf\cap\{E_i=E_j\}$, in the new MDCS instance $\Asf'=(\{X_1,\ldots,X_K\},\Emc',\Dmc',\Lbf',\Gmc')$, we will have:
\begin{enumerate}
\item $\Emc'=\Emc\setminus E_j$;
\item $\Dmc'=\Dmc\setminus \{D_l|\exists D_k\in \Dmc, k\neq l, D_k\in {\rm Fan}_{\Asf}(E_j), D_l\in {\rm Fan}_{\Asf}(E_i), ({\rm Fan}_{\Asf}(D_k)\setminus \{E_i,E_j\}) \subseteq ({\rm Fan}_{\Asf}(D_l)\setminus \{E_i,E_j\})$ and $\mathrm{Lev}_{\Asf}(D_l)\leq \mathrm{Lev}_{\Asf}(D_k)\}$;
\item For $\Lbf'$, ${\rm Lev}_{\Asf'}(D_d)={\rm Lev}_{\Asf}(D_d), \forall D_d\in \Dmc'$;
\item $\Gmc'=\{(E_e,D_d)|E_e\in \Emc', D_d\in \Dmc', (E_e,D_d)\in \Gmc\}\cup \{(E_i,D_d)|D_d\in {\rm Fan}_{\Asf}(E_j)\cap \Dmc'\}$.
\end{enumerate}
\end{definition}

After the unification of $E_j$, all decoders who have access to $E_j$ will have access to $E_i$.  Fig.\ \ref{fig:exampled} demonstrates the unification of an encoder. As it shows, when encoder $E_4$ is unified with $E_3=E_4$, all decoders which have access to $E_4$, i.e., fan of $E_4$, will have access to $E_3$. Decoder $D_3$ becomes redundant because if $D_4,D_5$ are given access to $E_3$ instead of $E_4$, they both will supersede $D_3$.

\subsection{Operation Order does not Matter}
Next, we consider the order of different operations. It is not difficult to see that if a collection of sources are deleted, it does not matter which source is deleted first. Similarly, if a collection of encoders are contracted, deleted or unified, the operation order on different elements does not matter. Next we would like to show that different orders of operations give equivalent results.

\begin{theorem}\label{thm:order}
Let $\Omc_1,\Omc_2$ be two different operations on different elements, among the four operations defined in Definition \ref{def:sourdel}-- Definition \ref{def:encuni}. Applying $\Omc_1$ first and $\Omc_2$ second on a MDCS instance $\Asf$ is equivalent to applying $\Omc_2$ first and $\Omc_1$ second. That is, $\Asf\cap \Omc_1\cap \Omc_2=\Asf\cap \Omc_2\cap \Omc_1$.
\end{theorem}
\begin{IEEEproof}
We need to consider the $\binom{4}{2}=6$ combinations of operations.

\textbf{Encoder deletion and encoder contraction:} Let $\Asf'=(\Asf\backslash E_1) \slash E_2$ and $\Asf''=(\Asf\slash E_2)\backslash E_1$. We need to show $\Asf'=\Asf''$.  If $E_1$ is deleted at first, we have that $\forall D_i \in {\rm Fan}(E_1)$, the decoder $D_i$ will have access to encoders in ${\rm Fan}(D_i)\setminus E_1$ and ${\rm Lev}_{\Asf'}(D_i)={\rm Lev}_{\Asf}(D_i)$. If $\exists D_j$ such that ${\rm Fan}_{\Asf}(D_j)={\rm Fan}_{\Asf}(D_i)\setminus E_1$ and ${\rm Lev}_{\Asf}(D_j)<{\rm Lev}_{\Asf}(D_i)$, $D_j\notin\Asf'$.
We only need to consider the case when $E_2$ is a fan of $D_i$ and/or $D_j$. If $E_2\in {\rm Fan}_{\Asf}(D_i)$ and $E_2\in {\rm Fan}_{\Asf}(D_j)$, no matter which operation is first, both $D_i$ and $D_j$ are gone. If $E_2\in {\rm Fan}_{\Asf}(D_j)$, no matter deletion or contraction is done first, $D_j$ will be gone. If $E_2\in {\rm Fan}_{\Asf}(D_i)$, we must have $E_2\in {\rm Fan}_{\Asf}(D_j)$ since ${\rm Fan}_{\Asf}(D_j)={\rm Fan}_{\Asf}(D_i)\setminus E_1$ and we assume that $E_1\neq E_2$. Thus, $\Asf'=\Asf''$.

\textbf{Encoder deletion and source deletion:} Let $\Asf'=(\Asf\backslash E_1) \backslash X_k$ and $\Asf''=(\Asf\backslash X_k)\backslash E_1$. We need to show $\Asf'=\Asf''$.  If $E_1$ is deleted first, we have that $\forall D_i \in {\rm Fan}_{\Asf}(E_1)$, ${\rm Fan}_{\Asf'}(D_i)={\rm Fan}_{\Asf}(D_i)\setminus E_1$ and ${\rm Lev}_{\Asf'}(D_i)={\rm Lev}(D_i)$. If $\exists D_j$ such that ${\rm Fan}_{\Asf}(D_j)={\rm Fan}_{\Asf}(D_i)\setminus E_1$ and ${\rm Lev}_{\Asf}(D_j)<{\rm Lev}_{\Asf}(D_i)$, $D_j\notin \Asf'$.
We only need to consider the case when $X_k$ is an output of $D_i$ and/or $D_j$. If $X_k\in {\rm Out}_{\Asf}(D_i)$ and $X_k\in {\rm Out}_{\Asf}(D_j)$, the deletion of $X_k$ does not affect the deletion of $E_1$. Hence the order does not matter. If $X_k\in {\rm Out}_{\Asf}(D_i)$, no matter which operation is done first, $D_j$ will be gone. If $X_k\in {\rm Out}_{\Asf}(D_j)$, we must have $X_k\in {\rm Out}_{\Asf}(D_i)$ since ${\rm Out}_{\Asf}(D_j)\subseteq {\rm Out}_{\Asf}(D_i)$. Thus, $\Asf'=\Asf''$.

\textbf{Encoder contraction and source deletion:} Let $\Asf'=(\Asf\backslash X_k)\slash E_1$ and $\Asf''=(\Asf\slash E_1) \backslash X_k$. We need to show $\Asf'=\Asf''$.   If $X_k$ is deleted at first, we have that $\forall D_i$ such that $X_k \in {\rm Out}_{\Asf}(D_i)$, ${\rm Out}_{\Asf'}(D_i)={\rm Out}(D_i)\setminus X$ and if $\exists D_j$ such that ${\rm Fan}_{\Asf}(D_j)\subseteq {\rm Fan}_{\Asf}(D_i)$ and ${\rm Out}_{\Asf}(D_j)={\rm Out}_{\Asf'}(D_i)$, $D_i\notin \Asf'$.
We only need to consider the case when $E_1$ is a fan of $D_i$ and/or $D_j$. If $E_1\in {\rm Fan}_{\Asf}(D_i)$ and $E_1\in {\rm Fan}_{\Asf}(D_j)$, the contraction of $E_1$ will make both encoders gone no matter which operation is first. If $E_1\in {\rm Fan}_{\Asf}(D_i)$ only, no matter which operation is done first, $D_i$ will be gone. If $E_1\in {\rm Fan}_{\Asf}(D_j)$, we must have $E_1\in {\rm Fan}_{\Asf}(D_i)$ since ${\rm Fan}_{\Asf}(D_j)\subseteq {\rm Fan}_{\Asf}(D_i)$. Thus, $\Asf'=\Asf''$.

\textbf{Encoder unification and encoder contraction:} Let $\Asf'=(\Asf\cap\{E_2=E_3\})\slash E_1$ and $\Asf''=(\Asf\slash E_1) \cap\{E_2=E_3\}$. We need to show $\Asf'=\Asf''$.  In unification of $E_3$ with $E_2=E_3$, all decoders having access to $E_3$ will also have access to $E_2$ and then $E_3$ is removed. A decoder $D_l$ is deleted if $\exists D_k,D_l\in \Dmc, k\neq l$, such that $D_k\in {\rm Fan}_{\Asf}(E_3)$, $D_l\in {\rm Fan}_{\Asf}(E_2)$, ${\rm Fan}_{\Asf}(D_k)\setminus \{E_2, E_3\} \subseteq {\rm Fan}(D_l)_{\Asf}\setminus \{E_2, E_3\}$ and $\mathrm{Lev}_{\Asf}(D_l)\leq \mathrm{Lev}_{\Asf}(D_k)$. If $E_1\in {\rm Fan}_{\Asf}(D_l)$, clearly contraction of $E_1$ will also delete $D_l$. If $E_1\in {\rm Fan}_{\Asf}(D_k)$, we will also have $E_1\in {\rm Fan}_{\Asf}(D_l)$ since ${\rm Fan}_{\Asf}(D_k)\setminus \{E_2,E_3\} \subseteq {\rm Fan}_{\Asf}(D_l)\setminus \{E_2,E_3\}$. No matter $E_1$ is contracted first or not, the resulting MDCS instance will be the same.  That is, $\Asf'=\Asf''$.

\textbf{Encoder unification and encoder deletion:} Let $\Asf'=(\Asf\cap\{E_2=E_3\})\backslash E_1$ and $\Asf''=(\Asf\backslash E_1) \cap\{E_2=E_3\}$. We need to show $\Asf'=\Asf''$.  In unification of $E_3$ with $E_2=E_3$, all decoders having access to $E_3$ will also have access to $E_2$ and then $E_3$ is removed. A decoder $D_l$ is deleted if $\exists D_k,D_l\in \Dmc, k\neq l$, such that $D_k\in {\rm Fan}_{\Asf}(E_3)$, $D_l\in {\rm Fan}_{\Asf}(E_2)$, ${\rm Fan}_{\Asf}(D_k)\setminus \{E_2, E_3\} \subseteq {\rm Fan}(D_l)_{\Asf}\setminus \{E_2, E_3\}$ and $\mathrm{Lev}_{\Asf}(D_l)\leq \mathrm{Lev}_{\Asf}(D_k)$. If $E_1\in {\rm Fan}_{\Asf}(D_l)$, we can see that when $E_1$ is deleted first, even though $D_l$ could survive after deleting $E_1$, it will also be deleted after the unification of $E_3$ because we have ${\rm Fan}_{\Asf}(D_k)\setminus \{E_2, E_3\} \subseteq {\rm Fan}_{\Asf}(D_l)\setminus \{E_1,E_2, E_3\}$. If $E_1\in {\rm Fan}_{\Asf}(D_k)$, we will also have $E_1\in {\rm Fan}_{\Asf}(D_l)$ since ${\rm Fan}_{\Asf}(D_k)\setminus \{E_2, E_3\} \subseteq {\rm Fan}_{\Asf}(D_l)\setminus \{E_2, E_3\}$. No matter $E_1$ is deleted first or not, the resulting MDCS instance will be the same.  That is, $\Asf'=\Asf''$.

\textbf{Encoder unification and source deletion:} Let $\Asf'=(\Asf\backslash X_k)\cap\{E_1=E_2\}$ and $\Asf''=(\Asf\cap\{E_1=E_2\}) \backslash X_k$. We need to show $\Asf'=\Asf''$.  If $X_k$ is deleted at first, we have that $\forall D_i$ such that $X_k \in {\rm Out}_{\Asf}(D_i)$, ${\rm Out}_{\Asf'}(D_i)={\rm Out}_{\Asf}(D_i)\setminus X_k$ and if $\exists D_j$ such that ${\rm Fan}_{\Asf}(D_j)\subseteq {\rm Fan}_{\Asf}(D_i)$ and ${\rm Out}_{\Asf}(D_j)={\rm Out}_{\Asf}(D_i)$, $D_i\notin \Asf'$.
We only need to consider the case when $E_2$ is a fan of $D_i$ and/or $D_j$. If $E_2\in {\rm Fan}_{\Asf}(D_i)$ and $E_2\in {\rm Fan}_{\Asf}(D_j)$ and the unification of $E_2$ is conducted first, $D_i,D_j$ will have access to $E_1$. This unification does not affect the subset relationship between ${\rm Fan}_{\Asf}(D_j)$ and ${\rm Fan}_{\Asf}(D_i)$. Thus, $D_j$ is deleted no matter which operation is conducted first. Similarly, if $E_2\in {\rm Fan}_{\Asf}(D_i)$ only or $E_2\in {\rm Fan}_{\Asf}(D_i)$ only, no matter which operation is done first, $D_j$ will be gone because the unification does not affect the conditions of deleting $D_j$. Thus, $\Asf'=\Asf''$.
\end{IEEEproof}

Based on these operations and Theorem \ref{thm:order}, we can define an {\it embedded} MDCS instance.

\begin{definition}[Embedded MDCS instances] \label{def:embedded}
An MDCS instance $\Asf'$ is said {\it embedded} in MDCS instance $\Asf$, i.e., $\Asf'$ is a {\it minor} of $\Asf$, denoted as $\Asf' \prec \Asf$, if $\Asf'$ can be obtained by a series of operations of source deletion, encoder deletion/ contraction/ unification on $\Asf$. Equivalently, we say that $\Asf$ is an extension of $\Asf'$, denoted as $\Asf \succ \Asf'$.
\end{definition}

Note that, as will be discussed in \S\ref{sec:results}, we consider all four operations for the preservation of the insufficiency of $\Fbb_q$ vector linear codes (\S\ref{sec:codeCons}) and the embedding relationship is denoted as $\Asf'\prec_v\Asf$ or $\Asf\succ_v\Asf'$.  When insufficiency of scalar linear codes (superposition coding) are considered, the encoder unification is not considered and the embedding relationship is denoted as $\Asf'\prec_s\Asf$ ($\Asf'\prec_{sp}\Asf$) or $\Asf\succ_s\Asf'$ ($\Asf\succ_{sp}\Asf'$).  Fig.\ \ref{fig:embedexample} demonstrates different operations and thus shows four examples of embedded MDCS instances.

\subsection{Inheritance of Code Class Sufficiency under Extension}

Recall the definitions of deletion/contraction of encoders/sources in \S \ref{sec:embmdcs}, which connect MDCS instances  for different $(K,|\Emc|)$ pairs. For example, if $K'\leq K,|\Emc|'\leq |\Emc|$, then for a $(K',|\Emc|')$ MDCS instance, $\Asf'$ , there exists a $(K,|\Emc|)$ MDCS instance $\Asf$ and a series of operations of source/encoder deletion/contraction such that $\Asf'$ can be obtained by applying these operations on $\Asf$, i.e., $\Asf'\prec \Asf$. 

This definition of an MDCS minor is motivated by the definition of a matroid minor, where if a matroid is not representable over $\Fbb_q$, then its extensions will also not be $\Fbb_q$-representable, because extensions also have the same forbidden minor(s) characterizing $\Fbb_q$-representability.  One interesting question is if there also exists similar forbidden minor characterizations for sufficiency of $\Fbb_q$ codes in MDCS instances  as the forbidden minor characterizations for representability of $\Fbb_q$ in matroids. Not surprisingly, we have the following theorem.

\begin{theorem} \label{thm:SrcDel}
Suppose a MDCS instance $\Asf'=(\Xbf_{\setminus k},\Emc',\Dmc',\Lbf',\Gmc')$ is obtained by deleting $X_k$ from another MDCS instance $\Asf=(\Xbf_{[[K]]},\Emc,\Dmc,\Lbf,\Gmc)$, then
\begin{equation}\label{eq:srcDeleq1}
\Rmc(\Asf')={\rm Proj}_{\Xbf_{\setminus k},\Rbf_{\Emc'}}\left(\Rmc(\Asf)\cap \left\{H(X_k)=0\right\}\right),
\end{equation}
\begin{equation}\label{eq:srcDeleq2}
\Rmc_q(\Asf')={\rm Proj}_{\Xbf_{\setminus k},\Rbf_{\Emc'}}\left(\Rmc_q(\Asf)\cap \left\{ H(X_k)=0\right\}\right),
\end{equation}
and similarly
\begin{equation}\label{eq:srcDeleq3}
\Rmc_{s,q}(\Asf')={\rm Proj}_{\Xbf_{\setminus k},\Rbf_{\Emc'}}\left(\Rmc_{s,q}(\Asf)\cap \left\{H(X_k)=0\right\}\right).
\end{equation}
\begin{equation}\label{eq:srcDeleq4}
\Rmc_{sp}(\Asf')={\rm Proj}_{\Xbf_{\setminus k},\Rbf_{\Emc'}}\left(\Rmc_{sp}(\Asf)\cap \left\{ H(X_k)=0\right\}\right).
\end{equation}
\end{theorem}
\begin{IEEEproof}
Select any point $\Rbf'\in \Rmc(\Asf')$, then there exist random variables $\{\Xbf_{\setminus k},U_i,i\in \Emc'\}$ such that their entropies satisfy all the constraints in \eqref{eq:generalrateregion} determined by $\Asf'$. Define $X_k$ to be the empty sources, $H(X_k)=0$. Then the entropies of random variables $\{\Xbf_{\setminus k},U_i,i\in \Emc'\}\cup X_k$ will satisfy the constraints in $\Asf$ with $H(X_k)=0$. Hence, the associated rate point $\Rbf\in \Rmc(\Asf)\cap \left\{H(X_k)=0\right\}$. Thus, we have $\Rmc(\Asf')\subseteq {\rm Proj}_{\Xbf_{\setminus k},\Rbf_{\Emc'}}(\Rmc(\Asf)\cap \{H(X_k)=0\})$. If $\Rbf'$ is achievable by $\Fbb_q$ codes or superposition coding, since letting $H(X_k)=0$ does not affect the other sources and codes, the same $\Fbb_q$ code, or superposition coding, will also achieve the point $\Rbf$ with $H(X_k)=0$. Thus, we have 
\begin{eqnarray}
\Rmc_q(\Asf')&\subseteq&{\rm Proj}_{\Xbf_{\setminus k},\Rbf_{\Emc'}}(\Rmc_q(\Asf)\cap \{H(X_k)=0\}),\\
\Rmc_{s,q}(\Asf')&\subseteq&{\rm Proj}_{\Xbf_{\setminus k},\Rbf_{\Emc'}}(\Rmc_{s,q}(\Asf)\cap \{H(X_k)=0\}),\\
\Rmc_{sp}(\Asf')&\subseteq&{\rm Proj}_{\Xbf_{\setminus k},\Rbf_{\Emc'}}(\Rmc_{sp}(\Asf)\cap \{H(X_k)=0\}).
\end{eqnarray}

On the other hand, if we select any point $\Rbf\in \Rmc(\Asf)\cap \{H(X_k)=0\}$, we can see that $\Rbf'={\rm Proj}_{\Xbf_{\setminus k},\Rbf_{\Emc'}}(\Rbf)\in \Rmc(\Asf')$ because $\Rbf'$ is still entropic and the entropies of $\{\Xbf_{\setminus k},U_i,i\in \Emc'\}$ satisfy all constraints determined by $\Asf'$. Thus, we have 
\begin{equation}
{\rm Proj}_{\Xbf_{\setminus k},\Rbf_{\Emc'}}(\Rmc(\Asf)\cap \{H(X_k)=0\})\subseteq\Rmc(\Asf').
\end{equation}

If $\Rbf$ is achievable by $\Fbb_q$ code $\Cbb$, then the code to achieve $\Rbf'$ could be the code $\Cbb$ with deletion of rows associated with source $X_k$, i.e., $\Cbb'=\Cbb_{\setminus I(X_k),:}$. Similarly, if $\Rbf$ is achievable by superposition coding, $\Rbf'$ can be achieved by same superposition coding without coding $X_k$. Thus, 
\begin{eqnarray}
{\rm Proj}_{\Xbf_{\setminus k},\Rbf_{\Emc'}}(\Rmc_q(\Asf)\cap \{H(X_k)=0\})&\subseteq&\Rmc_q(\Asf'),\\
{\rm Proj}_{\Xbf_{\setminus k},\Rbf_{\Emc'}}(\Rmc_{s,q}(\Asf)\cap \{H(X_k)=0\})&\subseteq&\Rmc_{s,q}(\Asf'),\\
{\rm Proj}_{\Xbf_{\setminus k},\Rbf_{\Emc'}}(\Rmc_{sp}(\Asf)\cap \{H(X_k)=0\})&\subseteq&\Rmc_{sp}(\Asf').
\end{eqnarray}
\end{IEEEproof}

\begin{theorem} \label{thm:EncDel}
Suppose a MDCS instance $\Asf'=(\Xbf_{[[K]]},\Emc',\Dmc',\Lbf',\Gmc')$ is obtained by contracting $E_e$ from another MDCS instance $\Asf=(\Xbf_{[[K]]},\Emc,\Dmc,\Lbf,\Gmc)$, i.e., $\Asf'=\Asf\setminus E_e$, then
\begin{eqnarray}
\Rmc(\Asf')&=&{\rm Proj}_{H(X_k),k\in[[K]],\Rbf_{\Emc'}}\Rmc(\Asf), \label{eq:encDeleq1}\\
\Rmc_q(\Asf')&=&{\rm Proj}_{H(X_k),k\in[[K]],\Rbf_{\Emc'}}\Rmc_q(\Asf), \label{eq:encDeleq2}\\
\Rmc_{s,q}(\Asf')&\supseteq&{\rm Proj}_{H(X_k),k\in[[K]],\Rbf_{\Emc'}}\Rmc_{s,q}(\Asf), \label{eq:encDeleq3}\\
\Rmc_{sp}(\Asf')&=&{\rm Proj}_{H(X_k),k\in[[K]],\Rbf_{\Emc'}}\Rmc_{sp}(\Asf), \label{eq:encDeleq4}
\end{eqnarray}
\end{theorem}
\begin{IEEEproof}
Select any point $\Rbf'\in \Rmc'$, then there exist random variables $\{\Xbf_{[[K]]},U_i,i\in \Emc'\}$ such that their entropies satisfy all the constraints in \eqref{eq:generalrateregion} determined by $\Asf'$. Define $U_e$ to be the concatenation of all sources, $U_e=\Xbf_{[[K]]}$. Then the entropies of random variables $\{\Xbf_{[[K]]},U_i,i\in \Emc'\}\cup U_e$ will satisfy the constraints in $\Asf$, and additionally obey $H(U_e)=\sum_{k=1}^K H(X_k)$. Hence, the associated rate point $\Rbf\in \Rmc(\Asf)\cap \left\{ H(U_e)\geq \sum_{k=1}^K H(X_k)\right\}$. Thus, we have 
\begin{equation} \label{eq:pfeq1encDel1}
\Rmc(\Asf')\subseteq{\rm Proj}_{H(X_k),k\in[[K]],\Rbf_{\Emc'}}(\Rmc(\Asf)\cap \{H(U_e)\geq \sum_{k=1}^K H(X_k)\})\subseteq {\rm Proj}_{H(X_k),k\in[[K]],\Rbf_{\Emc'}}\Rmc(\Asf).
\end{equation}

If $\Rbf'$ is achievable by general $\Fbb_q$ codes or superposition coding, since concatenation of all sources is a valid $\Fbb_q$ code and is superposition coding, we have
\begin{eqnarray}
\Rmc_q(\Asf')\subseteq{\rm Proj}_{H(X_k),k\in[[K]],\Rbf_{\Emc'}}(\Rmc_q(\Asf)\cap \{H(U_e)\geq \sum_{k=1}^K H(X_k)\})\subseteq {\rm Proj}_{H(X_k),k\in[[K]],\Rbf_{\Emc'}}\Rmc_q(\Asf),\label{eq:pfeq1encDel2}\\
\Rmc_{sp}(\Asf')\subseteq{\rm Proj}_{H(X_k),k\in[[K]],\Rbf_{\Emc'}}(\Rmc_{sp}(\Asf)\cap \{H(U_e)\geq \sum_{k=1}^K H(X_k)\})\subseteq {\rm Proj}_{H(X_k),k\in[[K]],\Rbf_{\Emc'}}\Rmc_{sp}(\Asf).\label{eq:pfeq1encDel4}
\end{eqnarray}

However, we cannot establish same relationship when scalar $\Fbb_q$ codes are considered, because for the point $\Rbf'$, the associated $\Rbf$ with $H(U_e)$ may not be scalar $\Fbb_q$ achievable. 

On the other hand, if we select any point $\Rbf\in \Rmc(\Asf)$, we can see that $\Rbf'={\rm Proj}_{H(X_k),k\in[[K]],\Rbf_{\Emc'}}\Rbf\in \Rmc(\Asf')$ because $\Rbf'$ is still entropic and the entropies of $\{\Xbf_{[[K]]},U_i,i\in \Emc'\}$ satisfy all constraints determined by $\Asf'$, since they are a subset of the constraints from $\Asf$. Thus, we have 
\begin{equation}\label{eq:pfeq2encDel1}
{\rm Proj}_{H(X_k),k\in[[K]],\Rbf_{\Emc'}}\Rmc(\Asf)\subseteq\Rmc(\Asf').
\end{equation}

If $\Rbf\in \Rmc(\Asf)$ is achievable by $\Fbb_q$ code $\Cbb$, then the code to achieve $\Rbf'={\rm Proj}_{H(X_k),k\in[[K]],\Rbf_{\Emc'}}\Rbf\in \Rmc(\Asf')$ could be the code $\Cbb$ with deletion of columns associated with encoder $E_e$, i.e., $\Cbb'=\Cbb_{:,\setminus E_e}$. Thus, we have 
\begin{equation}\label{eq:pfeq2encDel2}
{\rm Proj}_{H(X_k),k\in[[K]],\Rbf_{\Emc'}}\Rmc_q(\Asf)\subseteq\Rmc_q(\Asf').
\end{equation}

If $\Rbf$ is achievable by scalar $\Fbb_q$ code $\Cbb^1$, then the code to achieve $\Rbf'$ could be the code $\Cbb^{1}$ with deletion of the column associated with encoder $E_e$, i.e., $\Cbb^{1'}=\Cbb^{1}_{:,\setminus I(U_e)}$. Thus, we have 
\begin{equation}\label{eq:pfeq1encDel3}
{\rm Proj}_{H(X_k),k\in[[K]],\Rbf_{\Emc'}}\Rmc_{s,q}(\Asf)\subseteq\Rmc_{s,q}(\Asf').
\end{equation}

Similarly, if $\Rbf$ is achievable by superposition coding, the same superposition codes can achieve $\Rbf'$ without coding in $E_e$. Thus, we have
\begin{equation}\label{eq:pfeq1encDel4}
{\rm Proj}_{H(X_k),k\in[[K]],\Rbf_{\Emc'}}\Rmc_{sp}(\Asf)\subseteq\Rmc_{sp}(\Asf').
\end{equation}

Combination of \eqref{eq:pfeq1encDel1} and \eqref{eq:pfeq2encDel1} gives \eqref{eq:encDeleq1}. Combination of \eqref{eq:pfeq1encDel2} and \eqref{eq:pfeq2encDel2} gives \eqref{eq:encDeleq2}. \eqref{eq:pfeq1encDel3} indicates \eqref{eq:encDeleq3}.
\end{IEEEproof}

\begin{theorem} \label{thm:EncCont}
Suppose a MDCS instance $\Asf'=(\Xbf_{[[K]]},\Emc',\Dmc',\Lbf',\Gmc')$ is obtained by deleting $E_e$ from another MDCS instance $\Asf=(\Xbf_{[[K]]},\Emc,\Dmc,\Lbf,\Gmc)$, then
\begin{equation}\label{eq:encConteq1}
\Rmc(\Asf')={\rm Proj}_{H(X_k),k\in[[K]],\Rbf_{\Emc'}}(\Rmc(\Asf)\cap \{H(U_e)=0\}),
\end{equation}
\begin{equation}\label{eq:encConteq2}
\Rmc_q(\Asf')={\rm Proj}_{H(X_k),k\in[[K]],\Rbf_{\Emc'}}(\Rmc_q(\Asf)\cap \{H(U_e)=0\}),
\end{equation}
and similarly
\begin{equation}\label{eq:encConteq3}
\Rmc_{s,q}(\Asf')={\rm Proj}_{H(X_k),k\in[[K]],\Rbf_{\Emc'}}(\Rmc_{s,q}(\Asf)\cap \{H(U_e)=0\}),
\end{equation}
\begin{equation}\label{eq:encConteq4}
\Rmc_{sp}(\Asf')={\rm Proj}_{H(X_k),k\in[[K]],\Rbf_{\Emc'}}(\Rmc_{sp}(\Asf)\cap \{H(U_e)=0\}).
\end{equation}
\end{theorem}
\begin{IEEEproof}
Select any point $\Rbf'\in \Rmc(\Asf')$, then there exist random variables $\{\Xbf_{[[K]]},U_i,i\in \Emc'\}$ such that their entropies satisfy all the constraints in \eqref{eq:generalrateregion} determined by $\Asf'$. Let $U_e$ be empty set or encode all sources with the all-zero vector, $U_e=\emptyset$. Then the entropies of random variables $\{\Xbf_{[[K]]},U_i,i\in \Emc'\}\cup U_e$ will satisfy the constraints in $\Asf$, and additionally obey $H(U_e)=0$. Note that, even in the special case where there $\exists D_d,d\in \Dmc$ such that ${\rm Fan}(D_d)=E_e$, according to Definition \ref{def:EncCont}, after contraction of $E_e$, the decoder $D'_d$ will have no access to any decoder but needs to keep the same decoding capability, i.e., ${\rm Lev}(D'_d)={\rm Lev}(D_d)$. However, since ${\rm Fan}(D'_d)=\emptyset$, deletion of $E_e$ will result in deletion of all decoders of level ${\rm Lev}(D_d)$ or less, and deletion of sources $\Xbf_{1:{\rm Lev}(D_d)}$, i.e., $H(X_k)=0,k=1,\ldots,{\rm Lev}(D_d)$. Then it is still true that entropies of random variables $\{\Xbf_{[[K]]},U_i,i\in \Emc'\}\cup U_e$ will satisfy the constraints in $\Asf$. Hence, the associated rate point $\Rbf\in \Rmc(\Asf)\cap \left\{H(U_e)= 0\right\}$. Thus, we have 
\begin{equation}\label{eq:pfeq1encCont1}
\Rmc(\Asf')\subseteq{\rm Proj}_{H(X_k),k\in[[K]],\Rbf_{\Emc'}}(\Rmc(\Asf)\cap \{H(U_e)=0\}).
\end{equation}

If $\Rbf'$ is achievable by $\Fbb_q$ linear vector or scalar codes, or superposition coding,  since all-zero code is a valid $\Fbb_q$ code and superposition code, we have 
\begin{equation}\label{eq:pfeq1encCont2}
\Rmc_q(\Asf')\subseteq{\rm Proj}_{H(X_k),k\in[[K]],\Rbf_{\Emc'}}(\Rmc_q(\Asf)\cap \{H(U_e)=0\}),
\end{equation}
\begin{equation}\label{eq:pfeq1encCont3}
\Rmc_{s,q}(\Asf')\subseteq{\rm Proj}_{H(X_k),k\in[[K]],\Rbf_{\Emc'}}(\Rmc_{s,q}(\Asf)\cap \{H(U_e)=0\}),
\end{equation}
\begin{equation}\label{eq:pfeq1encCont4}
\Rmc_{sp}(\Asf')\subseteq{\rm Proj}_{H(X_k),k\in[[K]],\Rbf_{\Emc'}}(\Rmc_{sp}(\Asf)\cap \{H(U_e)=0\}).
\end{equation}

On the other hand, if we select any point $\Rbf\in \Rmc(\Asf)\cap \{H(U_e)=0\}$, we can see that $\Rbf'={\rm Proj}_{H(X_k),k\in[[K]],\Rbf_{\Emc'}}(\Rbf)\in \Rmc(\Asf')$ because $\Rbf'$ is still entropic and the entropies of $\{\Xbf_{[[K]]},U_i,i\in \Emc'\}$ satisfy all constraints determined by $\Asf'$, which is still true when the special case happens. Thus, we have 
\begin{equation}\label{eq:pfeq2encCont1}
{\rm Proj}_{H(X_k),k\in[[K]],\Rbf_{\Emc'}}(\Rmc(\Asf)\cap \{H(U_e)=0\})\subseteq \Rmc(\Asf').
\end{equation}

If $\Rbf$ is achievable by a $\Fbb_q$ code, vector or scalar, or a superposition code, $\Cbb$, then the code to achieve $\Rbf'$ could be the code $\Cbb$ with the deletion of the columns associated with encoder $E_e$, i.e., $\Cbb'=\Cbb_{:,\setminus I(U_e)}$, because $E_e$ is sending nothing. Thus, we have 
\begin{equation}\label{eq:pfeq2encCont2}
{\rm Proj}_{H(X_k),k\in[[K]],\Rbf_{\Emc'}}(\Rmc_q(\Asf)\cap \{H(U_e)=0\})\subseteq\Rmc_q(\Asf'),
\end{equation}
\begin{equation}\label{eq:pfeq2encCont3}
{\rm Proj}_{H(X_k),k\in[[K]],\Rbf_{\Emc'}}(\Rmc_{s,q}(\Asf)\cap \{H(U_e)=0\})\subseteq\Rmc_{s,q}(\Asf'),
\end{equation}
\begin{equation}\label{eq:pfeq2encCont4}
{\rm Proj}_{H(X_k),k\in[[K]],\Rbf_{\Emc'}}(\Rmc_{sp}(\Asf)\cap \{H(U_e)=0\})\subseteq\Rmc_{sp}(\Asf').
\end{equation}

Combination of \eqref{eq:pfeq1encCont1} and \eqref{eq:pfeq2encCont1} gives \eqref{eq:encConteq1}. Combination of \eqref{eq:pfeq1encCont2} and \eqref{eq:pfeq2encCont2} gives \eqref{eq:encConteq2}. Combination of \eqref{eq:pfeq1encCont3} and \eqref{eq:pfeq2encCont3} gives \eqref{eq:encConteq3}.
\end{IEEEproof}

\begin{theorem} \label{thm:EncUniCont}
Suppose a MDCS instance $\Asf'=(\Xbf_{[[K]]},\Emc',\Dmc',\Lbf',\Gmc')$ is obtained by unifying $E_f$ with $E_e$, from another MDCS instance $\Asf=(\Xbf_{[[K]]},\Emc,\Dmc,\Lbf,\Gmc)$. Let $\Lmc_1,\Lmc_2,\Lmc_5,\Lmc''_4$ be the constraints for $\Asf$ used in \eqref{eq:generalrateregionfree} to obtain $\Rmc(\Asf)$. Let 
\begin{equation}
\Lmc_4^0=\{(\mathbf{h}^T,\Rbf^T)^T\in\mathbb{R}_{+}^{2^N-1+|\Emc|}:R_{e'}\geq H(U_{e},U_f),R_i\geq H(U_i),i\in\Emc\setminus e\}
\end{equation} then
\begin{eqnarray}
\Rmc(\Asf')&=&{\rm Proj}_{H(X_k),k\in[[K]],R_{\Emc\setminus \{e,f\}},R_{e'}}\overline{\rm{con}(\Gamma_{N}^{*}\cap\mathcal{L}_{12})}\cap\mathcal{L}_5\cap \Lmc_4^0,\label{eq:encUniCont1}\\
\Rmc_q(\Asf')&=&{\rm Proj}_{H(X_k),k\in[[K]],R_{\Emc\setminus \{e,f\}},R_{e'}}(\Gamma_{N,\infty}^{q}\cap\mathcal{L}_{125}\cap \Lmc_4^0),\label{eq:encUniCont2}\\
\Rmc_{sp}(\Asf')&=&\mathrm{Proj}_{H(X_k),k\in[[K]],R_{\Emc\setminus \{e,f\}},R_{e'}}(\overline{\rm{con}(\Gamma_{N}^{*}\cap\mathcal{L}_{12})}\cap\mathcal{L}'_5\cap \mathcal{L}_4^0).\label{eq:encUniCont4}
\end{eqnarray} where the dimension $R_{e}$ is replaced with $R_{e'}$ in $\Lmc_4^0$ and the projection.
\end{theorem}
\begin{IEEEproof}
Select any point $\Rbf'\in \Rmc(\Asf')$, then there exist random variables $\{\Xbf_{[[K]]},U_i,i\in \Emc'\}$ such that their entropies satisfy all the constraints determined by $\Asf'$ and $R_i\geq H(U_i),i\in \Emc'$. Let $U_e=U_{e'},U_f=U_{e'}$, so that $H(U_e)=H(U_f)=H(U_{e'})=H(U_e,U_f)$ and $R_e=R_f=R_{e'}$. Then the entropies of random variables $\{\Xbf_{[[K]]},U_i,i\in \Emc'\setminus e'\}\cup \{U_e,U_f\}$ will satisfy the constraints in $\Asf$, and additionally $R_i,H(U_i),i\in \Emc$ will obey $\Lmc_4^0$. Hence, the associated rate point $\Rbf\in {\rm Proj}_{H(X_k),k\in[[K]],R_{\Emc\setminus \{e,f\}},R_{e'}}\overline{\rm{con}(\Gamma_{N}^{*}\cap\mathcal{L}_{12})}\cap\mathcal{L}_5\cap \Lmc_4^0$. Thus, we have 
\begin{equation}
\label{eq:pfeq1encUniCont1}
\Rmc(\Asf')\subseteq{\rm Proj}_{H(X_k),k\in[[K]],R_{\Emc\setminus \{e,f\}},R_{e'}}\overline{\rm{con}(\Gamma_{N}^{*}\cap\mathcal{L}_{12})}\cap\mathcal{L}_5\cap \Lmc_4^0.
\end{equation}

If $\Rbf'$ is achievable by $\Fbb_q$ vector codes or superposition coding, since $U_e,U_f$ are replicating $U_{e'}$ with exactly the same code, $\Rbf$ is also $\Fbb_q$ achievable or superposition coding achievable. Then we have 
\begin{eqnarray}\label{eq:pfeq1encUniCont2}
\Rmc_q(\Asf')&\subseteq&{\rm Proj}_{H(X_k),k\in[[K]],R_{\Emc\setminus \{e,f\}},R_{e'}}(\Gamma_{N,\infty}^{q}\cap\mathcal{L}_{125}\cap \Lmc_4^0),\\
\label{eq:pfeq1encUniCont4}
\Rmc_{sp}(\Asf')&\subseteq&{\rm Proj}_{H(X_k),k\in[[K]],R_{\Emc\setminus \{e,f\}},R_{e'}}\overline{\rm{con}(\Gamma_{N}^{*}\cap\mathcal{L}_{12})}\cap\mathcal{L}'_5\cap \Lmc_4^0.
\end{eqnarray}

On the other hand, if we select any point $\Rbf\in {\rm Proj}_{H(X_k),k\in[[K]],R_{\Emc\setminus \{e,f\}},R_{e'}}\overline{\rm{con}(\Gamma_{N}^{*}\cap\mathcal{L}_{12})}\cap\mathcal{L}_5\cap \Lmc_4^0$, there exist random variables $\{\Xbf_{[[K]]},U_i,i\in \Emc\}$ such that their entropies satisfy all the constraints determined by $\Asf$ together with $R_i\geq H(U_i),i\in \Emc$ and $R_{e'}\geq H(U_e,U_f)$. Let $U_{e'}$ be the concatenation of $U_e,U_f$ so that $H(U_{e'})=H(U_e,U_f)$. After unification, since all decoders that are fan of $E_e,E_f$ will have access to $E_{e'}$, $\Rbf'$ will satisfy all constraints determined in $\Asf'$. Thus, $\Rbf'\in \Rmc(\Asf')$. Hence, we have \begin{equation}
{\rm Proj}_{H(X_k),k\in[[K]],R_{\Emc\setminus \{e,f\}},R_{e'}}\overline{\rm{con}(\Gamma_{N}^{*}\cap\mathcal{L}_{12})}\cap\mathcal{L}_5\cap \Lmc_4^0\subseteq\Rmc(\Asf').
\end{equation} 
If $\Rbf$ is achievable by $\Fbb_q$ vector codes $\Cbb$ with time-sharing between basic solutions $\Cbb^{(1)},\Cbb^{(2)},\ldots,\Cbb^{(m)}$, then the odes to achieve $\Rbf'$ could be same time-sharing between basic solutions $\Cbb^{(1')},\Cbb^{(2')},\ldots,\Cbb^{(m')}$, where $\Cbb^{(i')}_{:,I_{i'}(U_j)}=\Cbb^{(i)}_{:,I_i(U_j)},j\in\Emc\setminus \{e,f\}$ and $\Cbb^{(i')}_{:,I_{i'}(U_e)}=\left[\Cbb^{(i)}_{:,I_i(U_e)} \ \Cbb^{(i)}_{:,I_i(U_f)\setminus \Bmc}\right],\Bmc=\left\{j\in I_i(U_f) \left|\Cbb^{(i)}_{:,j}\in {\rm Span}(\Cbb^{(i)}_{:,I_i(U_e)})\right.\right\},i\in\{1,2,\ldots,m\}$. Thus, 
\begin{equation}
{\rm Proj}_{H(X_k),k\in[[K]],R_{\Emc\setminus \{e,f\}},R_{e'}}\overline{\rm{con}(\Gamma_{N}^{q}\cap\mathcal{L}_{12})}\cap\mathcal{L}_5\cap \Lmc_4^0\subseteq\Rmc_q(\Asf').
\end{equation}

Similarly, if $\Rbf\in \mathrm{Proj}_{H(X_k),k\in[[K]],R_{\Emc\setminus \{e,f\}},R_{e'}}(\Gamma_{N}^{*}\cap\mathcal{L}_{12})\cap\mathcal{L}'_5\cap \mathcal{L}_4^0)$ is achieved by superposition coding, one can use entropy approaching codes, e.g., Huffman code with sufficient number of blocks, to jointly encode each component in $U_e,U_f$ for the sources so that $H(U_{e'}^{X_k})=H(U_e^{X_k},U_f^{X_k}),k\in[[K]]$. Thus, we have
\begin{equation}
{\rm Proj}_{H(X_k),k\in[[K]],R_{\Emc\setminus \{e,f\}},R_{e'}}\overline{\rm{con}(\Gamma_{N}^{*}\cap\mathcal{L}_{12})}\cap\mathcal{L}'_5\cap \Lmc_4^0\subseteq\Rmc_{sp}(\Asf').
\end{equation}

\end{IEEEproof}

The previous theorem presents a form of the rate regions which is best for proving the desired inheritance properties, but the form is not extendable to the scalar coding region.  We now present an alternate representation of the rate regions that also holds for the scalar coding region, but is not as useful for proving the desired inheritance properties.

\begin{theorem} \label{thm:EncUnif}
Suppose a MDCS instance $\Asf'=(\Xbf_{[[K]]},\Emc',\Dmc',\Lbf',\Gmc')$ is obtained by unifying $E_f$ with $E_e$, from another MDCS instance $\Asf=(\Xbf_{[[K]]},\Emc,\Dmc,\Lbf,\Gmc)$. Let $\Lmc_1,\Lmc_2,\Lmc_5,\Lmc''_4$ be the constraints for $\Asf$ used in \eqref{eq:generalrateregionfree} to obtain $\Rmc(\Asf)$. Let 
\begin{equation}
\Lmc_4^1=\{(\mathbf{h}^T,\Rbf^T)^T\in\mathbb{R}_{+}^{2^N-1+|\Emc|}:H(U_e)=H(U_f)= H(U_{e},U_f),R_i\geq H(U_i),i\in\Emc\},
\end{equation} then 
\begin{eqnarray}
\Rmc(\Asf')&=&{\rm Proj}_{H(X_k),k\in[[K]],R_e,e\in{\Emc'}}\overline{\rm{con}(\Gamma_{N}^{*}\cap\mathcal{L}_{12})}\cap\mathcal{L}_5\cap \Lmc_4^1,\\
\Rmc_q(\Asf')&=&{\rm Proj}_{H(X_k),k\in[[K]],R_e,e\in{\Emc'}}(\Gamma_{N,\infty}^{q}\cap\mathcal{L}_{125}\cap \Lmc_4^1),\\
\Rmc_{s,q}(\Asf')&=&{\rm Proj}_{H(X_k),k\in[[K]],R_e,e\in{\Emc'}}(\Gamma_{N}^{q}\cap\mathcal{L}_{125})\cap \Lmc_4^1),\\
\Rmc_{sp}(\Asf')&=&\mathrm{Proj}_{H(X_k),k\in[[K]],R_e,e\in{\Emc'}}
(\overline{\rm{con}(\Gamma_{N}^{*}\cap\mathcal{L}_{12})}\cap\mathcal{L}'_5\cap \mathcal{L}_4^1).
\end{eqnarray}
\end{theorem}
\begin{IEEEproof}
Select any point $\Rbf'\in \Rmc(\Asf')$, then there exist random variables $\{\Xbf_{[[K]]},U_i,i\in \Emc'\}$ such that their entropies satisfy all the constraints determined by $\Asf'$ and $R_i\geq H(U_i),i\in \Emc'$. Let $U_e=U_f,R_f=R_e$, so that $H(U_e)=H(U_f)=H(U_e,U_f)$. Then the entropies of random variables $\{\Xbf_{[[K]]},U_i,i\in \Emc'\}\cup U_f$ will satisfy the constraints in $\Asf$, and additionally $R_i,H(U_i),i\in \Emc$ will obey $\Lmc_4^1$. Hence, the associated rate point $\Rbf\in {\rm Proj}_{H(X_k),k\in[[K]],R_e,e\in{\Emc'}}\overline{\rm{con}(\Gamma_{N}^{*}\cap\mathcal{L}_{12})}\cap\mathcal{L}_5\cap \Lmc_4^1$. Thus, we have 
\begin{equation}\label{eq:pfeq1encUnif1}
\Rmc(\Asf')\subseteq{\rm Proj}_{H(X_k),k\in[[K]],R_e,e\in{\Emc'}}\overline{\rm{con}(\Gamma_{N}^{*}\cap\mathcal{L}_{12})}\cap\mathcal{L}_5\cap \Lmc_4^1.
\end{equation}

If $\Rbf'$ is achievable by a $\Fbb_q$ codes, scalar or vector, or a superposition code, since $U_f$ is replicating $U_{e}$ with exactly the same code, $\Rbf$ is also $\Fbb_q$ achievable. Then we have 
\begin{eqnarray}
\Rmc_q(\Asf')&\subseteq&{\rm Proj}_{H(X_k),k\in[[K]],R_e,e\in{\Emc'}}(\Gamma_{N,\infty}^{q}\cap\mathcal{L}_{125}\cap \Lmc_4^1),\label{eq:pfeq1encUnif2}\\
\Rmc_{s,q}(\Asf')&\subseteq&{\rm Proj}_{H(X_k),k\in[[K]],R_e,e\in{\Emc'}}(\Gamma_{N}^{q}\cap\mathcal{L}_{125}\cap \Lmc_4^1),\label{eq:pfeq1encUnif3}\\
\Rmc_{sp}(\Asf')&\subseteq&{\rm Proj}_{H(X_k),k\in[[K]],R_e,e\in{\Emc'}}\overline{\rm{con}(\Gamma_{N}^{*}\cap\mathcal{L}_{12})}\cap\mathcal{L}'_5\cap \Lmc_4^1.\label{eq:pfeq1encUnif4}
\end{eqnarray}

On the other hand, if we select any point $\Rbf\in {\rm Proj}_{H(X_k),k\in[[K]],R_e,e\in{\Emc'}}\overline{\rm{con}(\Gamma_{N}^{*}\cap\mathcal{L}_{12})}\cap\mathcal{L}_5\cap \Lmc_4^1$, there exist random variables $\{\Xbf_{[[K]]},U_i,i\in \Emc\}$ such that their entropies satisfy all the constraints determined by $\Asf$ together with $R_i\geq H(U_i),i\in \Emc$ and $H(U_e)=H(U_f)= H(U_e,U_f)$. After unification, since all decoders that are fan of $E_e,E_f$ will have access to $E_e$, $\Rbf'$ will satisfy all constraints determined in $\Asf'$ because $U_f$ is dependent on $U_e$. Thus, $\Rbf'\in \Rmc(\Asf')$. Hence, we have 
\begin{equation}
{\rm Proj}_{H(X_k),k\in[[K]],R_e,e\in{\Emc'}}\overline{\rm{con}(\Gamma_{N}^{*}\cap\mathcal{L}_{12})}\cap\mathcal{L}_5\cap \Lmc_4^1\subseteq\Rmc(\Asf').
\end{equation}

If $\Rbf$ is achievable by a $\Fbb_q$ code, scalar or vector, or a superposition code, $\Cbb$, then the codes to achieve $\Rbf'$ could be same as $\Cbb$ with deletion of columns associated with encoder $E_f$. Thus, 
\begin{eqnarray}
{\rm Proj}_{H(X_k),k\in[[K]],R_e,e\in{\Emc'}}(\Gamma_{N,\infty}^{q}\cap\mathcal{L}_{125}\cap \Lmc_4^1&\subseteq&\Rmc_q(\Asf'),\\
{\rm Proj}_{H(X_k),k\in[[K]],R_e,e\in{\Emc'}}(\Gamma_{N}^{q}\cap\mathcal{L}_{125}\cap \Lmc_4^1&\subseteq&\Rmc_{s,q}(\Asf'),\\
{\rm Proj}_{H(X_k),k\in[[K]],R_e,e\in{\Emc'}}\overline{\rm{con}(\Gamma_{N}^{*}\cap\mathcal{L}_{12})}\cap\mathcal{L}_5\cap \Lmc_4^1&\subseteq&\Rmc_{sp}(\Asf').
\end{eqnarray}
\end{IEEEproof}

\begin{corollary} \label{cor:embedded}
Given two MDCS instances $\Asf,\Asf'$ such that $\Asf'\prec_v\Asf$.  If $\Fbb_q$ linear vector codes suffice for $\Asf$, then $\Fbb_q$ linear vector codes suffice for $\Asf'$. Equivalently, if $\Fbb_q$ linear vector codes do not suffice for $\Asf'$, then $\Fbb_q$ linear vector codes do not suffice for $\Asf$. Equivalently, if $\Rmc_q(\Asf)=\Rmc(\Amc)$, then $\Rmc_q(\Amc')=\Rmc(\Amc')$.
\end{corollary}
\begin{IEEEproof}
From Definition \ref{def:embedded} we know that $\Asf'$ is obtained by a series of operations of source deletion, encoder deletion, encoder contraction, and encoder unified contraction. Theorem \ref{thm:order} indicates that the order of the operations does not matter. Thus, it suffices to show that the statement holds when $\Asf'$ can be obtained by one of the operations of source deletion, encoder deletion, encoder contraction, or encoder unification on $\Asf$. 

Suppose $\Fbb_q$ linear codes suffice to achieve every point in $\Rmc(\Asf)$, i.e., $\Rmc_q(\Asf)=\Rmc(\Asf)$. If $\Asf'$ is obtained by contracting one encoder in $\Asf$, \eqref{eq:encDeleq1} and \eqref{eq:encDeleq2} in Theorem \ref{thm:EncDel} indicate $\Rmc_q(\Asf')=\Rmc(\Asf)$. Similarly, same conclusion is obtained from \eqref{eq:srcDeleq1} and \eqref{eq:srcDeleq2} in Theorem \ref{thm:SrcDel},  \eqref{eq:encConteq1} and \eqref{eq:encConteq2} in Theorem \ref{thm:EncCont},  when $\Asf'$ is obtained by source deletion or deletion deletion. 

When $\Asf'$ is obtained by unifying two encoders in $\Asf$ and $\Rmc_q(\Asf)=\Rmc(\Asf)$. Trivially, $\Rmc_q(\Asf')\subseteq \Rmc(\Asf')$ since $\Gamma_{N,\infty}^q\subseteq \bar{\Gamma}_N^*$. For the other direction, we pick a point $\Rbf'\in \Rmc(\Asf')$. Note that $\Rmc(\Asf')\subseteq {\rm Proj}_{R_e,H(X_k),R_{e'},e\in\Emc\setminus \{e,f\},k\in[[K]]}\Rmc(\Asf)$ since $H({U_e,U_f})\geq H(U_e)$ and $\Lmc_4^0$ makes the region smaller than $\Rmc(\Asf)$.  Since $\Rmc_q(\Asf)=\Rmc(\Asf)$, we see that $\Rbf'\in {\rm Proj}_{R_e,H(X_k),R_{e'},e\in\Emc\setminus \{e,f\},k\in[[K]]}\Rmc_q(\Asf)$, which means that there exist $\hbf\in \Gamma_{N,\infty}^q$ and $R_i,i\in \Emc$ such that their entropies satisfy constraints determined by $\Asf$ and $R_{e'}\geq H(U_e,U_f)$. The $\Fbb_q$ code for $U_{e'}$ is the concatenation (with removal of redundancy) of $\Fbb_q$ codes for $U_e,U_f$. That is, $\Rbf'\in \Rmc_q(\Asf')$. Thus, $\Rmc(\Asf')\subseteq \Rmc_q(\Asf')$. Therefore, we have $\Rmc(\Asf')=\Rmc_q(\Asf')$.

We see that for one-step operations, sufficiency of $\Fbb_q$ codes is preserved. The statement holds in general.
\end{IEEEproof}


Note that scalar codes are a spacial class of general linear codes. If we only consider scalar linear codes and operations of source deletion, encoder deletion/ contraction, we will have the following similar corollary.

\begin{corollary}
Given two MDCS instances $\Asf,\Asf'$ such that $\Asf'\prec_s\Asf$. If $\Fbb_q$ scalar linear codes suffice for $\Asf$, then $\Fbb_q$ scalar linear codes suffice for $\Asf'$. Equivalently, if $\Fbb_q$ scalar linear codes do not suffice for $\Asf'$, then $\Fbb_q$ scalar linear codes do not suffice for $\Asf$. Equivalently, if $\Rmc_{s,q}(\Asf)=\Rmc(\Amc)$, then $\Rmc_{s,q}(\Amc')=\Rmc(\Amc')$.
\end{corollary}
\begin{IEEEproof}
If $\Rmc_{s,q}(\Asf)=\Rmc(\Amc)$, we can get $\Rmc_{s,q}(\Amc')=\Rmc(\Amc')$ from \eqref{eq:encConteq1} and \eqref{eq:encConteq3} in Theorem \ref{thm:EncCont} (\eqref{eq:srcDeleq1} and \eqref{eq:srcDeleq3} in Theorem \ref{thm:SrcDel}), if $\Asf'$ is obtained by deleting an encoder (source) from $\Asf$. 

Now we consider the case that $\Asf'$ is obtained by contracting an encoder from $\Asf$. If $\Rmc_{s,q}(\Asf)=\Rmc(\Amc)$, the projections ${\rm Proj}_{H(X_k),k\in[[K]],\Rbf_{\Emc'}}\Rmc_{s,q}(\Asf)={\rm Proj}_{H(X_k),k\in[[K]],\Rbf_{\Emc'}}\Rmc(\Amc)$. Together with \eqref{eq:encDeleq1} and \eqref{eq:encDeleq3}, we will have $\Rmc_{s,q}(\Amc')\supseteq\Rmc(\Amc')$. It is trivial that $\Rmc_{s,q}(\Amc')\subseteq\Rmc(\Amc')$ because $\Gamma_N^q\subseteq \bar{\Gamma}_N^*$. Thus, we have $\Rmc_{s,q}(\Amc')=\Rmc(\Amc')$.
\end{IEEEproof}

Similarly, if superposition coding is considered, we have the following corollary.
\begin{corollary}
Given two MDCS instances $\Asf,\Asf'$ such that $\Asf'\prec_{sp}\Asf$. If superposition coding suffices for $\Asf$, then it also suffices for $\Asf'$. Equivalently, if superposition coding does not suffice for $\Asf'$, then it does not suffice for $\Asf$. Equivalently, if $\Rmc_{sp}(\Asf)=\Rmc(\Amc)$, then $\Rmc_{sp}(\Amc')=\Rmc(\Amc')$.
\end{corollary}
\begin{IEEEproof}
Suppose superposition coding suffices to achieve every point in $\Rmc(\Asf)$, i.e., $\Rmc_q(\Asf)=\Rmc(\Asf)$. If $\Asf'$ is obtained by contracting one encoder in $\Asf$, \eqref{eq:encDeleq1} and \eqref{eq:encDeleq4} in Theorem \ref{thm:EncDel} indicate $\Rmc_q(\Asf')=\Rmc(\Asf)$. Similarly, same conclusion is obtained from \eqref{eq:srcDeleq1} and \eqref{eq:srcDeleq4} in Theorem \ref{thm:SrcDel},  \eqref{eq:encConteq1} and \eqref{eq:encConteq4} in Theorem \ref{thm:EncCont},  when $\Asf'$ is obtained by source deletion or encoder deletion. 

When $\Asf'$ is obtained by unifying two encoders in $\Asf$ and $\Rmc_{sp}(\Asf)=\Rmc(\Asf)$. Trivially, 
$\Rmc_{sp}(\Asf')\subseteq \Rmc(\Asf')$,
since a point achieved by superposition coding must be in the rate region. For the other direction, we pick a point $\Rbf'\in \Rmc(\Asf')$. Note that $\Rmc(\Asf')\subseteq {\rm Proj}_{R_e,H(X_k),R_{e'},e\in\Emc\setminus \{e,f\},k\in[[K]]} \Rmc(\Asf)$ since $H({U_e,U_f})\geq H(U_e)$ and $\Lmc_4^0$ makes the region smaller than $\Rmc(\Asf)$. Then we have $\Rbf'\in {\rm Proj}_{R_e,H(X_k),R_{e'},e\in\Emc\setminus \{e,f\},k\in[[K]]}\Rmc(\Asf)$. Since $\Rmc_{sp}(\Asf)=\Rmc(\Asf)$, we see that $\Rbf'\in {\rm Proj}_{R_e,H(X_k),R_{e'},e\in\Emc\setminus \{e,f\},k\in[[K]]}\Rmc_{sp}(\Asf)$, which means that there exist $\hbf\in \Gamma_{N}^*$ and $R_e=\sum_{i=1}^{K}R_e^{X_k},k\in[[K]], e\in \Emc$ such that their entropies satisfy constraints determined by $\Asf$ with $\Lmc'_5$ in equation \eqref{eq:decsp} and $R_{e'}\geq H(U_e,U_f)$. One can use an entropy approaching code, e.g., Huffman code with a sufficient large block length, to code the concatenation $U_e^{X_k},U_f^{X_k}$ for each $k\in[[K]]$ to obtain an overall superposition code for $U_{e'}$. That is, $\Rbf'\in \Rmc_{sp}(\Asf')$. Thus, $\Rmc(\Asf')\subseteq \Rmc_{sp}(\Asf')$. Therefore, we have $\Rmc(\Asf')=\Rmc_{sp}(\Asf')$.
\end{IEEEproof}

\section{Rate region results on MDCS problems}
\label{sec:results}


In this section, experimental results on thousands of MDCS instances are presented. We investigate rate regions for $7360$ non-isomorphic MDCS instances which represent $134617$ isomorphic instances including the cases when $(K,|\Emc|)=(2,2),(2,3),(2,4),(3,2),(3,3),(3,4)$. For each non-isomorphic MDCS instance, we calculated the bounds on its rate region using Shannon outer bound $\Gamma_N$, scalar binary representable matroid inner bound $\Gamma_N^2$, scalar ternary representable matroid inner bound $\Gamma_N^3$, vector binary representable matroid inner bounds $\Gamma_{N,N+1}^2,\Gamma_{N,N+2}^2,\Gamma_{N,N+3}^2$. If the outer bound on rate region obtained from Shannon outer bound matches with any inner bound from the corresponding inner bound on region of entropic vectors, we not only know the exact rate region but also know the codes which suffice to achieve any point in it. 

A summary of results can be found in Table \ref{MDCSresult}. Since there are thousands of networks, it is impossible to state the rate regions one by one. Interested readers are referred to our website \cite{EntVecSoft} for the complete list of rate regions and other interesting results such as the enumeration of MDCS instances and converse proofs generated by computer. 

Some interpretations of the results from several perspectives with some example networks are presented as follows.

\begin{table*}
\caption{\label{MDCSresult}Sufficiency of codes for MDCS instances: Columns 2--7 show the number of instances that the rate region inner bounds match with the Shannon outer bound.}
\begin{center}
\begin{tabular}{| p{0.8cm}| p{1.5cm} | p{1.5cm} | p{1.5cm} | p{1.5cm} | p{1.5cm} |p{1.5cm}|p{1.5cm} |}
\hline
$(K,|\Emc|)$ &$|\Mmc|$& $\Rmc_{s,2}(\Asf)$ & $\Rmc_{s,3}(\Asf)$ & $\Rmc_{2}^{N,N+1}(\Asf)$ & $\Rmc_{2}^{N,N+2}(\Asf)$  & $\Rmc_{2}^{N,N+3}(\Asf)$  & $\Rmc_{sp}(\Asf)$ \\ \hline
$(1,2)$ & 1 & 1 & 1 & 1 &1 & 1& 1\\ \hline
$(1,3)$ & 3 & 2 & 2 & 3 &3 & 3& 3\\ \hline
$(1,4)$ & 13 & 5 & 5 & 9 &11 & 12& 13\\ \hline
$(2,2)$ & 3 & 3 & 3 & 3 &3 & 3& 3\\ \hline
$(2,3)$ & 23 & 17 & 17 & 23 &23 & 23& 21 \\ \hline
$(2,4)$ & 445 & 152&152 & 317  &388 & 429& 315 \\ \hline
$(3,2)$ & 1 & 1 & 1 & 1 &1 & 1& 1\\ \hline
$(3,3)$ & 68 & 55 & 55 & 68 &68 & 68& 56\\ \hline
$(3,4)$ & 6803 & 1692 &1692 & 4336 & 5766 & 6326 & 3094\\ \hline
\end{tabular}
\end{center}
\end{table*}

\subsection{Tightness of Shannon outer bound}

The first question we are interested in is if the Shannon outer bound (i.e., the LP bound) is tight for the considered MDCS rate regions, i.e., if non-Shannon type inequalities are necessary in order to get the rate regions. Trivially, for $(1,2),(1,3),(1,4)$ MDCS problems, the Shannon outer bound is tight since the rate region for single source problems can be determined by the min-cut bound.
Our earlier results presented in \cite{CongduanAllerton2012,CongduanNetCod2013} proved that rate regions obtained from the Shannon outer bound are tight for all the cases for $(2,2),(2,3),(3,2),(3,3)$ MDCS instances as well (the Shannon outer bound turns out to be tight for the two cases miss-counted in \cite{HauThesis1995} as $(3,3)$ MDCS instances). For $(2,4)$ MDCS, we have proven that the Shannon outer bound is tight for 429 out of 455 non-isomorphic instances by using up to $\Gamma_{6,9}^2$. Similarly, for 3-level 4-encoder MDCS, we have proven that the Shannon outer bound is tight for 6326 out of 6803 non-isomorphic instances by using up to $\Gamma_{7,10}^2$. By grouping more variables on representable matroid inner bounds over various fields $\Fbb_q$, say using $\Gamma_{6,10}^2,\Gamma_{6,10}^3,\Gamma_{7,11}^2,\Gamma_{7,11}^3,$, etc., we may expect that the Shannon outer bound will be tight for additional 2-level and 3-level 4-encoder MDCS instances. As will be discussed later, for these instances where the Shannon outer bound on the region of entropic vectors is tight, the rate regions can be achieved by various codes, such as superposition and scalar/vector binary/ternary codes.

\subsection{Sufficiency of superposition}

In superposition coding, i.e., source separation, the data sources are encoded separately and the output from
an encoder is just the concatenation of those separated codewords. 
In this manner, each encoder can be viewed as a combination of several
sub-encoders, and thus the coding rate of an encoder is the sum of
coding rate of each sub-encoder. If every point in the rate region can be achieved by superposition coding, we say superposition coding suffices.

When there is only one source in the network, there is no distinguish between superposition and linear coding. Therefore, superposition suffices for all $(1,2),(1,3),(1,4)$ MDCS instances, as shown in Table \ref{MDCSresult}.

For the 2-level 2-encoder MDCS instances, superposition coding suffices. However, it is shown in \cite{HauThesis1995,HauYeungISIT1997} that superposition coding is not sufficient for all the 100 non-isomorphic 3-encoder MDCS instances\footnote{actually, after correcting the errors we discussed in \S\ref{sec:mdcs}, we found 95, including 3 cases for $(2,2)$, 1 case for $(3,2)$, 23 cases for $(2,3)$ and 68 cases for $(3,3)$ MDCS instances}. There are only 86 out of them\footnote{actually 81 out of 95} are achievable by superposition coding \cite{HauThesis1995,HauYeungISIT1997}. 
The remaining instances have rate regions for which every point can be achieved by linear coding between sources.  We found that superposition suffices for 315 out of the 455 non-isomorphic $(2,4)$ instances, and suffices for 3094 out of 6803 non-isomorphic $(3,4)$ MDCS instances. Superposition suffices for a significant fraction of all non-isomorphic MDCS instances in these classes.

\subsection{Sufficiency of scalar codes}

When coding across sources is necessary, we first would like to see if simple codes suffice. \cite{CongduanAllerton2012,CongduanNetCod2013} showed that scalar binary codes are insufficient for 6 out of the 23 cases and 15 instances out of the 68 cases (at the time of these publications, we believed the number of $(2,3)$ and $(3,3)$ MDCS instances to be the same as found in \cite{HauThesis1995}).
The 6 instances for $(2,3)$ can be found in Table \ref{tab:SixCases}. The 15 instances for $(2,3)$ MDCS include numbers 8, 14, 28, 32, 37, 42, 47, 49, 53, 55, 57, 59, 63, 65, 69 from the list in \cite{HauThesis1995}. Binary codes turn out to suffice for the two instances missed in \cite{HauThesis1995}. 

One natural question is whether scalar linear codes over a larger field size can eliminate the gap in any of the cases where scalar linear binary codes were insufficient.  Our calculations showed that exactly the same achievable rate regions for 2-level and 3-level MDCS instances with 3 encoders MDCS instances are obtained by considering the larger inner bound of matroids, i.e. by replacing $\Gamma_{N}^{2}$ with $\Gamma_{N}^{3}$ and $\Gamma_{N}^{{\text mat}}$ for $N\in\{5,6\}$, where $\Gamma_{N}^{{\text mat}}$ is the conic hull of all matroid ranks on $N$ elements. Since for $N\in\{5,6,7\}$, all matroids are representable in some field, $\Gamma_{N}^{{\text mat}}$ is also an inner bound on $\bar{\Gamma}_N^*$ but tighter than $\Gamma_N^2$ and $\Gamma_N^3$. 

For 2-level and 3-level 4-encoder MDCS instances, we can still observe that if scalar binary codes suffice, then scalar ternary will also be sufficient and if scalar binary codes do not suffice then neither will scalar ternary codes.  
In addition, we observe that for all MDCS instances we considered, the scalar ternary inner bounds match exactly with the matroid inner bound. Therefore, we have the following observation:
\begin{observation}
If there exists some field size such that scalar linear codes over that field obtain the entire rate region then in all $7360$ MDCS instances we considered, that field size may be taken to be binary.
\end{observation}

However, ternary codes do not give same rate regions as binary codes for some cases when neither of them suffice, since some networks (or points in the rate region) can be achievable by scalar ternary codes but not by scalar binary codes. 

\begin{example}One example network where ternary codes give tighter inner bound is shown in Fig.\ \ref{fig:bmtnm}.

The outer bound on the rate region $\Rmc_{{\rm out}}$ is 
\begin{equation}
\Rmc_{\rm out}=\left\{ \Rbf : 
\begin{array}{rcl}
R_1 & \geq & H(X)\\
R_2 & \geq & H(X)+H(Y)\\
R_1+R_2 & \geq & 2H(X)+H(Y)+H(Z) \\ 
R_3+R_4 & \geq & H(X)+H(Y)+H(Z) \\ 
R_1+R_3 & \geq & H(X)+H(Y)+H(Z)  \\ 
R_1+R_4 & \geq & H(X)+H(Y)+H(Z)  \\
R_2+R_4 & \geq & H(X)+H(Y)+H(Z)  \\
R_2+R_3 & \geq & H(X)+H(Y)+H(Z) 
\end{array} 
\right\}.
\end{equation}

\begin{figure}[t]

\centering

\includegraphics[scale=0.55]{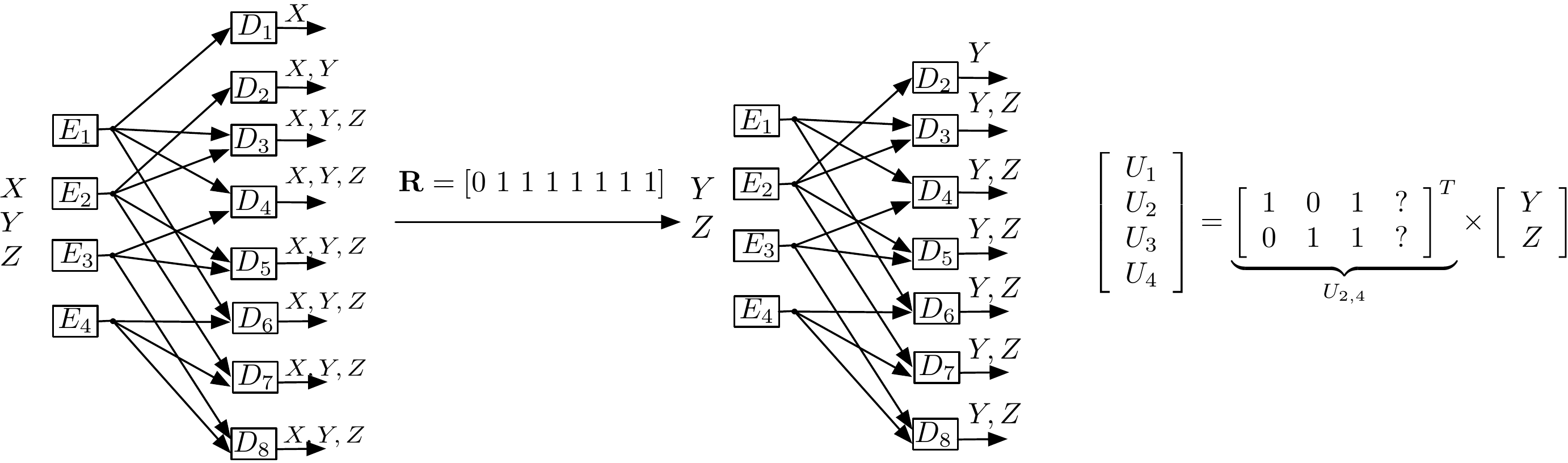}
\caption{\label{fig:bmtnm} A $(3,4)$ MDCS instance for which neither scalar binary codes nor scalar ternary codes suffice, but scalar ternary codes give a tighter inner bound than do scalar binary codes. The extreme ray which is not scalar binary achievable is shown. It is not scalar binary achievable because in the simplified network, a code associated with $U_{2,4}$, which is the forbidden minor for a matroid to be binary achievable, is required.}
\end{figure}

The binary achievable rate region $\Rmc_{\rm bin}$ is
\begin{equation}
\Rmc_{\rm bin}=\Rmc_{\rm out}\cap\left\{ \Rbf : 
\begin{array}{rcl}
R_1+R_2+R_3 & \geq & 2H(X)+2H(Y)+2H(z)\\
R_1+R_2+R_4 & \geq & 2H(X)+2H(Y)+2H(z)\\
R_1+R_3+R_4 & \geq & 2H(X)+2H(Y)+2H(z)\\
R_2+R_3+R_4 & \geq & 3H(X)+3H(Y)+3H(z)\\
R_1+R_2+R_3+R_4 & \geq & 2H(X)+2H(Y)+2H(z)
\end{array} 
\right\},
\end{equation}
while the ternary achievable rate region $\Rmc_{\rm ter}$, which is tighter than binary achievable rate region, is
\begin{equation}
\Rmc_{\rm ter}=\Rmc_{\rm out}\cap\left\{ \Rbf : 
\begin{array}{rcl}
R_1+R_2+R_3 & \geq & 2H(X)+H(Y)+2H(z)\\
R_1+R_2+R_4 & \geq & 2H(X)+H(Y)+2H(z)\\
R_1+R_3+R_4 & \geq & 2H(X)+H(Y)+2H(z)\\
R_1+R_3+R_4 & \geq & 2H(X)+2H(Y)+H(z)\\
2R_1+R_3+R_4 & \geq & 3H(X)+2H(Y)+2H(z)\\
R_2+R_3+R_4 & \geq & 2H(X)+H(Y)+2H(z)\\
R_1+R_2+R_3+R_4& \geq & 3H(X)+H(Y)+3H(z)
\end{array}
\right\}. 
\end{equation}

The extreme ray in the ternary bound that violates additional inequalities in the binary bound is 
\[\mathbf{R}=[H(X)\  H(Y)\  H(Z)\ R_1\ R_2\ R_3\ R_4 ]=[0\ 1\ 1\ 1\ 1\ 1\ 1].\] 
It is not hard to see why this extreme ray is not binary achievable but is ternary achievable. We can assign the entropies in the extreme ray $\Rbf$ to the variables in the network shown in Fig.\ \ref{fig:bmtnm}. 
Since $H(X)=0$, it is equivalent to delete source $X$. The network is then simplified to a $(2,4)$ MDCS instance where the six decoders receiving messages from size-two subsets of encoders demand the two sources $Y,Z$. It is known that sources $Y,Z$ are independent, so if a binary code achieves this extreme ray, every collection of two codewords must be able to decode $Y,Z$, i.e., the joint entropy of every two codewords is 2. Equivalently, the matroids associated with achieving codes must contain $U_{2,4}$ as a minor. However, it is known that any matroid containing $U_{2,4}$ is not binary representable, as shown in Fig.\ \ref{fig:bmtnm}. Therefore, this network is not binary achievable. 
\end{example} 

This example shows that for an extreme ray (or point) in the rate region of a network, if the matroids associated with its achieving codes are not binary representable, we conclude that the rate region is not scalar binary sufficient. In general, we have the following theorem.

\begin{theorem}
Let $\mathcal{R}$ be the rate region of a network with $N$ random variables and ${\rm Extr}(\Rmc)$ represent the (minimum integer) extreme rays of $\Rmc$. Linear codes in $\Fbb_q$ suffice to achieve $\Rmc$ if and only if $\forall \Rbf \in {\rm Extr}(\Rmc)$, $\exists\, \Rbf '\in \Gamma_{N,N'}^q$, for some $N'\geq N$, such that $\Rbf'$ satisfies all network constraints and $\Rbf={\rm Proj}_{h_{X_{k}},h_{U_e},k\in[[K]],e\in\mathcal{E}}(\Rbf ')$.

\end{theorem}

\begin{IEEEproof}
First we know that if achieving codes for two points in the rate region are given, time sharing between these two codes can achieve any point between theses two points. Therefore, it suffices to only consider the achievability of extreme rays (points) of the rate region.

For an extreme ray $\Rbf \in {\rm Extr}(\Rmc)$ scaled to minimum integer representation, if $\exists\ \Rbf '\in \Gamma_{N,N'}^q$, for some $N'\geq N$, such that $\Rbf'$ satisfies all network constraints and $\Rbf={\rm Proj}_{h_{U_e},h_{X_k},e\in \Emc, k\in [[K]]}\Rbf '$, we can construct the code to achieve it following the method in \S\ref{sec:codeCons} using the corresponding representation for $\Rbf '$. Note that when $N'=N$, we construct scalar codes and when $N'>N$, we construct vector codes. Thus, $\Rbf$ is achievable by $\Fbb_q$ linear codes and so is the entire region $\Rmc$.
\end{IEEEproof}

From this theorem we see that, in order to prove a network that is not $\Fbb_q$ representable, one just needs to show the non-existence of $\Fbb_q$ codes for one extreme ray in the rate region. 

Fig.\ \ref{fig:main} shows an example where scalar-binary codes are not optimal. Different from the proof in \S\ref{sec:codeCons}, one alternate proof for insufficiency of scalar-binary codes works as follows.

\begin{IEEEproof}[Alternate proof for scalar binary insufficiency of example in Fig.\ \ref{fig:main}]
We are given that the (Shannon outer bound on) rate region for this MDCS instance is $\Rmc$:
\begin{equation}
\Rmc = \left\{ \Rbf : 
\begin{array}{rcl}
R_1 & \geq & H(X) \\ 
R_2+R_3 & \geq & H(X)+H(Y) \\ 
R_1+R_3 & \geq & H(X)+H(Y) \\ 
R_1+R_2 & \geq & H(X)+H(Y) 
\end{array} 
\right\}. \\
\end{equation}

One extreme ray of it is $\Rbf=[0\  2\ 1\  1\  1]$ with the entries corresponding to $(H(X)\ H(Y)\ R_1\ R_2\ R_3)$ respectively. Since $H(Y)=2$, there does not exist a scalar binary code such that the coded messages have entropy 1. This completes the proof.
\end{IEEEproof}
Similarly, one can prove non-achievability of binary codes for other networks by showing the non-achievability of some extreme ray in the rate region. Due to the large number of cases and our limited space, only the proofs for the 6 cases in $(2,3)$ MDCS instances, where binary codes are not optimal, are shown in Table \ref{tab:SixCases} as examples.
\begin{table*}\tiny
\caption{\label{tab:SixCases}Six $(2,3)$ MDCS instances where scalar binary inner bound and Shannon
outer bound do not match. The inequalities where outer bound and inner
bound do not match are in bold. The listed extreme rays with entries corresponding to $[H(X)\ H(Y)\ R_1\ R_2\ R_3]$ violate bolded inequalities of binary inner bounds.}
\begin{tabular}{|C{3cm}|C{5.7cm}|C{5.2cm}|C{1.1cm}|}
\hline 
Case diagrams  & Exact rate regions  & Scalar binary inner Bounds & Violating extreme rays
\tabularnewline
\hline 
\includegraphics[width=2.8cm]{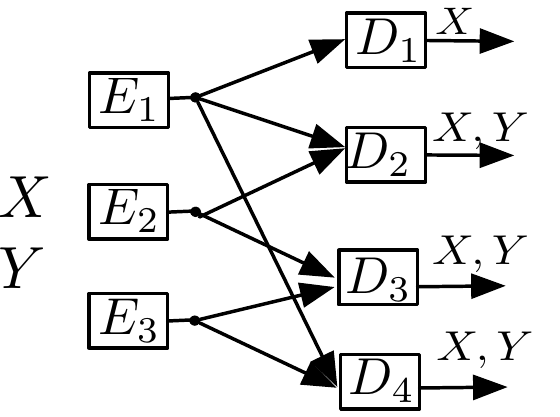}  & $\begin{array}[t]{rcl}
R_{1} & \geq & H(X)\\
R_{1}+R_{2} & \geq & H(X)+H(Y)\\
R_{1}+R_{3} & \geq & H(X)+H(Y)\\
R_{2}+R_{3} & \geq & H(X)+H(Y)
\end{array}$  & $\begin{array}[t]{rcl}
R_{1} & \geq & H(X)\\
R_{1}+R_{2} & \geq & H(X)+H(Y)\\
R_{1}+R_{3} & \geq & H(X)+H(Y)\\
R_{2}+R_{3} & \geq & H(X)+H(Y)\\
\mathbf{R_{1}+R_{2}+R_{3}} & \geq & \mathbf{H(X)+2H(Y)}
\end{array} $ & $[0\ 2\ 1\ 1\ 1]$\tabularnewline
\hline 
\includegraphics[width=3cm]{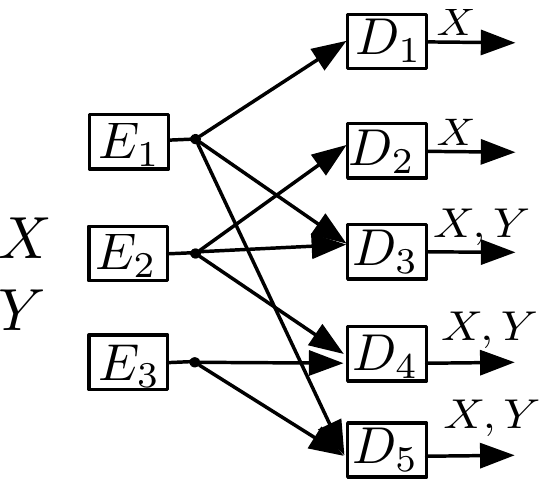}  & $\begin{array}[t]{rcl}
R_{1} & \geq & H(X)\\
R_{2} & \geq & H(X)\\
R_{1}+R_{2} & \geq & 2H(X)+H(Y)\\
R_{1}+R_{3} & \geq & H(X)+H(Y)\\
R_{2}+R_{3} & \geq & H(X)+H(Y)
\end{array}$  & $\begin{array}[t]{rcl}
R_{1} & \geq & H(X)\\
R_{2} & \geq & H(X)\\
R_{1}+R_{2} & \geq & 2H(X)+H(Y)\\
R_{1}+R_{3} & \geq & H(X)+H(Y)\\
R_{2}+R_{3} & \geq & H(X)+H(Y)\\
\mathbf{R_{1}+R_{2}+R_{3}} & \geq & \mathbf{2H(X)+2H(Y)}
\end{array}$ & $[0\ 2\ 1\ 1\ 1]$ \tabularnewline
\hline 
\includegraphics[width=3cm]{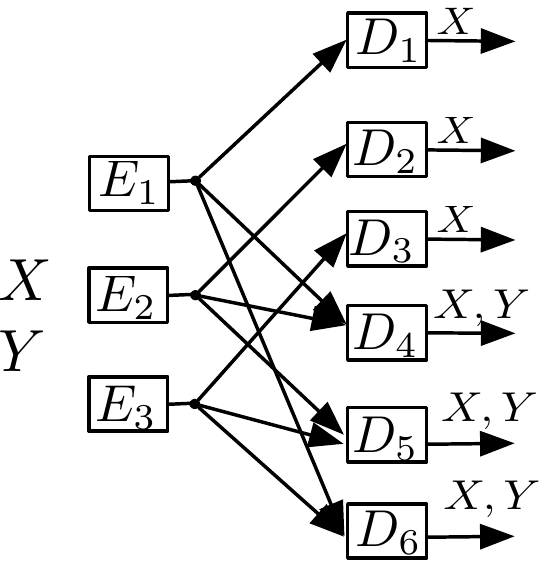}  & $\begin{array}[t]{rcl}
R_{1} & \geq & H(X)\\
R_{2} & \geq & H(X)\\
R_{3} & \geq & H(X)\\
R_{1}+R_{2} & \geq & 2H(X)+H(Y)\\
R_{1}+R_{3} & \geq & 2H(X)+H(Y)\\
R_{2}+R_{3} & \geq & 2H(X)+H(Y)
\end{array}$  & $\begin{array}[t]{rcl}
R_{1} & \geq & H(X)\\
R_{2} & \geq & H(X)\\
R_{3} & \geq & H(X)\\
R_{1}+R_{2} & \geq & 2H(X)+H(Y)\\
R_{1}+R_{3} & \geq & 2H(X)+H(Y)\\
R_{2}+R_{3} & \geq & 2H(X)+H(Y)\\
\mathbf{R_{1}+R_{2}+R_{3}} & \geq & \mathbf{3H(X)+2H(Y)}
\end{array}$ & $[0\ 2\ 1\ 1\ 1]$\tabularnewline
\hline 
\includegraphics[width=3cm]{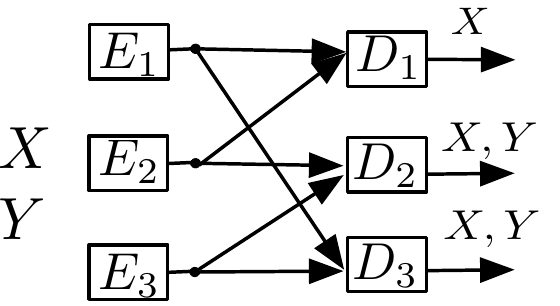}  & $\begin{array}[t]{rcl}
R_{1}+R_{2} & \geq & H(X)\\
R_{1}+R_{3} & \geq & H(X)+H(Y)\\
R_{2}+R_{3} & \geq & H(X)+H(Y)
\end{array}$  & $\begin{array}[t]{rcl}
R_{1}+R_{2} & \geq & H(X)\\
R_{1}+R_{3} & \geq & H(X)+H(Y)\\
R_{2}+R_{3} & \geq & H(X)+H(Y)\\
\mathbf{R_{1}+R_{2}+R_{3}} & \geq & \mathbf{2H(X)+H(Y)}
\end{array}$ &$[2\ 0\ 1\ 1\ 1]$ \tabularnewline
\hline 
\includegraphics[width=3cm]{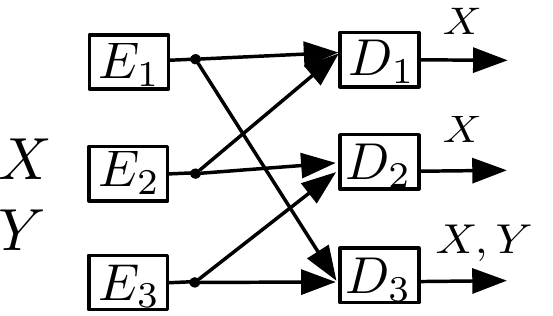}  & $\begin{array}[t]{rcl}
R_{1}+R_{2} & \geq & H(X)\\
R_{1}+R_{3} & \geq & H(X)+H(Y)\\
R_{2}+R_{3} & \geq & H(X)\\
\mathbf{R_{1}+2R_{2}+R_{3}} & \mathbf{\geq} & \mathbf{2H(X)+H(Y)}
\end{array}$  & $\begin{array}[t]{rcl}
R_{1}+R_{2} & \geq & H(X)\\
R_{1}+R_{3} & \geq & H(X)+H(Y)\\
R_{2}+R_{3} & \geq & H(X)\\
\mathbf{R_{1}+R_{2}+R_{3}} & \mathbf{\geq} & \mathbf{2H(X)+H(Y)}
\end{array}$ & $[2\ 0\ 1\ 1\ 1]$ \tabularnewline
\hline 
\includegraphics[width=3cm]{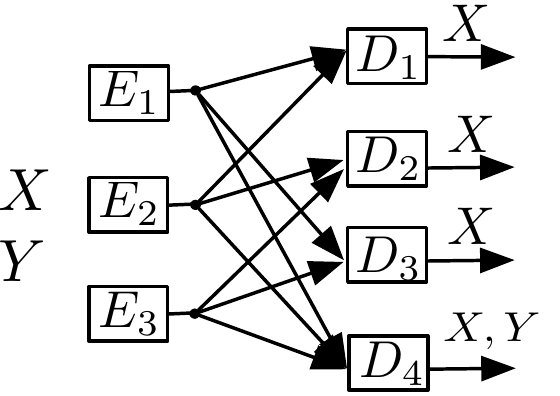}  & $\begin{array}[t]{rcl}
R_{1}+R_{2} & \geq & H(X)\\
R_{1}+R_{3} & \geq & H(X)\\
R_{2}+R_{3} & \geq & H(X)\\
\mathbf{R_{1}+R_{2}+2R_{3}} & \geq & \mathbf{2H(X)+H(Y)}\\
\mathbf{R_{1}+2R_{2}+2R_{3}} & \geq & \mathbf{2H(X)+H(Y)}\\
\mathbf{2R_{1}+R_{2}+R_{3}} & \mathbf{\geq} & \mathbf{2H(X)+H(Y)}\\
\mathbf{3R_{1}+2R_{2}+2R_{3}} & \geq & \mathbf{3H(X)+2H(Y)}
\end{array}$  & $\begin{array}[t]{rcl}
R_{1}+R_{2} & \geq & H(X)\\
R_{1}+R_{3} & \geq & H(X)\\
R_{2}+R_{3} & \geq & H(X)\\
\mathbf{R_{1}+R_{2}+R_{3}} & \geq & \mathbf{2H(X)+H(Y)}
\end{array}$ & $[2\ 0\ 1\ 1\ 1]$\tabularnewline
\hline 
\end{tabular}
\end{table*}

\subsection{Sufficiency of vector codes}

As mentioned above, simple scalar codes are not always sufficient to achieve every point in the rate region, even for these simple small MDCS instances. A natural alternative is to employ vector linear codes instead, which means encoding a group of outcomes of source variables for several time steps together. Recall that vector linear codes are corresponding to vector representable matroids by grouping elements together as one new variable. As we will show in \S\ref{sec:codeCons}, the construction of codes are based on the representation matrix of the original matroid before grouping elements. Therefore, we allow coding occurs not only between different portions of a source (i.e, superposition) but also portions of different sources (i.e, vector linear combinations).

Passing from scalar codes to vector codes, in this sense, by replacing $\Gamma_N^{2}$ with $\Gamma_{N,N+1}^{2}, N=5,6$ in our 2-level 3-encoder and 3-level 3-encoder achievable rate regions, closes all of gaps, hence proving that the exact rate regions for 2-level 3-encoder and 3-level 3-encoder MDCS instances are the same as that obtained from the Shannon outer bound.  This proves that vector linear codes (in the sense of \S \ref{sec:codeCons}) suffice to achieve all of the fundamental rate regions of 2-level 3-encoder and 3-level 3-encoder MDCS instances. Also note that, only one extra bit is necessary to use as vector binary codes to achieve the entire rate regions for 2-level and 3-level 3-encoder MDCS instances. 

With up to three extra bits, we are able to close gaps for almost all of the 455 $(2,4)$ and 6803 $(3,4)$ MDCS instances, except a small fraction of them, which can be further reduced if vector ternary inner bounds or vector binary inner bounds with more extra bits are applied. The example presented in \S\ref{sec:codeCons} shows the benefit of using vector linear codes instead of scalar codes in obtaining exact rate regions.

\subsection{Forbidden Minors for Code Class Sufficiency}
\begin{figure}
\centering
\subfigure [\label{fig:nestingbinary}Scalar binary codes]
{\includegraphics[scale=0.65]{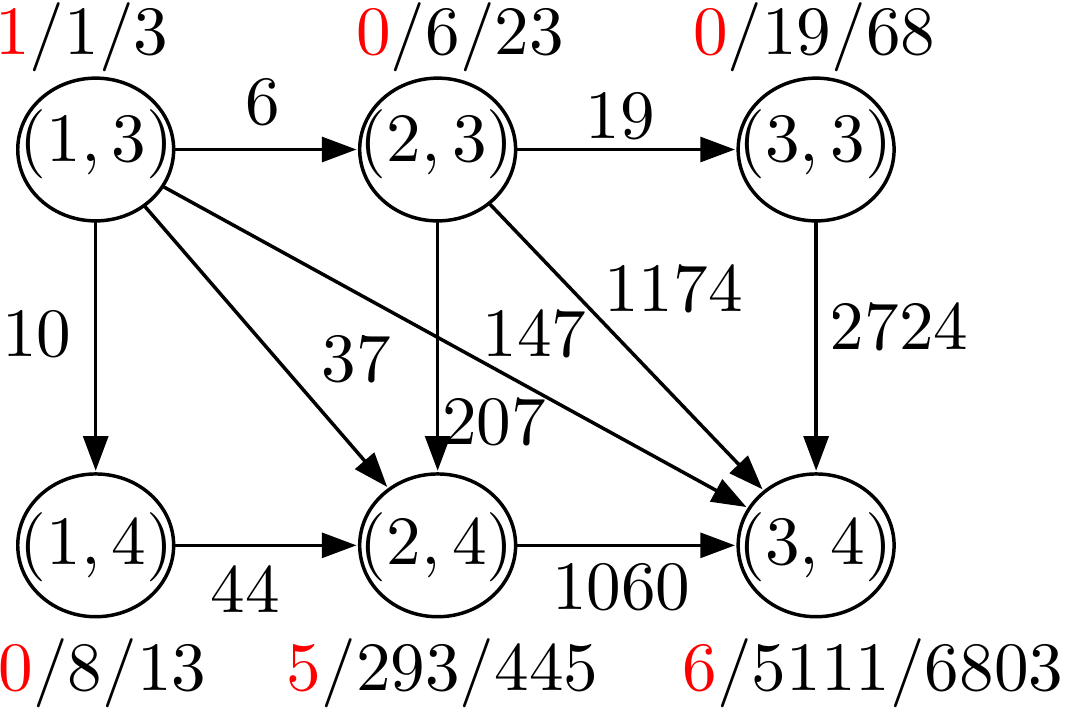}}
\subfigure [\label{fig:nestingsuperposition}Superposition coding]
{\includegraphics[scale=0.65]{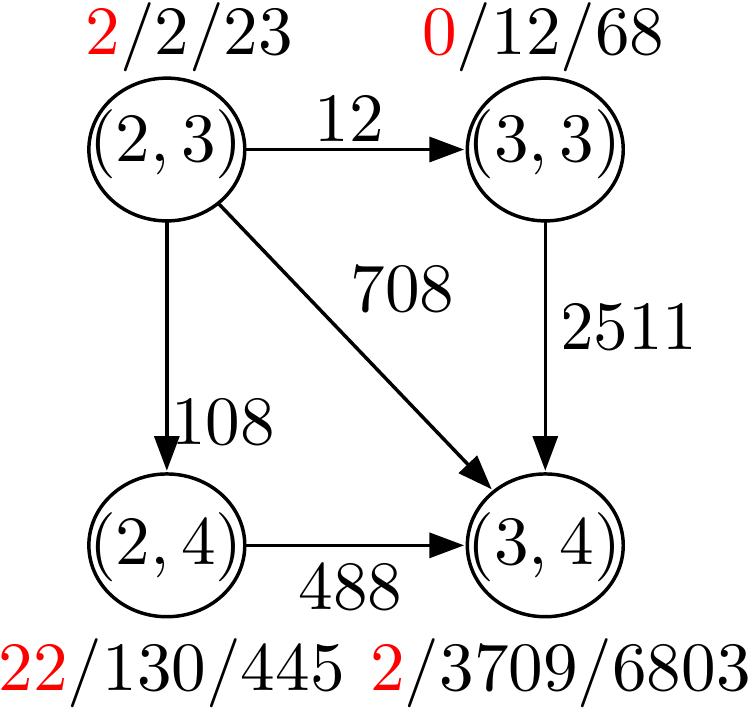}}

\caption{\label{fig:nesting} Nesting relationships between $(1,3),(1,4),(2,3),(3,3),(2,4),(3,4)$ MDCS instances regarding the sufficiency of classes of codes. The three numbers at each node indicate the number of forbidden minors, number of instances that the codes do not suffice, and the total number of non-isomorphic MDCS instances.  The numbers on every edge indicates how many scalar binary insufficient head MDCS instances have predecessors in the tail MDCS instances.}
\end{figure}

Fig.\ \ref{fig:nesting} summarizes our observations on the relationships between $(1,3),(1,4),(2,3),(2,4),(3,3),(3,4)$ MDCS instances in terms of the sufficiency of different classes of codes, including scalar binary codes and superposition coding.  

Fig.\ \ref{fig:nestingbinary} shows the relationships regarding the sufficiency of scalar binary codes. The operations we utilized were source deletion, encoder deletion, and encoder contraction.  We observed that all 19 binary insufficient $(3,3)$ MDCS instances have minors of one of the 6 scalar binary insufficient $(2,3)$ MDCS instances, which themselves all have the same predecessor, the scalar binary insufficient $(1,3)$ MDCS instance. All the 8 scalar binary insufficient $(1,4)$ MDCS instances also have the same predecessor, the scalar binary insufficient $(1,3)$ MDCS instance. However, for $(2,4)$ ($(3,4)$) MDCS, there are 5 (6) instances that we cannot find predecessors for them.  We list the 12 forbidden network minors for scalar binary sufficiency in Table \ref{tab:minors1} and Table \ref{tab:minors2}. If encoder unification is also considered, the number of forbidden minors can be reduced to 10 in total.

Fig.\ \ref{fig:nestingsuperposition} shows the relationships regarding the sufficiency of superposition codes. The operations we utilized include source deletion, encoder deletion, encoder contraction, and encoder unification.  We observed that all 12 superposition insufficient $(3,3)$ MDCS instances have minors of one of the 2 superposition insufficient $(2,3)$ MDCS instances.  However, for $(2,4)$ ($(3,4)$) MDCS, there are 22 (2) instances that we cannot find predecessors for them. The numbers on every edge indicates how many superposition insufficient head MDCS instances have predecessors in the tail MDCS instances. 
\begin{table*}
\centering
\tiny
\caption{\label{tab:minors1}The only one $(1,3)$ and 5 $(2,4)$ MDCS instances for which scalar binary codes do not suffice. The inequalities where outer bound and inner bound do not match are in bold. The listed extreme rays with entries corresponding to $[H(X_k),k\in[[K]]\  R_e,e\in\Emc]$ violate bolded inequalities of binary inner bounds.}
\begin{tabular}{|C{2.8cm}|C{5cm}|C{5.1cm}|C{1cm}|}
\hline 
{\small Case diagrams}  & {\small Exact rate regions}  & {\small Scalar binary inner Bounds} & {\small Violating rays}
\tabularnewline
\hline 
\includegraphics[width=3cm]{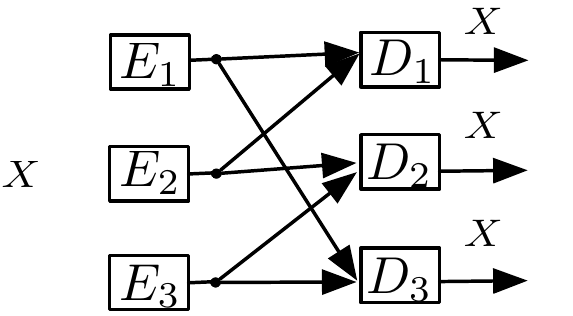}  & $\begin{array}[t]{rcl}
R_{1}+R_{2} & \geq & H(X)\\
R_{1}+R_{3} & \geq & H(X)\\
R_{2}+R_{3} & \geq & H(X)
\end{array}$  & $\begin{array}[t]{rcl}
R_{1}+R_{2} & \geq & H(X)\\
R_{1}+R_{3} & \geq & H(X)\\
R_{2}+R_{3} & \geq & H(X)\\
\mathbf{R_{1}+R_{2}+R_{3}} & \geq & \mathbf{2H(X)}
\end{array} $ & $\left[ \begin{array}{c} 2\\ 1\\ 1\\ 1 \end{array} \right]$\tabularnewline
\hline 
\includegraphics[width=3cm]{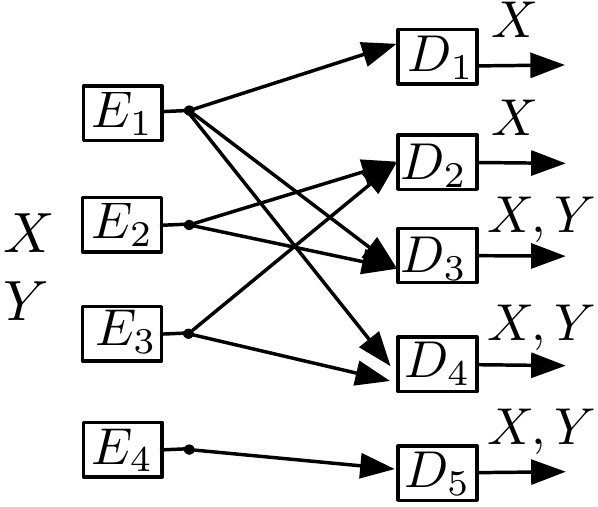}  & $\begin{array}[t]{rcl}
R_{1} & \geq & H(X)\\
R_{1}+R_{2} & \geq & H(X)+H(Y)\\
R_{1}+R_{3} & \geq & H(X)+H(Y)\\
R_{2}+R_{3} & \geq & H(X)\\
R_4&\geq& H(X)+H(Y)\\
R_1+R_{2}+R_{3} & \geq & 2H(X)+H(Y)
\end{array}$  & $\begin{array}[t]{rcl}
R_{1} & \geq & H(X)\\
R_{1}+R_{2} & \geq & H(X)+H(Y)\\
R_{1}+R_{3} & \geq & H(X)+H(Y)\\
R_{2}+R_{3} & \geq & H(X)\\
R_4&\geq& H(X)+H(Y)\\
\mathbf{2R_{1}+R_{2}+R_{3}} & \geq & \mathbf{3H(X)+2H(Y)}
\end{array}$ & $\left[\begin{array}{c} 1\\ 1\\ 1\\ 1\\ 1\\ 2 \end{array}\right]$ \tabularnewline
\hline 
\includegraphics[width=3cm]{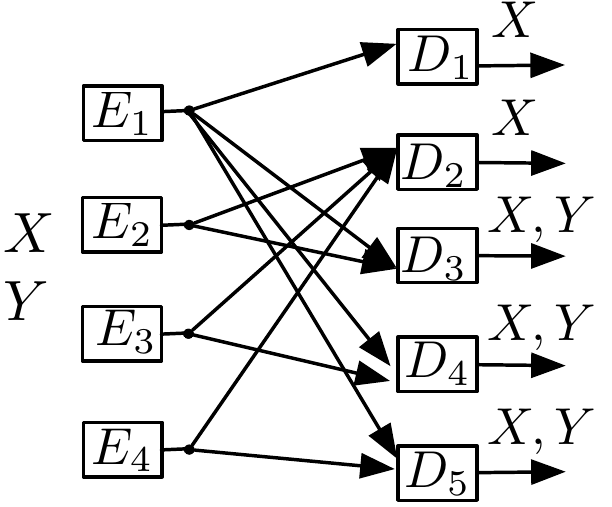}  & $\begin{array}[t]{rcl}
R_{1} & \geq & H(X)\\
R_{2}+R_3+R_4 & \geq & H(X)\\
R_{1}+R_{2} & \geq & H(X)+H(Y)\\
R_{1}+R_{3} & \geq & H(X)+H(Y)\\
R_{1}+R_{4} & \geq & H(X)+H(Y)\\
R_{1}+R_2+R_{3}+R_4 & \geq & 2H(X)+H(Y)
\end{array}$  & $\begin{array}[t]{rcl}
R_{1} & \geq & H(X)\\
R_{2}+R_3+R_4 & \geq & H(X)\\
R_{1}+R_{2} & \geq & H(X)+H(Y)\\
R_{1}+R_{3} & \geq & H(X)+H(Y)\\
R_{1}+R_{4} & \geq & H(X)+H(Y)\\
\mathbf{2R_{1}+R_{2}+R_{3}} & \geq & \mathbf{3H(X)+2H(Y)}
\end{array}$ & $\left[ \begin{array}{c} 2\\ 1\\ 2\\ 1\\ 1\\ 1 \end{array} \right]$\tabularnewline
\hline 
\includegraphics[width=3cm]{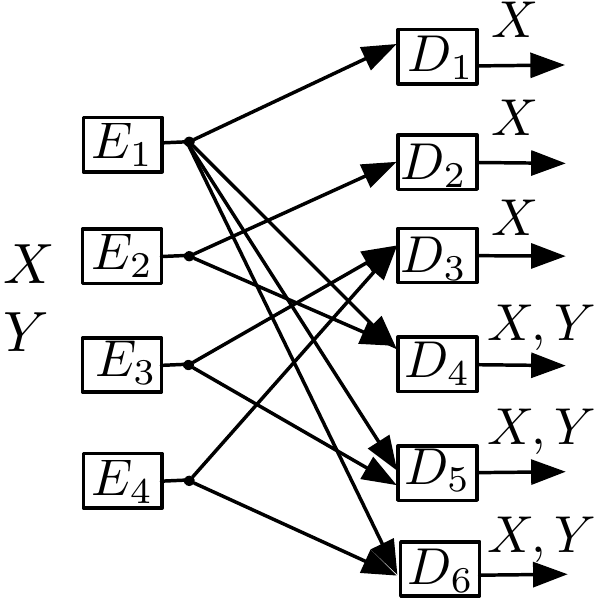}  & $\begin{array}[t]{rcl}
R_{1} & \geq & H(X)\\
R_{2} & \geq & H(X)\\
R_{3}+R_{4} & \geq & H(X)\\
R_{1}+R_{2} & \geq & 2H(X)+H(Y)\\
R_{1}+R_{3} & \geq & H(X)+H(Y)\\
R_{1}+R_{4} & \geq & H(X)+H(Y)\\
R_1+R_3+R_4&\geq&2H(X)+H(Y)
\end{array}$  & $\begin{array}[t]{rcl}
R_{1} & \geq & H(X)\\
R_{2} & \geq & H(X)\\
R_{3}+R_{4} & \geq & H(X)\\
R_{1}+R_{2} & \geq & 2H(X)+H(Y)\\
R_{1}+R_{3} & \geq & H(X)+H(Y)\\
R_{1}+R_{4} & \geq & H(X)+H(Y)\\
\mathbf{2R_{1}+R_{3}+R_{4}} & \geq & \mathbf{3H(X)+2H(Y)}
\end{array}$ &$\left[ \begin{array}{c} 1\\ 2\\ 1\\ 1\\ 1\\ 1 \end{array} \right]$ \tabularnewline
\hline 
\includegraphics[width=3cm]{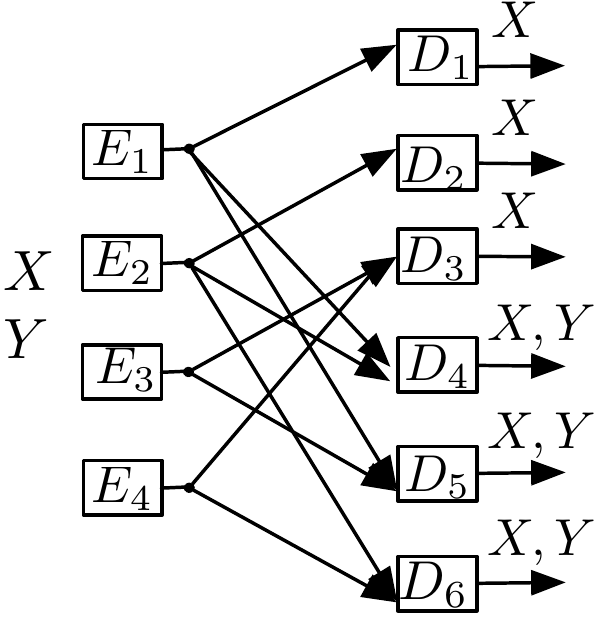}  & $\begin{array}[t]{rcl}
R_{1} & \geq & H(X)\\
R_{2} & \geq & H(X)\\
R_{3}+R_{4} & \geq & H(X)\\
R_1+R_3&\geq&H(X)+H(Y)\\
R_2+R_4&\geq&H(X)+H(Y)\\
R_{1}+R_{2} & \geq & 2H(X)+H(Y)\\
R_{1}+R_{3}+R_4 & \geq & 2H(X)+H(Y)\\
R_{2}+R_{3}+R_4 & \geq & 2H(X)+H(Y)\\
\end{array}$  & $\begin{array}[t]{rcl}
R_{1} & \geq & H(X)\\
R_{2} & \geq & H(X)\\
R_{3}+R_{4} & \geq & H(X)\\
R_1+R_3&\geq&H(X)+H(Y)\\
R_2+R_4&\geq&H(X)+H(Y)\\
R_{1}+R_{2} & \geq & 2H(X)+H(Y)\\
R_{1}+R_{3}+R_4 & \geq & 2H(X)+H(Y)\\
R_{2}+R_{3}+R_4 & \geq & 2H(X)+H(Y)\\
\mathbf{R_{1}+2R_{2}+R_{3}+R_4} & \geq & \mathbf{3H(X)+2H(Y)}
\end{array}$ & $\left[\begin{array}{c} 1\\ 2\\ 2\\ 2\\ 1\\ 1\end{array}\right]$ \tabularnewline
\hline 
\includegraphics[width=3cm]{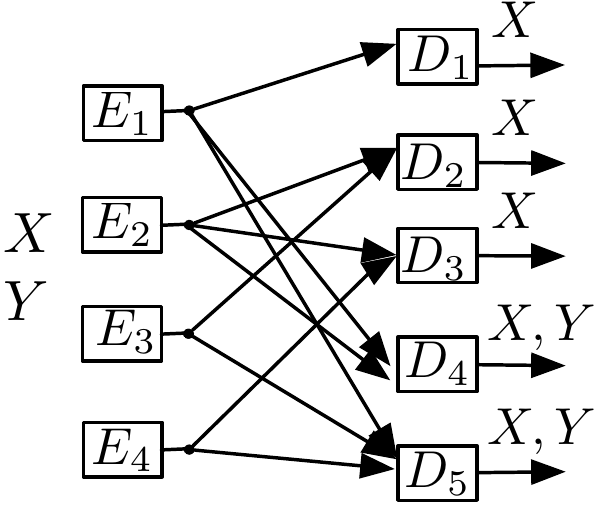}  & $\begin{array}[t]{rcl}
R_{1} & \geq & H(X)\\
R_{2}+R_3 & \geq & H(X)\\
R_{2}+R_{4} & \geq & H(X)\\
R_1+R_2&\geq&H(X)+H(Y)\\
R_1+R_3+R_4&\geq&H(X)+H(Y)\\
R_{1}+R_{2}+R_3 & \geq & 2H(X)+H(Y)\\
R_{1}+R_{2}+R_4 & \geq & 2H(X)+H(Y)\\
\end{array}$  & $\begin{array}[t]{rcl}
R_{1} & \geq & H(X)\\
R_{2}+R_3 & \geq & H(X)\\
R_{2}+R_{4} & \geq & H(X)\\
R_1+R_2&\geq&H(X)+H(Y)\\
R_1+R_3+R_4&\geq&H(X)+H(Y)\\
R_{1}+R_{2}+R_3 & \geq & 2H(X)+H(Y)\\
R_{1}+R_{2}+R_4 & \geq & 2H(X)+H(Y)\\
\mathbf{2R_{1}+R_{2}+R_{3}} & \geq & \mathbf{3H(X)+2H(Y)}
\end{array}$ & $\left[ \begin{array}{c} 1\\ 2\\ 1\\ 2\\ 1\\ 1 \end{array} \right]$\tabularnewline
\hline 
\end{tabular}
\end{table*}

\begin{table*}\tiny
\caption{\label{tab:minors2}The six $(3,4)$ MDCS instances for which scalar binary codes do not suffice. The inequalities where outer bound and inner
bound do not match are in bold. The listed extreme rays with entries corresponding to $[H(X_k),k\in[[K]]\  R_e,e\in\Emc]$ violate bolded inequalities of binary inner bounds.}
\begin{tabular}{|C{2.1cm}|C{5.9cm}|C{6.1cm}|C{.8cm}|}
\hline 
{\small Case diagrams } & {\small Shannon outer bounds}  & {\small Scalar binary inner Bounds } & {\small Violating ray}
\tabularnewline
\hline 
\includegraphics[width=2.3cm]{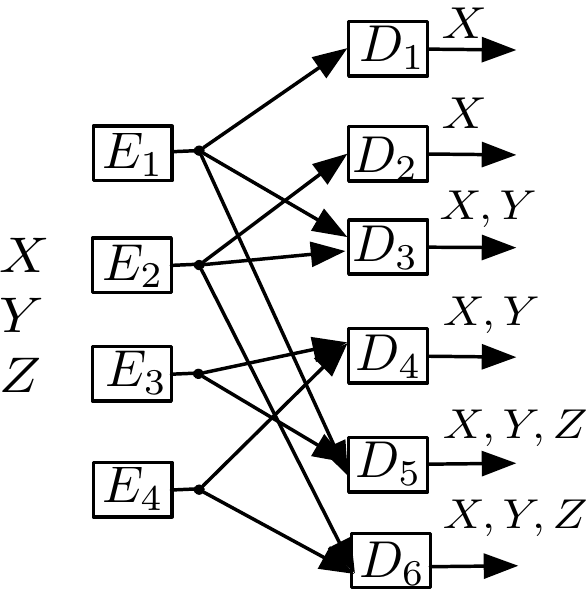}  & \vspace{-3mm} $\begin{array}[t]{rcl}
R_{1} & \geq & H(X)\\
R_2&\geq&H(X)\\
R_{1}+R_{2} & \geq & 2H(X)+H(Y)\\
R_{3}+R_{4} & \geq & H(X)+H(Y)\\
R_{1}+R_{3} & \geq & H(X)+H(Y)+H(Z)\\
R_{2}+R_{4} & \geq & H(X)+H(Y)+H(Z)\\
R_{1}+R_{3}+R_4 & \geq & 2H(X)+H(Y)+H(Z)\\
R_{2}+R_3+R_{4} & \geq & 2H(X)+H(Y)+H(Z)\\
R_1+R_{2}+R_3+R_{4} & \geq & 3H(X)+2H(Y)+H(Z)
\end{array}$   &   \vspace{-3mm} $\begin{array}[t]{rcl}
R_{1} & \geq & H(X)\\
R_2&\geq&H(X)\\
R_{1}+R_{2} & \geq & 2H(X)+H(Y)\\
R_{3}+R_{4} & \geq & H(X)+H(Y)\\
R_{1}+R_{3} & \geq & H(X)+H(Y)+H(Z)\\
R_{2}+R_{4} & \geq & H(X)+H(Y)+H(Z)\\
R_{1}+R_{3}+R_4 & \geq & 2H(X)+H(Y)+H(Z)\\
R_{2}+R_3+R_{4} & \geq & 2H(X)+H(Y)+H(Z)\\
\mathbf{R_{1}+R_{2}+2R_{3}+2R_4} & \geq & \mathbf{4H(X)+3H(Y)+2H(Z)}
\end{array} $ \vfill & $\left[\begin{array}{c} 1\\ 1\\ 1\\ 2\\ 2\\ 1\\ 1 \end{array}\right]$\tabularnewline
\hline 
\includegraphics[width=2.3cm]{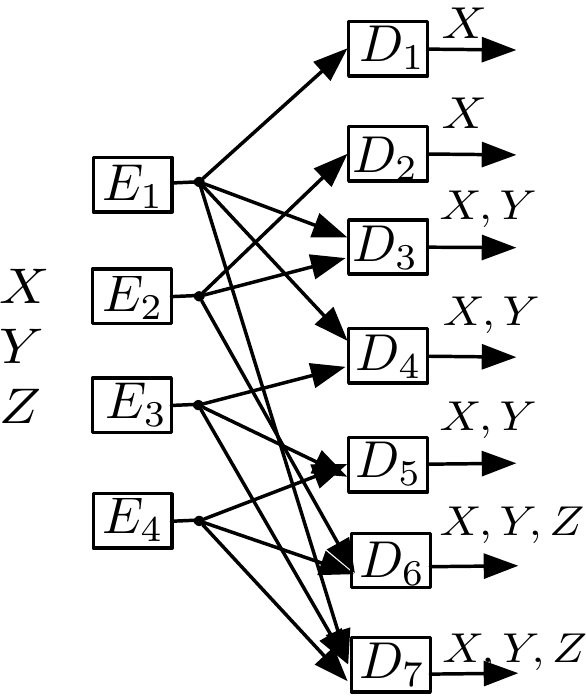}  & \vspace{-3mm}  $\begin{array}[t]{rcl}
R_{1} & \geq & H(X)\\
R_{2} & \geq & H(X)\\
R_{1}+R_{2} & \geq & 2H(X)+H(Y)\\
R_{1}+R_{3} & \geq & H(X)+H(Y)\\
R_{3}+R_{4} & \geq & H(X)+H(Y)\\
R_2+R_4&\geq&H(X)+H(Y)+H(Z)\\
R_1+R_3+R_4&\geq&2H(X)+H(Y)+H(Z)\\
R_1+2R_3+R_4&\geq&2H(X)+2H(Y)+H(Z)\\
R_2+R_3+R_4&\geq&2H(X)+H(Y)+H(Z)\\
R_1+R_2+R_3+R_4&\geq&3H(X)+2H(Y)+H(Z)\\
\end{array}$  & \vspace{-3mm}  $\begin{array}[t]{rcl}
R_{1} & \geq & H(X)\\
R_{2} & \geq & H(X)\\
R_{1}+R_{2} & \geq & 2H(X)+H(Y)\\
R_{1}+R_{3} & \geq & H(X)+H(Y)\\
R_{3}+R_{4} & \geq & H(X)+H(Y)\\
R_2+R_4&\geq&H(X)+H(Y)+H(Z)\\
R_1+R_3+R_4&\geq&2H(X)+H(Y)+H(Z)\\
R_1+2R_3+R_4&\geq&2H(X)+2H(Y)+H(Z)\\
R_2+R_3+R_4&\geq&2H(X)+H(Y)+H(Z)\\
\mathbf{R_{1}+R_{2}+R_{3}+R_4} & \geq & \mathbf{4H(X)+3H(Y)+2H(Z)}
\end{array}$ & $\left[\begin{array}{c} 1\\ 1\\ 1\\ 2\\ 2\\ 1\\ 1 \end{array} \right]$ \tabularnewline
\hline 
\includegraphics[width=2.3cm]{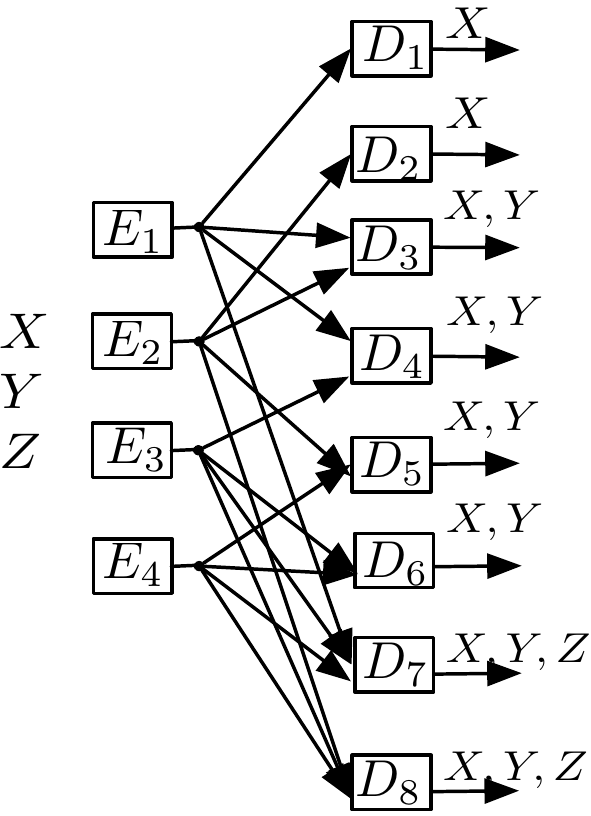}  &  \vspace{-3mm} $\begin{array}[t]{rcl}
R_{1} & \geq & H(X)\\
R_{2} & \geq & H(X)\\
R_{1}+R_{2} & \geq & 2H(X)+H(Y)\\
R_{1}+R_{3} & \geq & H(X)+H(Y)\\
R_{2}+R_{4} & \geq & H(X)+H(Y)\\
R_{3}+R_{4} & \geq & H(X)+H(Y)\\
R_{1}+R_3+R_{4} & \geq & 2H(X)+H(Y)+H(Z)\\
R_{1}+2R_3+R_{4} & \geq & 2H(X)+2H(Y)+H(Z)\\
R_{2}+R_3+R_{4} & \geq & 2H(X)+H(Y)+H(Z)\\
R_{2}+2R_3+R_{4} & \geq & 2H(X)+H(Y)+H(Z)\\
R_{1}+R_2+R_3+R_{4} & \geq & 3H(X)+2H(Y)+H(Z)\\
R_{1}+R_2+2R_3+R_{4} & \geq & 3H(X)+3H(Y)+2H(Z)
\end{array}$  &  \vspace{-3mm} $\begin{array}[t]{rcl}
R_{1} & \geq & H(X)\\
R_{2} & \geq & H(X)\\
R_{1}+R_{2} & \geq & 2H(X)+H(Y)\\
R_{1}+R_{3} & \geq & H(X)+H(Y)\\
R_{2}+R_{4} & \geq & H(X)+H(Y)\\
R_{3}+R_{4} & \geq & H(X)+H(Y)\\
R_{1}+R_3+R_{4} & \geq & 2H(X)+H(Y)+H(Z)\\
R_{1}+2R_3+R_{4} & \geq & 2H(X)+2H(Y)+H(Z)\\
R_{2}+R_3+R_{4} & \geq & 2H(X)+H(Y)+H(Z)\\
R_{2}+2R_3+R_{4} & \geq & 2H(X)+H(Y)+H(Z)\\
\mathbf{R_{1}+R_2+2R_3+2R_{4}} & \geq & \mathbf{4H(X)+3H(Y)+2H(Z)}
\end{array}$ & $\left[\begin{array}{c} 1\\ 1\\ 1\\ 2\\ 2\\ 1\\ 1 \end{array} \right]$\tabularnewline
\hline 
\includegraphics[width=2.3cm]{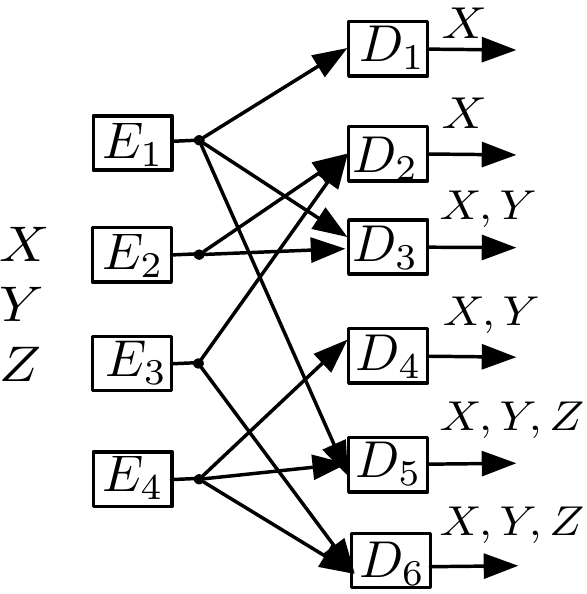}  & \vspace{-3mm} $\begin{array}[t]{rcl}
R_{1} & \geq & H(X)\\
R_{2}+R_3 & \geq & H(X)\\
R_{1}+R_{2} & \geq & H(X)+H(Y)\\
R_4 & \geq & H(X)+H(Y)\\
R_{1}+R_{4} & \geq & 2H(X)+H(Y)+H(Z)\\
R_{3}+R_{4} & \geq & H(X)+H(Y)+H(Z)\\
R_{1}+R_2+R_{3} & \geq & 2H(X)+H(Y)\\
R_{1}+R_2+R_{4} & \geq & 2H(X)+2H(Y)+H(Z)\\
R_{2}+R_3+R_{4} & \geq & 2H(X)+H(Y)+H(Z)\\
R_{1}+R_2+R_3+R_{4} & \geq & 3H(X)+2H(Y)+H(Z)
\end{array}$  & \vspace{-3mm} $\begin{array}[t]{rcl}
R_{1} & \geq & H(X)\\
R_{2}+R_3 & \geq & H(X)\\
R_{1}+R_{2} & \geq & H(X)+H(Y)\\
R_4 & \geq & H(X)+H(Y)\\
R_{1}+R_{4} & \geq & 2H(X)+H(Y)+H(Z)\\
R_{3}+R_{4} & \geq & H(X)+H(Y)+H(Z)\\
R_{1}+R_2+R_{3} & \geq & 2H(X)+H(Y)\\
R_{1}+R_2+R_{4} & \geq & 2H(X)+2H(Y)+H(Z)\\
R_{2}+R_3+R_{4} & \geq & 2H(X)+H(Y)+H(Z)\\
\mathbf{R_{1}+R_2+R_3+2R_{4}} & \geq & \mathbf{4H(X)+3H(Y)+2H(Z)}\\
\end{array}$ &$\left[\begin{array}{c} 1\\ 1\\ 1\\ 2\\ 1\\ 1\\ 2 \end{array} \right]$ \tabularnewline
\hline 
\includegraphics[width=2.3cm]{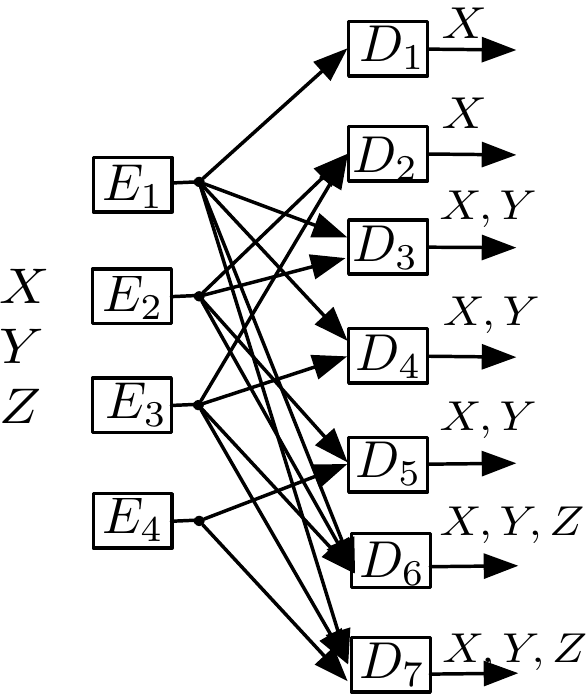}  & \vspace{-3mm} $\begin{array}[t]{rcl}
R_{1} & \geq & H(X)\\
R_{2}+R_3 & \geq & H(X)\\
R_{1}+R_{2} & \geq & H(X)+H(Y)\\
R_{1}+R_{3} & \geq & H(X)+H(Y)\\
R_{2}+R_{4} & \geq & H(X)+H(Y)\\
R_{1}+R_2+R_{3} & \geq & 2H(X)+H(Y)+H(Z)\\
R_{1}+R_3+R_{4} & \geq & H(X)+2H(Y)+H(Z)\\
2R_{1}+R_2+R_{3} & \geq & 2H(X)+2H(Y)+H(Z)\\
R_{1}+R_2+R_{3}+R_4 & \geq & 2H(X)+2H(Y)+H(Z)
\end{array}$  & \vspace{-3mm}  $\begin{array}[t]{rcl}
R_{1} & \geq & H(X)\\
R_{2}+R_3 & \geq & H(X)\\
R_{1}+R_{2} & \geq & H(X)+H(Y)\\
R_{1}+R_{3} & \geq & H(X)+H(Y)\\
R_{2}+R_{4} & \geq & H(X)+H(Y)\\
R_{1}+R_2+R_{3} & \geq & 2H(X)+H(Y)+H(Z)\\
R_{1}+R_3+R_{4} & \geq & H(X)+2H(Y)+H(Z)\\
2R_{1}+R_2+R_{3} & \geq & 2H(X)+2H(Y)+H(Z)\\
R_{1}+R_2+R_{3}+R_4 & \geq & 2H(X)+2H(Y)+H(Z)\\
\mathbf{3R_{1}+R_{2}+2R_{3}+R_4} & \geq & \mathbf{4H(X)+3H(Y)+2H(Z)}
\end{array}$ & $\left[ \begin{array}{c} 1\\ 1\\ 1\\ 1\\ 2\\ 1\\ 1 \end{array} \right]$ \tabularnewline
\hline 
\includegraphics[width=2.3cm]{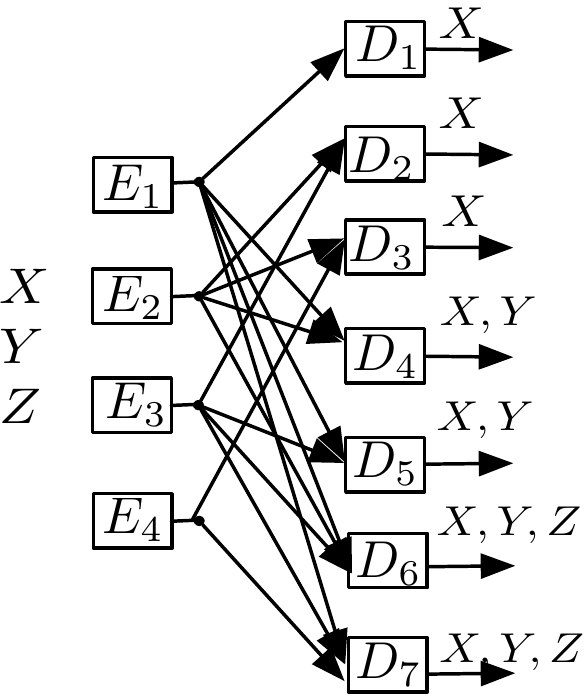}  & \vspace{-3mm} $\begin{array}[t]{rcl}
R_{1} & \geq & H(X)\\
R_{2}+R_3 & \geq & H(X)\\
R_{2}+R_{4} & \geq & H(X)\\
R_{1}+R_{2} & \geq & H(X)+H(Y)\\
R_{1}+R_{3} & \geq & H(X)+H(Y)\\
R_{1}+R_2+R_{4} & \geq & 2H(X)+H(Y)\\
R_{1}+R_2+R_{3} & \geq & 2H(X)+H(Y)+H(Z)\\
R_{1}+R_3+R_{4} & \geq & H(X)+H(Y)+H(Z)\\
2R_{1}+R_2+R_{3} & \geq & 2H(X)+2H(Y)+H(Z)\\
2R_{1}+R_2+R_{3}+R_4 & \geq & 3H(X)+2H(Y)+H(Z)
\end{array}$ \vfill & \vspace{-4mm}  $\begin{array}[t]{rcl}
R_{1} & \geq & H(X)\\
R_{2}+R_3 & \geq & H(X)\\
R_{2}+R_{4} & \geq & H(X)\\
R_{1}+R_{2} & \geq & H(X)+H(Y)\\
R_{1}+R_{3} & \geq & H(X)+H(Y)\\
R_{1}+R_2+R_{4} & \geq & 2H(X)+H(Y)\\
R_{1}+R_2+R_{3} & \geq & 2H(X)+H(Y)+H(Z)\\
R_{1}+R_3+R_{4} & \geq & H(X)+H(Y)+H(Z)\\
2R_{1}+R_2+R_{3} & \geq & 2H(X)+2H(Y)+H(Z)\\
\mathbf{3R_{1}+R_2+2R_{3}+R_4} & \geq & \mathbf{4H(X)+3H(Y)+2H(Z)}
\end{array}$ & $\left[ \begin{array}{c} 1\\ 1\\ 1\\ 1\\ 2\\ 1\\ 1 \end{array} \right]$\tabularnewline
\hline 
\end{tabular}
\end{table*}


\section{Computer aided converse proof}
\label{sec:cap}
Recall that an inequality in the rate region of a network is in terms of rate variables and source entropies. Sometimes, the inequalities in an MDCS rate region are easy to derive. However, for many MDCS instances, it is not easy to derive by hand all of the inequalities for the rate region. For instance, as was shown in the MDCS instances discussed in \S \ref{sec:mdcs}, it is tedious, and some instances difficult, to calculate the converses for all non-isomorphic MDCS instances of even small problem sizes manually.  Motivated by these observations, and inspired by a technique discussed in \cite{TianISIT2013,Tian433Journalversion}, in this section, we will show how to use a computer to generate converse proofs automatically.

As discussed in \S \ref{sec:rateReg}, the projection of the Shannon outer bound plus network constraints gives an outer bound on the coding rate region,  which is in principle equivalent to a converse proof. This can be seen from a property of the polar of a polyhedral cone (c.f. Proposition $B.16.d$ in \cite{BertsekasNLP} and page 122 in \cite{RockafellarConvex}): suppose a polyhedral cone $\Pmc$ is determined by $\mathbb{A}\mathbf{x}\geq \mathbf{0}$ and for an inequality implied by the projected cone, $\bbf^T\mathbf{x}\geq 0$, where $\bbf^T$ has zeros on eliminated dimensions, then there exists a vector $\boldsymbol{\lambda}\geq \mathbf{0}$ such that $\mathbb{A}^T\boldsymbol{\lambda}=\bbf$. The vector $\boldsymbol{\lambda}$ gives coefficients for the weighted sum of inequalities in $\mathcal{P}$.

In the calculation of the Shannon outer bound on rate regions, the Shannon outer bound and network constraints form a polyhedral cone $\mathcal{P}=\{\xbf|\Abb\xbf\geq \mathbf{0}\}$ in the dimension of $2^N-1+|\Emc|$, which is then projected onto only $|\Emc|+K$ dimensions, as shown in \eqref{eq:Shannonfree}. After this projection, the rate region, which must still be a polyhedral cone, has some inequality description $\mathbb{B}\mathbf{x}\geq \mathbf{0}$, where $\mathbb{B}_{:,j}=\mathbf{0}$, for all $j$-th columns such that $\xbf_j$ is eliminated in the projection. For each inequality $\bbf^T\mathbf{x}\geq 0$ in the rate region, the weighted sum $\bbf=\boldsymbol{\lambda}^T\Abb$ with coefficients $\boldsymbol\lambda$ is actually the converse proof for this inequality, deriving it as a sum of network constraints and Shannon information inequalities.  

Such a coefficient vector $\boldsymbol\lambda$ need not be unique. Hence, among the various possibilities for $\boldsymbol\lambda$ satisfying $\bbf=\boldsymbol{\lambda}^T\Abb$, we want to select one that gives the simplest proof, which means it involves the smallest number of inequalities and constraints.  This vector will thus be the sparsest vector $\boldsymbol\lambda$, having the fewest non-zero entries, i.e. the lowest $l_0$-norm.  Obtaining such a vector $\boldsymbol\lambda$ becomes the optimization problem\begin{equation} \label{eq:l0norm}
\begin{aligned}
& \underset{\boldsymbol{\boldsymbol\lambda}}{\text{minimize}}
& & \parallel \boldsymbol{\lambda} \parallel _0 \\
& \text{subject to}
& &\mathbb{A}^T\boldsymbol{\lambda}=\bbf \\
& & &\boldsymbol{\lambda}\geq \mathbf{0}.
\end{aligned}
\end{equation}

Because the $l_0$-norm is a complicated non-differentiable and non-linear function, this is a difficult optimization problem to solve directly. Hence we use the typical relaxation that replaces the $l_0$-norm with the $l_1$-norm.  Furthermore, since $\boldsymbol\lambda\geq \mathbf{0}$, we do not need the absolute values in the $l_1$-norm, and the problem in \eqref{eq:l0norm} is approximated by the linear program
\begin{equation} \label{eq:l1norm}
\begin{aligned}
& \underset{\boldsymbol{\lambda}}{\text{minimize}}
& & \parallel \boldsymbol{\lambda} \parallel _1 \\
& \text{subject to}
& &\mathbb{A}^T\boldsymbol{\lambda}=\bbf \\
& & &\boldsymbol{\lambda}\geq \mathbf{0},
\end{aligned}
\end{equation}
which is equivalent to
\begin{equation} \label{eq:l1normwcost}
\begin{aligned}
& \underset{\boldsymbol{\lambda}}{\text{minimize}}
& & \mathbf{1}^T\boldsymbol{\lambda} \\
& \text{subject to}
& &\mathbb{A}^T\boldsymbol{\lambda}=\bbf \\
& & &\boldsymbol{\lambda}\geq \mathbf{0}.
\end{aligned}
\end{equation}

If the goal is solely to calculate the rate region, when projecting Shannon outer bound and network constraints, it is best to start with only the non-redundant inequalities to reduce the complexity of the projection calculation. However, if we want to generate the human readable proof, by solving the problem in \eqref{eq:l1normwcost}, including many redundant information inequalities can aid us in finding a sparser vector $\boldsymbol\lambda$ than the process without redundant inequalities.  In particular, the non-redundant form of the Shannon outer bound only involves inequalities of the form $I(X_i,X_j|\Xbf_\Kmc)$ and $H(X_i|\Xbf_{\setminus i})\geq 0$. Rather than expressing other Shannon information inequalities as sums of these elemental inequalities, it is preferable when solving \eqref{eq:l1normwcost} to add them to the list of inequalities in the Shannon outer bound, even though they are redundant, because they can lead to sparser $\boldsymbol\lambda$, and hence simpler human readable proofs.  Similarly, some redundant inequalities from the network constraints are also helpful. For instance, for a constraint $H(\Amc|\Bmc)=0$, the redundant equalities $H(\Cmc|\Bmc)=0,\Cmc\subseteq \Amc$ can be very useful in generating short converse proofs.

After the vector $\boldsymbol\lambda$ is obtained, let $\boldsymbol\eta$ be the non-zeros entries in $\boldsymbol\lambda$. We know that the weighted sum of the inequalities associated with $\boldsymbol\eta$ will give the desired inequality in the rate region to prove.  However, such a potentially large weighted sum is still not a form that is easily interpreted by a human.  In order to get a conventional converse proof as would have been done by hand, we also need to select an order to apply the inequalities identified by $\boldsymbol\eta$ step by step.

An algorithm to determine the order on involved coefficients and inequalities is shown in Algorithm \ref{ag:order}. This algorithm always finishes because every term not shown in the target inequality must be cancelled out by the weighted sum of all of the yet unselected inequalities involving this term. 

\begin{algorithm}
\SetAlgoLined
 \KwIn{Target inequality $\bbf^T\xbf\geq 0$, coefficient vector $\boldsymbol\eta$, involved original inequalities $\Abb_{\boldsymbol\eta}$} with row index set $\{1,2,\ldots n\}$.
 
 \KwOut{Ordered index set of $\{1,2,\ldots n\}$ that applying associated inequalities in $\Abb_{\boldsymbol\eta}$ gives human readable proof step by step.}
 \BlankLine
 \textbf{Initialization:} find inequalities which have non-zero entries at rate variables, put their indices to step 1\;
 mark these indices selected\;
 calculate the sum of inequalities in step 1 as the current inequality\;
 \While{exists unselected indices}{
  get terms in current inequality that are not shown in target inequality\;
  find inequalities which have non-zero entries at these terms\;
  put their indices to next step and marked them as selected\;
  calculate the weighted sum of current inequality and the newly selected inequalities to update current inequality\;
 }
 \caption{Determination of the order of application of inequalities for a human readable proof.}
 \label{ag:order}
\end{algorithm}

Next, we would like to use two examples to show the procedure of obtaining the human readable converse proofs.

\begin{figure}
\centering \includegraphics[scale=0.65]{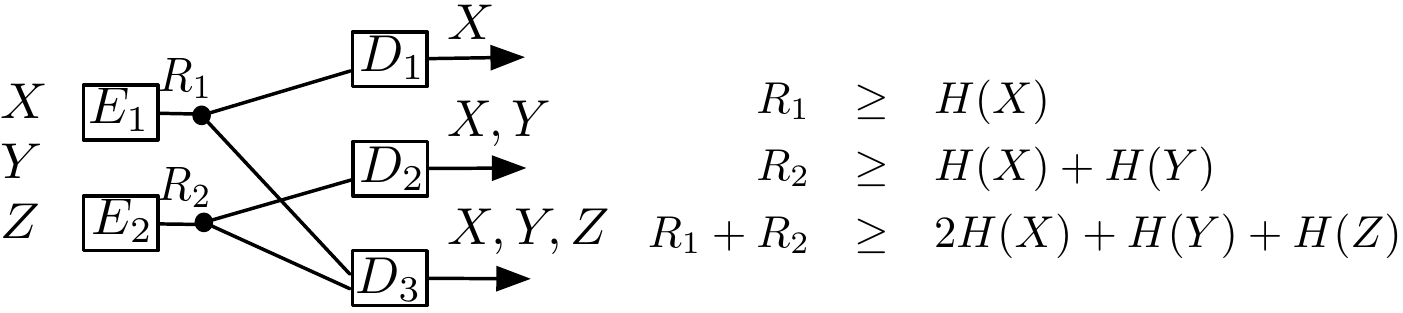} \caption{ Block diagram and rate region for Example \ref{ex:1}: a 3-level 2-encoder MDCS instance}

\label{fig:CAPexample1} 
\end{figure}

\begin{table}
\caption{\label{ineqtable1}Ordered inequalities with coefficients given by computer for Example \ref{ex:1}}
\begin{center}
\begin{tabular}{| p{.8cm}| p{1.2cm} | p{5cm} |}
\hline
Step & Coefficients & Inequality or equality \\ \hline
$1$ & 1 & $R_1\geq H(U_1)$\\ \hline
$1$ & 1 &  $R_2\geq H(U_2)$ \\ \hline
$2$ & 1 &  $H(X|U_1)= 0$\\ \hline
$2$ & 1 &  $H(X,Y|U_2)=0$\\ \hline
$3$ & 1 & $I(U_1;YU_2|X)\geq 0$\\ \hline
$4$ & 1 & $H(X,Y|U_1,U_2)=0$\\ \hline
$5$ & 1 & $H(X,Y,Z|U_1,U_2)=0$\\ \hline
$6$ & 1 & $H(X,Y,Z)=H(X)+H(Y)+H(Z)$\\ \hline
\end{tabular}

\end{center}
\end{table}

\begin{example} \label{ex:1}
A 3-level 2-encoder MDCS instance with block diagram and rate region shown in Fig.\ \ref{fig:CAPexample1}.
The first two inequalities in the rate region are easy to prove and only very few inequalities are needed for the converse proof. For the last inequality
\begin{equation} \label{eq:ex1target}
R_1+R_2\geq 2H(X)+H(Y)+H(Z),
\end{equation}
a computer, after running $LP$ solver and Algorithm \ref{ag:order}, could give inequalities with coefficients, and the order in Table \ref{ineqtable1}. We can get the conventional converse proof by the steps given in Table \ref{ineqtable1}.

Step 1, we start with
\begin{equation}
R_1+R_2\geq H(U_1)+H(U_2).
\end{equation}

Then in step 2, we apply two equalities
\begin{eqnarray}
H(X|U_1)&=&0,\\
H(X,Y|U_2)&=&0
\end{eqnarray}
according to the decoding constraints of decoder $D_1,D_2$ and can get
\begin{equation}
R_1+R_2\geq H(U_1,X)+H(U_2,X,Y).
\end{equation}

Next step, we use the trivial inequality
\begin{equation}
I(U_1;Y,U_2|X)\geq 0
\end{equation}
to get 
\begin{equation}
R_1+R_2\geq H(X)+H(X,Y,U_1,U_2).
\end{equation}

Since $U_1,U_2$ can decode $X,Y,Z$ in decoder $D_3$, we can apply 
\begin{eqnarray}
H(X,Y|U_1,U_2)&=&0\\
H(X,Y,Z|U_1,U_2)&=&0
\end{eqnarray}
respectively in step 4 and step 5 to get
\begin{equation}
R_1+R_2\geq H(X)+H(X,Y,Z).
\end{equation}

Finally, due to the independence of sources, we apply
\begin{equation}
H(X,Y,Z)= H(X)+H(Y)+H(Z)
\end{equation}
to get the desired inequality and complete the proof.
\end{example}

\begin{example} \label{ex:2}
A 2-level 3-encoder MDCS instance with block diagram and rate region shown in Fig.\ \ref{fig:CAPexample2}.

\begin{figure}
\centering \includegraphics[scale=.65]{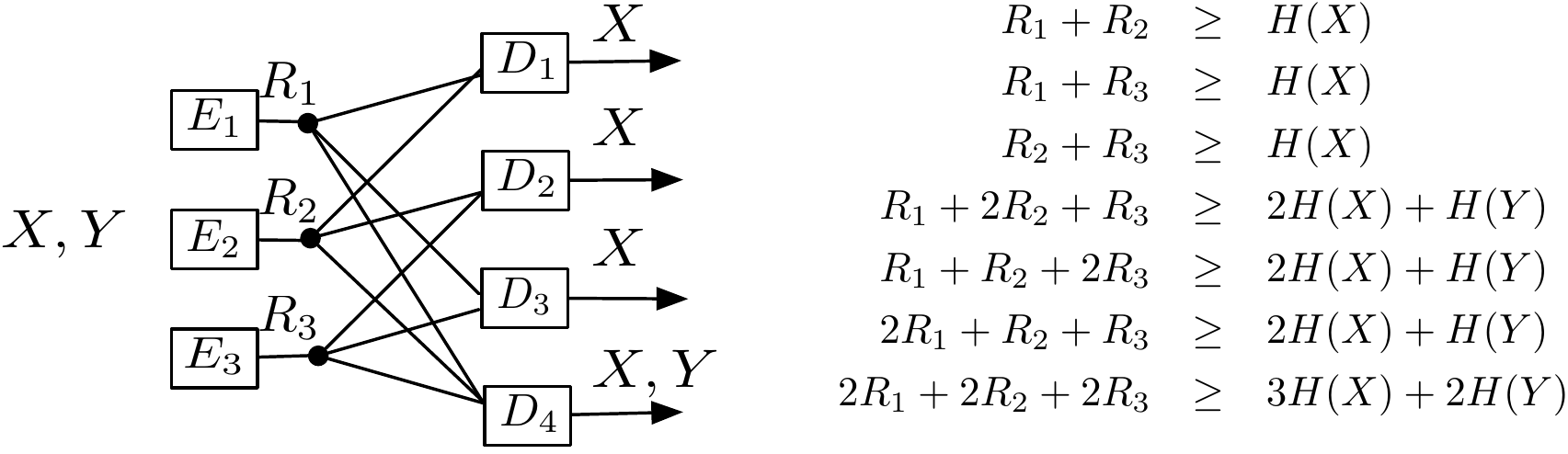} \caption{ Block diagram and rate region for Example \ref{ex:2}: a 2-level 3-encoder MDCS instance}

\label{fig:CAPexample2} 
\end{figure}

\begin{table}
\caption{\label{ineqtable2}Ordered inequalities with coefficients given by computer for Example \ref{ex:2}}
\begin{center}
\begin{tabular}{| p{1.5cm}| p{2cm} | p{4cm} |}
\hline
Step & Coefficients & Inequality or equality \\ \hline
$1$ & 2 & $R_1\geq H(U_1)$\\ \hline
$1$ & 2 &  $R_2\geq H(U_2)$ \\ \hline
$1$ & 2 &  $R_3\geq H(U_3)$ \\ \hline
$2$ & 1 & $I(U_1;U_2)\geq 0$\\ \hline
$2$ & 1 & $I(U_1;U_3)\geq 0$\\ \hline
$2$ & 1 & $I(U_2;U_3)\geq 0$\\ \hline
$3$ & 1 &  $H(X|U_1,U_2)= 0$\\ \hline
$3$ & 1 &  $H(X|U_1,U_3)= 0$\\ \hline
$3$ & 1 &  $H(X|U_2,U_3)= 0$\\ \hline
$4$ & $1/3$ &  $I(U_1;U_2,U_3|X)\geq0$\\ \hline
$4$ & $1/3$ &  $I(U_1;U_2,U_3|X)\geq0$\\ \hline
$4$ & $1/3$&  $I(U_2;U_1,U_3|X)\geq0$\\ \hline
$4$ & $1/3$&  $I(U_1;U_2|X,U_3)\geq0$\\ \hline
$4$ & $1/3$&  $I(U_1;U_3|X,U_2)\geq0$\\ \hline
$4$ & $1/3$ &  $I(U_2;U_3|X,U_1)\geq0$\\ \hline
$5$ & 2 & $H(X|U_1,U_2,U_3)= 0$\\ \hline
$6$ & 2 & $H(X,Y|U_1,U_2,U_3)=0$\\ \hline
$7$ & 2 & $H(X,Y)=H(X)+H(Y)$\\ \hline
\end{tabular}

\end{center}
\end{table}

We still pick the last inequality in the rate region as an example. The order and inequalities with coefficients for the converse proof are given in Table \ref{ineqtable2}.

Step 1, we start with inequalities with coefficients
\begin{eqnarray}
2R_1+2R_2+2R_3&\geq&2H(U_1)+2H(U_2)+H(U_3)\nonumber \\
& =&(H(U_1)+H(U_2))+(H(U_1) \nonumber \\ 
&&+H(U_3))+(H(U_2)+H(U_3)). \nonumber
\end{eqnarray}

Then in step 2, for each combined term, we apply the following three inequalities
\begin{eqnarray}
I(U_1;U_2)&\geq&0,\\
I(U_1;U_3)&\geq&0,\\
I(U_2;U_3)&\geq&0
\end{eqnarray}
respectively to get
\begin{equation}
2R_1+2R_2+2R_3\geq H(U_1,U_2)+H(U_1,U_3)+H(U_2,U_3).
\end{equation}

Due to the decoding functions of $D_1,D_2,D_3$, we can use the three equalities
\begin{eqnarray}
H(X|U_1,U_2)&=& 0,\\
H(X|U_1,U_3)&=& 0,\\
H(X|U_2,U_3)&=& 0
\end{eqnarray}
to get
\begin{equation} \label{eq:sub}
2R_1+2R_2+2R_3\geq H(X,U_1,U_2)+H(X,U_1,U_3)+H(X,U_2,U_3).
\end{equation}

Next, we notice that
\begin{eqnarray}
I(U_1;U_2,U_3|X)=H(X,U_1)+H(X,U_2,U_3)-H(X,U_1,U_2,U_3)-H(X)&\geq& 0,\label{eq:6eq1}\\
I(U_1;U_2,U_3|X)=H(X,U_1)+H(X,U_2,U_3)-H(X,U_1,U_2,U_3)-H(X)&\geq& 0,\label{eq:6eq2}\\
I(U_2;U_1,U_3|X)=H(X,U_2)+H(X,U_1,U_3)-H(X,U_1,U_2,U_3)-H(X)&\geq& 0,\label{eq:6eq3}\\
I(U_1;U_2|X,U_3)=H(X,U_1,U_3)+H(X,U_2,U_3)-H(X,U_1,U_2,U_3)-H(X,U_3)&\geq& 0,\label{eq:6eq4}\\
I(U_1;U_3|X,U_2)=H(X,U_1,U_2)+H(X,U_2,U_3)-H(X,U_1,U_2,U_3)-H(X,U_2)&\geq& 0,\label{eq:6eq5}\\
I(U_2;U_3|X,U_1)=H(X,U_1,U_2)+H(X,U_1,U_3)-H(X,U_1,U_2,U_3)-H(X,U_1)&\geq& 0.\label{eq:6eq6}
\end{eqnarray}

The summation of \eqref{eq:6eq1} -- \eqref{eq:6eq6} gives
\begin{equation}
3(H(X,U_1,U_2)+H(X,U_1,U_3)+H(X,U_2,U_3))\geq 3(H(X)+2H(X,U_1,U_2,U_3))\label{eq:sum6eq}
\end{equation}

Multiplying \eqref{eq:sum6eq} with a coefficient $1/3$, we get
\begin{equation} \label{eq:6eq}
H(X,U_1,U_2)+H(X,U_1,U_3)+H(X,U_2,U_3)\\ 
\geq H(X)+2H(X,U_1,U_2,U_3).
\end{equation}

Applying \eqref{eq:6eq} to \eqref{eq:sub}, we can get
\begin{equation}
2R_1+2R_2+2R_3\geq H(X)+2H(X,U_1,U_2,U_3).
\end{equation}

Similarly, as $U_1,U_2,U_3$ can decode $X,Y$ in decoder $D_4$, we can apply 
\begin{eqnarray}
H(X|U_1,U_2,U_3)&=&0\\
H(X,Y|U_1,U_2,U_3)&=&0
\end{eqnarray}
respectively in step 5 and step 6 to get
\begin{equation}
2R_1+2R_2+2R_3\geq H(X)+2H(X,Y).
\end{equation}

Finally, by applying the source independence equality $H(X,Y)=H(X)+H(Y)$, we get
\begin{equation}
2R_1+2R_2+2R_3\geq 3H(X)+2H(Y)
\end{equation}
to complete the proof.
\end{example}

\section{Conclusion}
\label{sec:conclusion}

This paper investigated the rate regions of $7360$ non-isomorphic MDCS instances, which represent $135043$ isomorphic instances.  Although the exact rate region of a general network coding or distributed storage instance is, in general, only expressible in terms of the (non-polyhedral) region of entropic vectors, for the class of MDCS problems considered in this paper we are able to bypass this difficulty through the use of both $i)$ the polyhedral Shannon outer bound, and $ii)$ several polyhedral inner bounds (associated with binary matroid, ternary matroid, representable matroid, and superposition regions), by showing that the associated inner and outer bounds on the rate region are equal.  Part of our contribution is the explicit construction of superposition, scalar, and vector codes over various fields using the polyhedral inner bounds.  Inspired by the notion of a forbidden matroid minor, we introduced several MDCS contraction, deletion, and unification operations and demonstrated that these operations can be used to identify a set of forbidden MDCS instances for the achievability of various classes of codes, in the sense that any extension of such a forbidden MDCS instance will likewise have a rate region for which that class of codes is insufficient.  Finally, we demonstrated that we can automate the creation of ``human readable proofs'' for the rate regions for these $7360$ non-isomorphic MDCS instances by posing a suitable optimization problem asking for the smallest cardinality set of Shannon-type inequalities required to step-by-step construct the rate region ``by hand''.  

Several remaining intriguing questions remain as future work.  One important question is whether or not the Shannon outer bound is tight for the class of MDCS instances.  For all the instances considered in this paper this has been the case, although we note there are $492$ 
 instances for which we have not yet established the rate region.  A second important question is whether or not there exists a finite list of forbidden minors for the classes of codes considered in this paper.  That is, our results have established some forbidden minor MDCS instances, but we have not established for any class of codes if we have all such forbidden minors for that class.  


\bibliographystyle{IEEEtran}
\bibliography{TransIT2013}

\end{document}